\documentclass[10pt,journal,compsoc]{IEEEtran}
\usepackage[table,xcdraw]{xcolor}

\usepackage{amssymb}
\usepackage[algoruled,vlined,linesnumbered]{algorithm2e}
\usepackage{amsthm}

\theoremstyle{definition}

\theoremstyle{remark}

\usepackage{bm}
\usepackage{cite} 
\usepackage{amsmath}
\usepackage{float}
\usepackage{graphicx}
\usepackage{subfigure}
\usepackage{xcolor}
\usepackage{hyperref}
\usepackage{algorithmic}

\usepackage{booktabs}
\usepackage{multirow}
\usepackage{array}
\newcolumntype{C}[1]{>{\centering\let\newline\\\arraybackslash\hspace{0pt}}m{#1}}
\newcolumntype{L}[1]{>{\raggedright\let\newline\\\arraybackslash\hspace{0pt}}m{#1}}

\usepackage{bbding}
\usepackage{utfsym}
\usepackage{booktabs}
\usepackage{colortbl}

\begin{document}
\title{
A Lightweight Deep Exclusion Unfolding Network for Single Image Reflection Removal
}

\author{
Jun-Jie Huang$^{\dagger}$~\IEEEmembership{Member,~IEEE},
Tianrui Liu$^{\dagger *}$,
Zihan Chen,
Xinwang Liu~\IEEEmembership{Senior Member,~IEEE}, \\
Meng Wang~\IEEEmembership{Fellow,~IEEE}, 
and Pier Luigi Dragotti~\IEEEmembership{Fellow,~IEEE}

\IEEEcompsocitemizethanks{\IEEEcompsocthanksitem J.-J. Huang, T. Liu,  Z. Chen, and X. Liu are affiliated with the College of Computer Science and Technology, National University of Defense Technology, Changsha, China. E-mail: \{jjhuang, trliu, chenzihan21, xinwangliu@nudt.edu.cn\}.
\IEEEcompsocthanksitem M. Wang is affiliated with the School of Computer Science and Information Engineering, Hefei University of Technology, Hefei, China. E-mail: eric.mengwang@gmail.com.
\IEEEcompsocthanksitem P.L. Dragotti is affiliated with the Department of Electrical and Electronic Engineering, Imperial College London, London, UK. E-mail: p.dragotti@imperial.ac.uk.}
\thanks{$^{\dagger}$These authors contributed equally to this work. $^{*}$Corresponding author.}

}

\markboth{SUBMITTED TO IEEE Trans. on Pattern Analysis and Machine Intelligence}
{}

\IEEEtitleabstractindextext{
\begin{abstract}
Single Image Reflection Removal (SIRR) is a canonical blind source separation problem and refers to the issue of separating a reflection-contaminated image into a transmission and a reflection image. The core challenge lies in minimizing the commonalities among different sources. Existing deep learning approaches either neglect the significance of feature interactions or rely on heuristically designed architectures. 
In this paper, we propose a novel Deep Exclusion unfolding Network (DExNet), a lightweight, interpretable, and effective network architecture for SIRR. DExNet is principally constructed by unfolding and parameterizing a simple iterative Sparse and Auxiliary Feature Update (i-SAFU) algorithm, which is specifically designed to solve a new model-based SIRR optimization formulation incorporating a general exclusion prior. This general exclusion prior enables the unfolded SAFU module to inherently identify and penalize commonalities between the transmission and reflection features, ensuring more accurate separation. The principled design of DExNet not only enhances its interpretability but also significantly improves its performance.
Comprehensive experiments on four benchmark datasets demonstrate that DExNet achieves state-of-the-art visual and quantitative results while utilizing only approximately 8\% of the parameters required by leading methods.
\end{abstract}

}

\maketitle

\IEEEpeerreviewmaketitle

\section{Introduction}
\label{sec:intro}

Blind Source Separation (BSS)~\cite{BSS97ICNN,BSS_PIEEE1998, HBSS10SPM,UBSS17TNNLS} 
represents a foundational challenge within the realms of signal and image processing, where the objective is to decompose
a mixed signal into its constituent source signals. 
It arises from a spectrum of real-world applications
including speech enhancement~\cite{BSSAudio_TASLP2006}, image enhancement~\cite{Kaiming_TPAMI2011}, medical data analysis~\cite{BSSMedic_2000}, and remote sensing~\cite{BSSHyper_TSP2014}.
In real-world scenarios, uncovering the underlying sources from a mixture is highly ill-posed due to the insufficient observation of the sources. It is therefore essential to exploit both the prior knowledge and the interaction mechanism of the source signals for effective source separation.

Single Image Reflection Removal (SIRR)~\cite{SIR2iccv17, Benchmark2022} is a typical BSS problem, which aims to decompose a reflection-contaminated image $\mathbf{I}$ captured through a glass into a transmission image $\mathbf{T}$ and a reflection image $\mathbf{R}$. The transmission image refers to the image content originating from the target scene on the other side of the glass, and the reflection image refers to the image content from a distinct scene that is mirrored off the glass surface.
Its image formation model~\cite{Wang23TMM} is complex
due to the multitude of interacting effects,
including light absorption, selective absorption, inner reflections, 
that come into play during
the light transport process.

%-----------------------------------------------------------
\begin{figure}[t]
\center
    {\includegraphics[width=0.88\columnwidth]{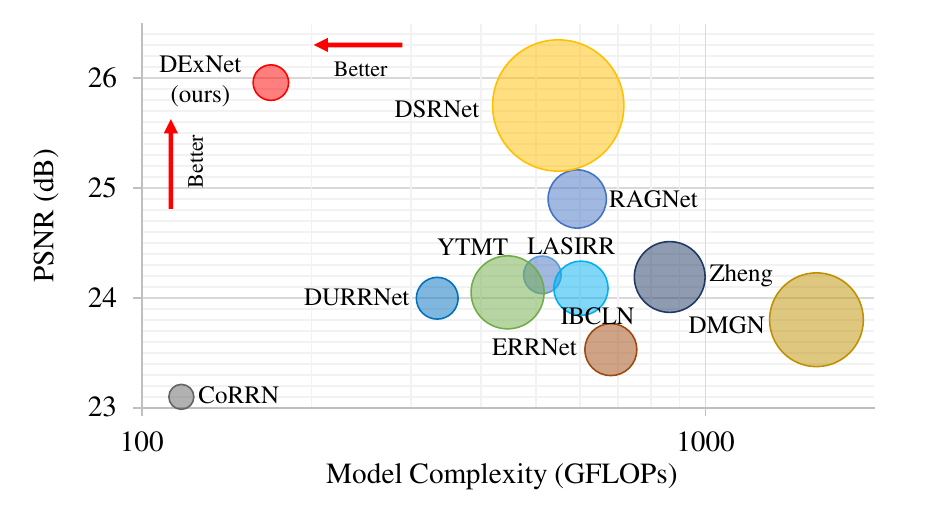}}
    \caption{Comparison of leading image reflection removal methods on the \textit{Real20} and \textit{SIR$^2$} dataset. The results are evaluated in terms of average PSNR (dB), number of parameters (M) and complexity (GFLOPs). The model size is depicted as the area of the ball.}
\label{fig:Performance}
\end{figure}
%-----------------------------------------------------------

The main objective of SIRR is to precisely decompose the reflection-contaminated image into transmission and reflection components. Such a task involves the identification and minimization of the common image contents between the decomposed images, and is known to be highly ill-posed~\cite{levin2004separating}.  
It is essential to employ a ``simple but not simpler" image formation model~\cite{Kong13TPAMI, RC_TPAMI2023, beyond_linear_2019, Zhang20CVPR, Wang23TMM, Chen24TIP} to precisely account for the nonlinear effects and develop a mechanism that can naturally suppress the emergence of common contents in the estimated transmission and reflection images.

In this paper, we introduce Deep Exclusion unfolding Network (DExNet) a lightweight, interpretable, and effective deep network architecture for single image reflection removal.  
The DExNet architecture is designed by deep unfolding an innovative iterative algorithm tailored for a new SIRR optimization problem that incorporates a general exclusion prior. This prior identifies and penalizes commonalities between the features of the two image layers, thereby enhancing feature disentanglement. The working principles of DExNet’s key network units align with those of the model-based iterative algorithm, leading to enhanced interpretability and controllability. Specifically, we initially formulate the SIRR problem with Convolutional Sparse Coding (CSC), augmented with the general exclusion prior. While the additional exclusion prior is effective in minimizing feature commonality between transmission and reflection images, it also introduces optimization difficulties.
To address this, we introduce an iterative Sparse and Auxiliary Feature Update (i-SAFU) algorithm, enhancing sparse codes and auxiliary features through proximal gradient descent. Subsequently, we extend i-SAFU into a multi-scale DExNet, featuring a key Sparse and Auxiliary Feature Update (SAFU) module fostering rich feature interactions between transmission and reflection images.
Facilitated by the model-driven deep network structure, the proposed DExNet enables accurate image formation modelling, effective feature interaction, and robust prior learning. Therefore, DExNet is not only more effective, but is also with a high degree of interpretability.

Fig.~\ref{fig:Performance} compares the average PSNR (dB), model complexity, and number of parameters of different SIRR methods evaluated on 474 images from \textit{Set20}~\cite{perceptual_loss_2018} dataset and \textit{SIR$^2$}~\cite{SIR2iccv17} dataset. We can see that the proposed DExNet achieves better performance with a smaller model and lower complexity compared to other state-of-the-art single image reflection removal methods.

The contribution of this paper is three-fold:
\begin{itemize}
    \item This paper introduces a new model-based SIRR optimization formulation enriched with a general exclusion prior to promote accurate image estimation and commonalities reduction, and proposes to solve it using a novel iterative Sparse and Auxiliary Feature Update (i-SAFU) algorithm.

    \item We derive a lightweight DExNet from i-SAFU algorithm in a principled manner through deep unfolding. The model-inspired blocks in the key SAFU module effectively enhance the model performance with low parameter footprint.

    \item Extensive experiments demonstrate that the proposed lightweight DExNet, with approximately 8\% of the leading method's model parameters, achieves superior performance over existing SOTA methods in terms of average PSNR, model complexity, and parameter efficiency. 
   
\end{itemize}

A preliminary version of this work, titled Deep Unfolded Reflection Removal Network (DURRNet), was published at ICASSP 2024~\cite{Huang24DURRNet}. Beyond the conference version, we present significant improvements on optimization formulation, deep network design and model performance. In terms of optimization formulation, DExNet improves the formation modeling with a nonlinear residual image and incorporates a general exclusion prior instead of a wavelet-based exclusion prior. Therefore, DExNet has a lighter network for feature interactions because it does not require the use of Invertible Neural Networks as learnable wavelet transform. As a result, the proposed DExNet outperform DURRNet by around 2dB in PSNR while it has only around 80\% of model parameters and 50\% of model complexity in GFLOPs with respect to DURRNet.

The rest of the paper is organized as follows: Section~\ref{sec:RelatedWork} discusses related works in Single Image Reflection Removal and Deep Unfolding. 
Section~\ref{sec:method} presents the details of Deep Exclusion  unfolding Network (DExNet) including model formulation, optimization algorithm design and the complete DExNet architecture. Section~\ref{sec:results} introduces the implementation details and presents the experimental results and comparisons. Finally in Section~\ref{sec:conclusion}, we conclude the paper  and discuss the future directions.

\section{Related Work}
\label{sec:RelatedWork}
\subsection{Model-based SIRR Methods}

Model-based SIRR methods~\cite{levin2004separating,levin2007user,li2014single,guo2014robust,ghost_cues_2015,reflect_suppression2017,fast_convex_2019,wan2017sparsity,RegionAware2018Wan, han2017reflection,Mirror2019Daneil} formulate the single image reflection removal problem as a tractable optimization problem and design efficient and robust iterative algorithms to solve it. 
The optimization problem usually consists of a reconstruction term, and in order to ease the optimization a linear image formation model is usually adopted.
To constrain the solutions, different priors have been exploited, including, for example, sparsity prior, relative smoothness prior, low-rank prior and non-local image prior.

The gradient sparsity prior of natural images has been exploited in~\cite{levin2004separating,levin2007user} by assuming that edges and corners are sparse in natural images. The relative smoothness prior has been proposed in~\cite{li2014single} since the reflected image is usually more blurred than the transmission image. 
Guo \textit{et al.}~\cite{guo2014robust} propose a robust method for reflection separation from {multiple images} by exploiting the correlation across multiple images, and the sparsity prior and independence prior of two image layers’ edge responses.
Li and Brown~\cite{fast_convex_2019} propose a convex model, which implies a partial differential equation with gradient thresholding, to suppress the reflection from a single input image. The Laplacian fidelity prior and the $l_0$ gradient sparsity prior have been used in~\cite{reflect_suppression2017} to formulate the optimization problem for reflection suppression. In~\cite{ghost_cues_2015}, Gaussian Mixture Model (GMM) has been applied for modelling patch prior to exploit the ghosting effects on reflection. 
Wan~\textit{et. al.}~\cite{wan2017sparsity} propose a reflection removal method based on a sparse representation framework with non-local image prior.
Wan \textit{et al.}~\cite{RegionAware2018Wan} propose a region-aware single image reflection removal method with a unified content and gradient prior.
Han \textit{et al.}~\cite{han2017reflection} propose a reflection removal algorithm based on low-rank matrix completion for separating reflection in multiple images taken at near camera locations.
Heydecker \textit{et al.}~\cite{Mirror2019Daneil} propose an optimization-based single image reflection removal algorithm with simple user interaction scheme and a H$^2$ fidelity term to encourage recovering details and enforcing global color similarity.

In general, model-based image reflection removal methods lead to interpretable mathematical formulations but the final results may not be satisfactory when strong and complex reflections are present, especially when compared with recent methods based on deep learning.

\subsection{Learning-based SIRR Methods}

Recently, learning-based approaches~\cite{wan2017benchmarking,RRCNN2019Chang,kim2020single,hong2023PAR2Net,wan2019corrn,generic_smooth_2017,wan2018crrn,wei2019single,prasad2021v,song2023robust,
perceptual_loss_2018,beyond_linear_2019,yang2018seeing,dong2021location,FengTIP2021,LiAI2023,cascaded_Refine_2020,hu2021trash,Shao2021MoGSIRR,zhang2022content,Hu_2023_ICCV,Huang24DURRNet}, especially those that employ Deep Neural Networks (DNNs), have emerged to complement traditional model-based single image reflection removal methods and achieve better performances. 
Novel deep network architectures have been proposed by taking different estimation strategies including 
i) directly estimating the transmission and reflection images,
ii) alternatively estimating the two images,
and iii) cooperatively estimating the two images.

Wan \textit{et al.}~\cite{wan2018crrn} propose a Concurrent Reflection Removal Network (CRRN) with two cooperative gradient inference network and image inference network.
Enhanced Reflection Removal Network (ERRNet)~\cite{wei2019single} proposes to learn image reflection removal network with Channel Attention to estimate directly the transmission image. Furthermore, ERRNet utilizes an alignment-invariant loss to learn from misaligned training data. 
Prasad \textit{et al.}~\cite{prasad2021v} propose a light-weight VDESIRR model with a scale sub-network and a progressive inference stage to process higher resolution scales.
Song~\textit{et al.} \cite{song2023robust} investigate the robust deep learning based single image reflection removal methods against adversarial attacks. 
Wen \textit{et al.}~\cite{beyond_linear_2019} propose to inject non-linearity into the image formation model by using a synthetic network to synthesize the reflection-contaminated image using a alpha blending mask and to a reflection removal network to jointly estimate the transmission and reflection images.
Zhang \textit{et al.}~\cite{perceptual_loss_2018} propose a convolutional network with a perceptual loss and an exclusion loss for single image reflection removal.

Bi-Directional Network (BDN)~\cite{yang2018seeing} consists of a cascaded deep network and has been proposed to alternatively estimate the reflection image and then the transmission image. 
Iterative Boost Convolutional LSTM Network (IBCLN)~\cite{cascaded_Refine_2020} has been proposed to progressively separate the reflection-contaminated image into two image layers. 
Dong \textit{et al.}~\cite{dong2021location} propose a Location-Aware Single Image Reflection Removal (LASIRR) method by leveraging a learned Laplacian feature to locate and remove reflections. 
Feng~\textit{et al.}~\cite{FengTIP2021} propose a Deep-Masking Generative Network (DMGN) for single image reflection removal using a coarse-to-fine approach with a novel Residual Deep-Masking Cell to control the information propagation.
Li~\textit{et al.}~\cite{LiAI2023} propose a two-stage Reflection-Aware Guidance Network (RAGNet) by first estimating the reflection layer then the transmission layer.

Hu \textit{et al.}~\cite{hu2021trash} propose a two-stage ``Your Trash is My Treasure" (YTMT) method with a dual-stream decomposition network to enable information exchange at different branches. 
Most recently, He \textit{et al.}~\cite{Hu_2023_ICCV} proposed a Dual-stream Semantic-aware network with Residual correction (DSRNet) with a novel learnable residual module  to compensate the non-linearity nature of the forward model and a Mutually-Gated Interactive (MuGI) block to enable feature interaction of two image layers.
There are also deep unfolding methods for single image reflection removal~\cite{Shao2021MoGSIRR,zhang2022content,Huang24DURRNet}.

\subsection{Deep Unfolding Methods}

Deep unfolding (or deep unrolling) methods~\cite{gregor2010learning, monga2021algorithm} aim to combine the merits of model-based and deep-learning methods by embedding the physical prior of the specific task into deep neural architecture, therefore leading to effective and more interpretable models.
In the existing deep unfolding methods, the proximal operator of the prior term for each individual variable is usually parameterized as a deep neural network to enable learning from training data.

In the seminal work \cite{gregor2010learning}, Gregor and LeCun propose to convert the Iterative Shrinkage-Thresholding Algorithm (ISTA) into a deep network by setting the dictionaries in ISTA as learnable parameters for sparse coding. Thereafter, many methods which use deep unfolding for solving inverse problems in image/signal processing tasks emerged.
ADMM-Net~\cite{yang2016deep} has been proposed to unfold the Alternating Direction Method of Multipliers (ADMM) algorithm for compressive sensing Magnetic Resonance Imaging (MRI) reconstruction. 
Deep Unfolding single image Super-Resolution Network (USRNet)~\cite{zhang2020deep} has been proposed for single image super-resolution task by unfolding Maximum A Posteriori (MAP) formulation via a Half-Quadratic Splitting (HQS) algorithm. 
Deep Unrolling method for Blind Deblurring (DUBLID) network~\cite{li2020efficient} has been proposed which unfolds a Total Variation (TV) based blind deconvolution algorithm with a small number of learnable filter only. 
For the rain removal task, Wang \textit{et al.}~\cite{model_driven_rain_2020} propose a model-inspired Rain Convolutional Dictionary Network (RCDNet) based on proximal gradient descent to simplify the computation for convolution operators.
Zhang~\textit{et al.}~\cite{AMPNetTIP2021} propose AMP-Net by unrolling the iterative denoising process of the Approximate Message Passing (AMP) algorithm for visual image Compressive Sensing (CS) problem.
Pu \textit{et al.}~\cite{pu2022mixed} propose a self-supervised deep unfolding network for separating X-ray images of artworks into a surface X-ray image and a hidden X-ray image.
Aberdam~\textit{et al.}~\cite{AdaLISTA_TPAMI2022} propose an adaptive deep unfolding learned solver for sparse coding dubbed Ada-LISTA with unfixed dictionary.
Huang~\textit{et al.}~\cite{DGSM_TPAMI2023} propose to unfold Maximum a Posterior (MAP) estimation algorithm with a learned Gaussian Scale Mixture prior for image restoration.
Deng~\textit{et al.}~\cite{DeepM2CDL_TPAMI2023} propose a novel DeepM$^2$CDL network by unfolding a multi-scale multi-modal convolutional dictionary learning model for multi-modal image restoration and fusion.
Liu and Dragotti~\cite{V2E2V_TPAMI2023} propose a deep unfolding convolutional ISTA network with Long Short-term Temporal Consistency (LSTC) constraints for Events-to-Video reconstruction.

There are also deep unfolding methods for reflection removal, however, these methods tackled reflection removal task with a traditional approach for deep unfolding by modeling independent priors for each image layer and somewhat overlooked the importance of the interplay between the common image contents in two image layers.
Shao \textit{et al.}~\cite{Shao2021MoGSIRR} propose a Model-Guided SIRR network (MoG-SIRR) by unfolding a reflection removal formulation with a
non-local autoregressive prior and a dereflection prior. 
Zhang \textit{et al.}~\cite{zhang2022content} propose a Content and Gradient-guided Deep Network (CGDNet) for SIRR by alternatively updating the content and the gradient of the transmission and reflection image. Huang~\textit{et al.}~\cite{Huang24DURRNet} propose a Deep Unfolded Reflection Removal Network (DURRNet) which leverages Invertible Neural Networks~\cite{huang2021linn,huang2021winnet} for evaluating the exclusion condition.

\section{Proposed Method}
\label{sec:method}

%-----------------------------------------------
\begin{table}[]
    \caption{{The notations used in this paper. {We use letters for scalars, bold lowercase letters for vectors and bold capital letters for matrices or tensors.} }}
    \centering
    \begin{tabular}{|c|L{6cm}|}
    \hline
        \textbf{Symbol} & \textbf{Definition} \\ \hline 
        $\otimes$                                   & Convolution operator  \\ 
        $\otimes^T$                                 & Transposed convolution operator  \\ 
        $\odot$                                     & Element-wise multiplication \\
        $\copyright$                                & Concatenate operator \\                          
        $\mathbf{I}$                                & Input reflection-contaminated image\\
        $\mathbf{T},\mathbf{R},\mathbf{N}$                     & Transmission, reflection and nonlinear residual image\\
        $\{ \mathbf{D}_{i} \}_{i=\mathbf{T},\mathbf{R},\mathbf{N}}$           & Convolutional dictionary \textit{w.r.t.} $\mathbf{T},\mathbf{R},\mathbf{N}$\\
        $\mathbf{W}$           & Convolutional filter for common feature extraction\\
        $w$                     & Kernel size of the convolutional filters\\
        $n$                     & Number of convolutional filters in $\mathbf{D}_{i}$\\
        $m$                     & Number of convolutional filters in $\mathbf{W}$\\
        $\{ \mathbf{M}_{i} \}_{i=\mathbf{T},\mathbf{R}}$           & Convolutional dictionary \textit{w.r.t.} $\mathbf{T},\mathbf{R}$ with $\mathbf{W} \otimes \mathbf{D}_{i} $\\
        $\{ \mathbf{z}_{i} \}_{i=\mathbf{T},\mathbf{R},\mathbf{N}}$           & Sparse feature \textit{w.r.t.} the $\mathbf{T},\mathbf{R},\mathbf{N}$\\
        $ \mathbf{z}_{A}$           & Feature of the auxiliary variable\\
        $\{ {p}_{i}(\cdot) \}_{i=\mathbf{T},\mathbf{R},\mathbf{N}}$           & Prior term \textit{w.r.t.} $\mathbf{T},\mathbf{R},\mathbf{N}$\\
        $ {p}_{E}(\cdot)$           & Exclusion prior term \\
        $\text{prox}_{\lambda}(\cdot)$                     & Proximal operator\\
        $S$     & Number of scales of SAFU module in DExNet \\
        $K$     & Number of stages in each scale\\
        $\mathbf{X}_{i}^{k}$     & The representation $\mathbf{X}$ \textit{w.r.t.} image $i \in \{\mathbf{T},\mathbf{R},\mathbf{N}\}$ at $k$-th stages. \\
        \hline
    \end{tabular}
    
    \label{tab:notation}
\end{table}
%-----------------------------------------------

The goal of this work is to design an effective deep architecture for the typical blind source separation problem, \textit{i.e.,} single image reflection removal (SIRR). 
The essence of blind source separation lies in the minimization of commonalities
among the estimated sources. Therefore, enabling interactions to reduce common contents between image layers is essential for effective reflection separation. This is something which is often overlooked by the existing deep unfolding approaches~\cite{Shao2021MoGSIRR, zhang2022content}. 
To close this gap,
we propose a highly effective and parameter efficient Deep Exclusion  unfolding Network (DExNet).
DexNet is rooted in
an iterative algorithm which is specifically designed for a Convolutional Sparse Coding problem with the incorporation of a general exclusion prior which facilitates the minimization of common features. 
Through unfolding, the key network modules in DExNet enable a frequent and 
dynamic interplay of the common contents between image layers, which is instrumental in achieving a more precise and reliable reflection separation.

In what follows, we first introduce the proposed model-based optimization formulation for single image reflection separation, then design an iterative algorithm for the optimization problem based on proximal gradient descent, finally we present the proposed DExNet and the training strategy. For clarity, we highlight some frequently used notations in Table \ref{tab:notation}.

\subsection{Model Formulation and Optimization Problem}
\label{sec:formulation}

\subsubsection{Model Formulation with a Residual Image}

A common and simplified formation model for SIRR assumes that the reflection-contaminated image is a linear combination of a transmission image and a reflection image~\cite{wan2019corrn, hu2021trash, Shao2021MoGSIRR, Huang24DURRNet}. This formulation, however, does not capture in full the non-linear optical interactions due to glass reflection.
Inspired by Hu \textit{et al.}~\cite{Hu_2023_ICCV}, in this work we use a ``simple but not simpler" image formation model with an additional
nonlinear residual image $\mathbf{N}$ term which compensates the non-linear effects in real-world scenarios. Hence, the image formation model can be expressed as:
\begin{equation}
    \mathbf{I} = \mathbf{T} + \mathbf{R} + \mathbf{N},
    \label{eq:model}
\end{equation}
where $\mathbf{I} \in \mathbb{R}^{W \times H \times 3}$ is the reflection-contaminated image, $\mathbf{T} \in \mathbb{R}^{W \times H \times 3}$ is the transmission image, $\mathbf{R} \in \mathbb{R}^{W \times H \times 3}$ is the reflection image, $\mathbf{N} \in \mathbb{R}^{W \times H \times 3}$ is the residual image, with $W$ and $H$ denoting the width and height of the image.

Sparse representation provides an effective prior for inverse problems.
By imposing a sparsity prior on the representations, we can recast the image estimation problem as a more tractable sparse coding problem
over a convolutional dictionary.
Normally, the transmission image and reflection image exhibit distinct image characteristics. The camera primarily captures the transmission image with greater clarity, manifesting in sharper edges and more details,
while in contrast the reflection image is often obscured by the glass surface and tends to have a more blurred appearance. 
Considering that, we model $\mathbf{T}, \mathbf{R}, \mathbf{N}$ as each having a sparse representation over a corresponding convolutional dictionary. Hence, Eqn. (\ref{eq:model}) can be represented as:
\begin{equation}
    \mathbf{I} = \sum_{i \in \{ \mathbf{T}, \mathbf{R}, \mathbf{N} \}} \mathbf{D}_i \otimes \mathbf{z}_i,
\end{equation}
where $\mathbf{D}_i \in \mathbb{R}^{w \times w \times 3 \times n} $ are convolutional dictionaries with spatial kernel size $w \times w$ and $\mathbf{z}_i \in \mathbb{R}^{W \times H \times n}$ are the sparse features for $i \in \{ \mathbf{T}, \mathbf{R}, \mathbf{N} \}$. The width, height and number of channels of $\mathbf{z}_i$ are denoted as $W$, $H$, and $n$, respectively. Here, $\otimes$ denotes the convolution operator.

The sparse representation for the transmission, reflection and residual image, \textit{i.e.} $\mathbf{z}_\mathbf{T}, \mathbf{z}_\mathbf{R}, \mathbf{z}_\mathbf{N}$, can be estimated via a Convolutional Sparse Coding (CSC)~\cite{papyan2017convolutional,bristow2013fast} optimization as follows:
\begin{equation}
    \underset{ \{ \mathbf{z}_\mathbf{T}, \mathbf{z}_\mathbf{R}, \mathbf{z}_\mathbf{N} \}}{\arg\min} \frac{1}{2} \left\Vert \mathbf{I} - \sum_{i \in \{ \mathbf{T}, \mathbf{R}, \mathbf{N} \}} \mathbf{D}_i \otimes \mathbf{z}_i \right\Vert_F^2 + \sum_{i \in \{ \mathbf{T}, \mathbf{R}, \mathbf{N} \}} \lambda_i p_i(\mathbf{z}_i),
    \label{eq:csc_p0}
\end{equation}
where $\lambda_i$ are regularization parameters and $p_i(\cdot)$ is a sparse prior term, for example, $L_1$ norm, used to model the sparsity of feature $\mathbf{z}_i$.

\subsubsection{Optimization with Exclusion Prior}

The optimization problem in Eqn. (\ref{eq:csc_p0}) only considers a sparse prior for each individual image. 
For blind source separation problems, it is essential to 
penalize the commonality 
between estimated sources. 
The exclusion loss~\cite{perceptual_loss_2018} has proven to be an effective unsupervised training loss term to facilitate commonality reduction. 
However, it is non-trivial to incorporate the exclusion loss in the optimization problem in Eqn. (\ref{eq:csc_p0}) due to the nonlinear $\tanh(\cdot)$ function. 

Here, we propose to incorporate a general Exclusion Prior~\cite{pu2022mixed} (which has a form similar to the independence prior in \cite{guo2014robust}) in the optimization problem $p_{E} \left( \left( \mathbf{W}  \otimes \mathbf{T} \right)  \odot \left( \mathbf{W} \otimes \mathbf{R}  \right) \right)$,
where $\mathbf{W} \in \mathbb{R}^{w \times w \times m \times 3}$ represents learnable convolution filters that generalize the gradient operator, and $p_{E}(\cdot)$ is a penalty function to penalize the common features.
The exclusion prior extracts and identifies common features from two images through convolution and element-wise multiplication. Minimizing over the exclusion prior encourages a minimization of the common image contents which is the essence of SIRR.

With this general exclusion prior, the optimization problem can then be reformulated as:
\begin{equation}
    \begin{aligned}
     \underset{ \{\mathbf{z}_\mathbf{T}, \mathbf{z}_\mathbf{R}, \mathbf{z}_\mathbf{N} \}}{\arg\min} & \frac{1}{2} \left\Vert \mathbf{I} - \sum_{i \in \{ \mathbf{T}, \mathbf{R}, \mathbf{N} \}} \mathbf{D}_i \otimes \mathbf{z}_i \right\Vert_F^2 + \sum_{i \in \{ \mathbf{T}, \mathbf{R}, \mathbf{N} \}} \lambda_i p_i(\mathbf{z}_i)\\
    & + \kappa p_{E} \left( \left( \mathbf{M}_\mathbf{T} \otimes \mathbf{z}_\mathbf{T} \right) \odot \left( \mathbf{M}_\mathbf{R} \otimes \mathbf{z}_\mathbf{R} \right) \right),
    \end{aligned}
    \label{eq:opt-model}
\end{equation}
where $\kappa$ is the regularization parameter, and $\mathbf{M}_i = \mathbf{W} \otimes \mathbf{D}_i$ ($i \in \{ {\mathbf{T}, \mathbf{R}} \}$).

In contrast to the conventional deep unfolding approaches that incorporate prior term focused solely on a single variable, the exclusion prior involves both $\mathbf{z}_\mathbf{T}$ and $\mathbf{z}_\mathbf{R}$ and considers interactions between the two variables. 
This fact makes the optimization problem more challenging and necessitates a carefully designed strategy for effective resolution.

To address this issue, we introduce an auxiliary variable $\mathbf{z}_\mathbf{A}$$= \left( \mathbf{M}_\mathbf{T} \otimes \mathbf{z}_\mathbf{T} \right) \odot \left( \mathbf{M}_\mathbf{R} \otimes \mathbf{z}_\mathbf{R} \right)$ that decouples $\mathbf{z}_\mathbf{T}$ and $\mathbf{z}_\mathbf{R}$ from the exclusion prior. This allows us to design an effective iterative algorithm that solves simpler sub-problems with closed-form solutions while retains rich feature interactions. 
Based on Half-Quadratic Splitting (HQS) algorithm, we reformulate the problem in Eqn. (\ref{eq:opt-model}) as a minimization problem involving $\mathbf{z}_\mathbf{T}, \mathbf{z}_\mathbf{R}, \mathbf{z}_\mathbf{N}$ and $\mathbf{z}_\mathbf{A}$, accompanied by an additional auxiliary equality term:
\begin{equation}
    \begin{aligned}
        \underset{ \{\mathbf{z}_\mathbf{T}, \mathbf{z}_\mathbf{R}, \mathbf{z}_\mathbf{N}, \mathbf{z}_\mathbf{A} \} }{\arg\min} & 
        \underbrace{ \frac{1}{2} \Vert \mathbf{I} - \sum_{i \in \{ {\mathbf{T}, \mathbf{R}, \mathbf{N}} \}} \mathbf{D}_i \otimes \mathbf{z}_i \Vert_F^2}_{ \text{$g(\cdot)$: reconstruction term}}
         \\
        & + \underbrace{\frac{\tau}{2} \Vert \mathbf{z}_\mathbf{A} - \left( \mathbf{M}_\mathbf{T} \otimes \mathbf{z}_\mathbf{T} \right) \odot \left( \mathbf{M}_\mathbf{R} \otimes \mathbf{z}_\mathbf{R} \right) \Vert_F^2}_{\text{$q(\cdot)$: auxiliary equality term}}\\
        &+ \sum_{i \in \{ {\mathbf{T}, \mathbf{R}, \mathbf{N}} \}} \lambda_i p_i(\mathbf{z}_i) + \kappa p_{E}(  \mathbf{z}_\mathbf{A} ),  \\
    \end{aligned}
    \label{eq:HQS}
\end{equation}
where $\tau$ is a regularization parameter.

%-----------------------------------------------------------
\begin{algorithm}[t]
    \SetAlgoLined
    \textbf{Input:} Input image $\mathbf{I}$, dictionaries $\{ \mathbf{D}_{i} \}_{i=\mathbf{T},\mathbf{R},\mathbf{N}}$ and $\{ \mathbf{W}_{i} \}_{i=\mathbf{T},\mathbf{R}}$;
    
    \textbf{Initialize:} $k=1$;
    
    \While{halting criterion false}{
        \textcolor{green}{\# Sparse Features Update}
        
        Update $\mathbf{z}_\mathbf{T}^{{k+1}}$ with fixed $\mathbf{z}_\mathbf{T}^{{k}}, \mathbf{z}_\mathbf{N}^{{k}}, \mathbf{z}_\mathbf{A}^{{k}}$;% \textit{w.r.t.} Eq. (\ref{eq:solveZT});

        Update $\mathbf{z}_\mathbf{R}^{{k+1}}$ with fixed $\mathbf{z}_\mathbf{T}^{{k+1}}, \mathbf{z}_\mathbf{N}^{{k}}, \mathbf{z}_\mathbf{A}^{{k}}$;%\textit{w.r.t.} Eqn. (\ref{eq:solveZR});

        Update $\mathbf{z}_\mathbf{N}^{{k+1}}$ with fixed $\mathbf{z}_\mathbf{T}^{{k+1}}, \mathbf{z}_\mathbf{R}^{{k}}$;%\textit{w.r.t.} Eqn. (\ref{eq:solveT}); , \mathbf{z}_\mathbf{A}^{k}

        \textcolor{green}{\# Auxiliary Variable Update }

        Update $\mathbf{z}_\mathbf{A}^{{k+1}}$ with fixed $\mathbf{z}_\mathbf{T}^{{k+1}}, \mathbf{z}_\mathbf{R}^{{k+1}}$;%\textit{w.r.t.} Eq. (\ref{eq:solveR}); , \mathbf{z}_\mathbf{N}^{k+1}

        $k \leftarrow k + 1$ \;
    }
    
    $\hat{\mathbf{T}}, \hat{\mathbf{R}}, \hat{\mathbf{N}}  \leftarrow  \mathbf{D}_\mathbf{T} \otimes \mathbf{z}_\mathbf{T}^{{k}}, \mathbf{D}_\mathbf{R} \otimes \mathbf{z}_\mathbf{R}^{{k}}, \mathbf{D}_\mathbf{N} \otimes \mathbf{z}_\mathbf{N}^{{k}}$
    
    \textbf{Output:} Estimated transmission image $\hat{\mathbf{T}}$ and reflection image $\hat{\mathbf{R}}$.
 \caption{Proposed i-SAFU Algorithm for SIRR}
 \label{algorithm}
\end{algorithm}
%-----------------------------------------------------------

%-----------------------------------------------------------
\begin{figure*}[t]
    \centering
    \includegraphics[width=0.9\textwidth]{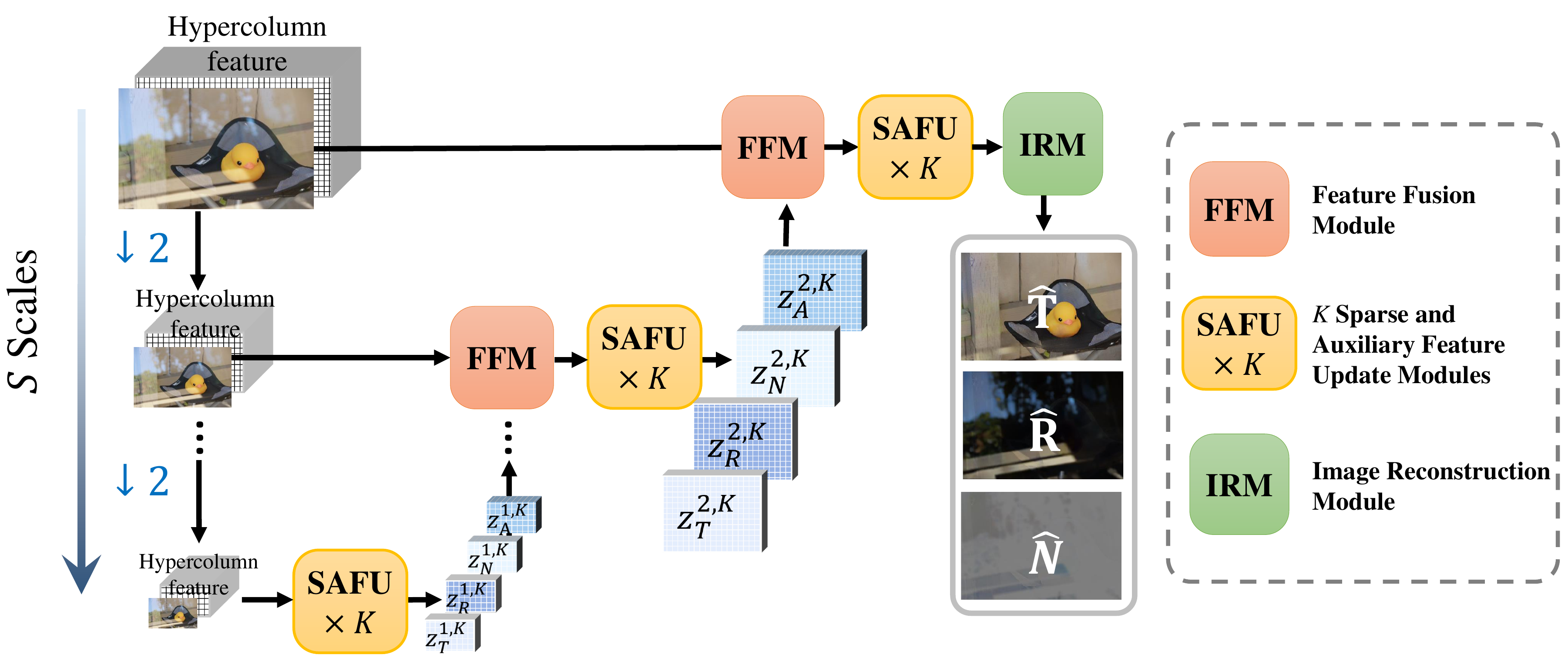} 
\caption{Overview of the proposed Deep Exclusion  unfolding Network (DExNet). DExNet consists of $S$ progressively refinement scales to estimate the transmission and the reflection images in a coarse-to-fine manner.
Each scale contains $K$ stages of Sparse and Auxiliary Feature Update (SAFU) module whose network architecture has one-to-one correspondence to the proposed alternating minimization i-SAFU algorithm and therefore is with high model interpretability. 
For each scale $s>1$, the features from the lower scale are fused with the counterparts of the upper scale with a Feature Fusion Module (FFM).
Finally, the features of the transmission image, the reflection image and the nonlinear residual image are mapped to image domain with an Image Reconstruction Module (IRM). 
Here $\downarrow 2$ denotes bilinear interpolation by a factor of 0.5}.
\label{fig:DExNet_overview}
\end{figure*}
%-----------------------------------------------------------

\subsubsection{Iterative Sparse and Auxiliary Feature Update (I-SAFU) Algorithm}
\label{sec:i-SAFU}
We propose an iterative Sparse and Auxiliary Feature Update (i-SAFU) algorithm to sequentially update the sparse features $\mathbf{z}_\mathbf{T}, \mathbf{z}_\mathbf{R}, {\mathbf{z}}_\mathbf{N}$ and the auxiliary variable $\mathbf{z}_\mathbf{A}$ based on the Proximal Gradient Descent (PGD) approach \cite{beck2009fast,model_driven_rain_2020}. The pseudo-code of the proposed i-SAFU algorithm for single image reflection removal is given in \textbf{Algorithm \ref{algorithm}}.

We will see in the next section that the introduction of the general exclusion prior and the auxiliary variable will inherently induce a deep network architecture with rich interactions between transmission and reflection features through deep unfolding.

\subsection{Deep Exclusion Unfolding SIRR Network}
\label{sec:network}

In this section, we present DExNet, a deep network architecture derived from unfolding the i-SAFU algorithm. 
In the following, Section 3.2.1 provides an overview of the DExNet framework. Section 3.2.2 details the Sparse and Auxiliary Feature Update (SAFU) module, which unfolds the i-SAFU algorithm into a network module with four update units for transmission, reflection, nonlinear residual, and auxiliary features. Finally, Section 3.2.3 and Section 3.2.4 describes the Feature Fusion Module (FFM) and the Image Reconstruction Module (IRM), respectively. 

{\subsubsection{Framework Overview}

Fig.~\ref{fig:DExNet_overview} illustrates the overview of DExNet, which employs $S$ progressive refinement scales.
At each scale, the framework leverages $K$ cascaded SAFU modules. Each scale $s>1$ also incorporates a FFM to adaptively integrate lower-scale features with higher-scale counterparts. The SAFU modules serve as the core modules which facilitate accurate image formation modeling, effective feature interaction, and robust prior learning.
Each SAFU module mathematically unfolds one iteration of the i-SAFU algorithm to perform alternating updates on the sparse features, \textit{i.e.}, $\mathbf{z}_\mathbf{T}^{s, k}$, $\mathbf{z}_\mathbf{R}^{s, k}$, ${\mathbf{z}}_\mathbf{N}^{s, k}$, and the auxiliary variable ${\mathbf{z}}_\mathbf{A}^{s, k}$.  
The superscripts $s,k$ denote the $k$-th iteration at scale $s \in [1, S]$, with $\mathbf{z}^{s,0}$ representing initial input features.
The final reconstruction phase utilizes an IRM to transform the updated features to the estimated transmission image $\mathbf{\hat{T}}$, reflection image $\mathbf{\hat{R}}$ and nonlinear residual image $\mathbf{\hat{N}}$.

\subsubsection{Sparse and Auxiliary Feature Update (SAFU) Module} 
\label{sec:SAFU}
The SAFU module is designed by unfolding one iteration of
i-SAFU algorithm.
An overview of $K$ SAFU modules is depicted in Fig.~\ref{fig:SAFU module}. 
There are four basic units, namely a Transmission Feature Update (TFU) unit, a Reflection Feature Update (RFU) unit, a Nonlinear Feature Update (NFU) unit and an Auxiliary Feature Update (AFU) unit, to alternatively update the transmission feature $\mathbf{z}_\mathbf{T}$, the reflection feature $\mathbf{z}_\mathbf{R}$, the nonlinear residual feature $\mathbf{z}_\mathbf{N}$, and the auxiliary feature $\mathbf{z}_\mathbf{A}$, respectively.

In the following, we delve into the design principles and the functionality of each basic unit of the SAFU module. Note that here we introduce the SAFU module of a particular scale therefore we omit the scale notation $s$ used in this section since the approach is similar for other scales.

\textbf{Transmission Feature Update (TFU) unit:} 
TFU unit is designed to update the features from the transmission image $\mathbf{z}_\mathbf{T}$ by fixing the other three variables, \textit{i.e.} $\mathbf{z}_\mathbf{R}$, $\mathbf{z}_\mathbf{N}$ and $\mathbf{z}_\mathbf{A}$. By keeping the terms related to $\mathbf{z}_\mathbf{T}$ from Eqn. (\ref{eq:HQS}), we establish a sub-problem for optimizing $\mathbf{z}_\mathbf{T}$ as follow:
\begin{equation}
    \begin{aligned}
        \underset{\mathbf{z}_\mathbf{T}}{\min} 
        &\frac{1}{2} \Vert \mathbf{I} - \sum_{i \in \{ {{\mathbf{T}, \mathbf{R}, \mathbf{N}}} \}} \mathbf{D}_i \otimes \mathbf{z}_i \Vert_F^2 + \lambda_\mathbf{T} p_\mathbf{T}({\mathbf{z}}_\mathbf{T}) \\
        & + \frac{\tau}{2} \Vert \mathbf{z}_\mathbf{A} - \left( \mathbf{M}_\mathbf{T} \otimes \mathbf{z}_\mathbf{T} \right) \odot \left( \mathbf{M}_\mathbf{R} \otimes \mathbf{z}_\mathbf{R} \right) \Vert_F^2.
    \end{aligned}
    \label{eq:sub-zT}
\end{equation}

The transmission feature $\mathbf{z}_\mathbf{T}$ can be updated by solving the quadratic approximation~\cite{beck2009fast} of the problem in Eqn.~(\ref{eq:sub-zT}):
\begin{equation}
    \underset{\mathbf{z}_\mathbf{T}}{\min} \frac{1}{2} \Vert \mathbf{z}_\mathbf{T} - \left( \mathbf{z}_\mathbf{T}^{k} - \eta_\mathbf{T} \nabla f(\mathbf{z}_\mathbf{T}^{k})  \right) \Vert_F^2 +  \frac{\eta_\mathbf{T} \lambda_\mathbf{T}}{\tau}  p_\mathbf{T}(\mathbf{z}_\mathbf{T}),
\end{equation}
with 
\begin{equation}
    \begin{aligned}
    f(\mathbf{z}_\mathbf{T}) &= \frac{1}{2} \Vert \mathbf{I} - \sum_{i \in \{ {{\mathbf{T}, \mathbf{R}, \mathbf{N}}} \}} \mathbf{D}_i \otimes \mathbf{z}_i \Vert_F^2\\
    &+ \frac{\tau}{2} \Vert \mathbf{z}_\mathbf{A} - \left( \mathbf{M}_\mathbf{T} \otimes \mathbf{z}_\mathbf{T} \right) \odot \left( \mathbf{M}_\mathbf{R} \otimes \mathbf{z}_\mathbf{R} \right) \Vert_F^2,
    \end{aligned}
    \nonumber
\end{equation}
where $\eta_\mathbf{T}$ denotes the step-size for updating.

The solution is used to construct the Transmission Feature Update unit and can be expressed as:
\begin{equation}
    \mathbf{z}_\mathbf{T}^{k+1} = \text{prox}_{\eta_\mathbf{T} \lambda_\mathbf{T} / \tau} \left( \mathbf{z}_\mathbf{T}^{k} - \eta_\mathbf{T} \nabla f(\mathbf{z}_\mathbf{T}^{k}) \right),
    \label{eq:solveZT}
\end{equation}
where $\text{prox}_{\eta_\mathbf{T} \lambda_\mathbf{T} / \tau}(\cdot)$ is the proximal operator corresponding to the prior term $p_\mathbf{T}(\cdot)$, and 
\begin{equation}
    \begin{aligned}
        &\nabla f(\mathbf{z}_\mathbf{T}^{k}) = 
        \underbrace{- \mathbf{D}_\mathbf{T}^{k} \otimes^{T} \left( \mathbf{I} - \sum_{i \in \{ {{\mathbf{T}, \mathbf{R}, \mathbf{N}}} \}} \mathbf{D}_i^{k} \otimes \mathbf{z}_i^{k} \right)}
        _{\nabla g(\mathbf{z}_\mathbf{T}^{k}): \; \text{gradient of $\mathbf{z}_\mathbf{T}$ \textit{w.r.t.} reconstruction term}}\\
        &\underbrace{- \tau^{k} \mathbf{M}_\mathbf{T}^{k} \otimes^{T} \left( \left( \mathbf{M}_\mathbf{R}^{k} \otimes \mathbf{z}_\mathbf{R}^{k} \right) \odot \left(\mathbf{z}_\mathbf{A}^{k} - \mathbf{P}^{k}\right) \right)}
        _{\nabla q(\mathbf{z}_\mathbf{T}^{k}): \; \text{gradient of $\mathbf{z}_\mathbf{T}$ \textit{w.r.t.} auxiliary equality term}},
    \end{aligned}
    \nonumber
\end{equation}
where $\mathbf{P}^{k} = \left( \mathbf{M}_\mathbf{T}^{k} \otimes \mathbf{z}_\mathbf{T}^{k} \right) \odot \left( \mathbf{M}_\mathbf{R}^{k} \otimes \mathbf{z}_\mathbf{R}^{k} \right)$
and $\otimes^{T}$ denotes the transposed convolution\footnote{The operation $\otimes^{T}$ can be implemented using the function “torch.nn.ConvTransposed2d” in PyTorch.}. 

Following Eqn.~(\ref{eq:solveZT}), TFU unit first updates the sparse feature $\mathbf{z}_{\mathbf{T}}^{k}$ towards the gradient descent direction of $f(\mathbf{z}_\mathbf{T}^{k})$, and then non-linearly transforms the sparse features by the proximal operator of the prior term $p_{\mathbf{T}}(\cdot)$. The gradient descent step is driven by the model-based problem formulation with the exclusion prior, and it helps to update $\mathbf{z}_{\mathbf{T}}^{k}$ to better meet the formation model as well as the exclusion condition. We leverage a learnable proximal operator, which is parameterized by a deep network, to learn from training data.

%-----------------------------------------------------------
\begin{figure*}[t]
\begin{centering}
\center\includegraphics[width=1.8\columnwidth]{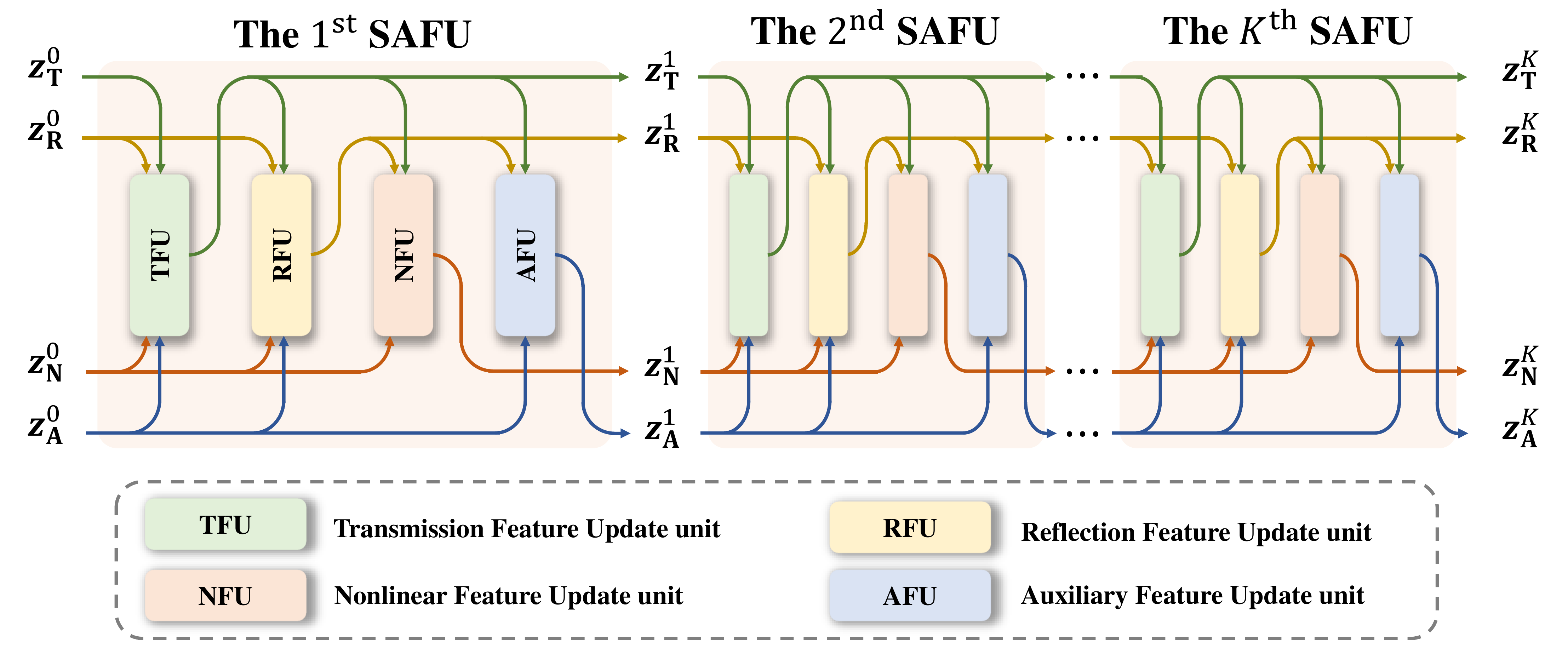}
\par\end{centering}
\caption{Overview of $K$ Sparse and Auxiliary Feature Update (SAFU) modules. Each SAFU module alternatively updates the transmission feature $\mathbf{z}_\mathbf{T}^{k}$, reflection feature $\mathbf{z}_\mathbf{R}^{k}$, nonlinear residual feature $\mathbf{z}_\mathbf{N}^{k}$, and auxiliary feature $\mathbf{z}_\mathbf{A}^{k}$ with a Transmission Feature Update (TFU) unit,a Reflection Feature Update (RFU) unit, a Nonlinear Feature Update (NFU) unit and an Auxiliary Feature Update (AFU) unit, respectively.
}
\label{fig:SAFU module}
\end{figure*}
%-----------------------------------------------------------

\textbf{Reflection Feature Update (RFU) unit:} 
RFU unit is used to update the sparse feature of the reflection image $\mathbf{z}_\mathbf{R}$.
The principle and updating rule of RFU unit for the sparse feature $\mathbf{z}_\mathbf{R}$ is similar to that of $\mathbf{z}_\mathbf{T}$ in Eqn.~(\ref{eq:solveZT}) and can be expressed as:
\begin{equation}
    \mathbf{z}_\mathbf{R}^{{k+1}} = \text{prox}_{\eta_\mathbf{R} \lambda_\mathbf{R} / \tau} \left( \mathbf{z}_\mathbf{R}^{k} - \eta_\mathbf{R} \nabla f(\mathbf{z}_\mathbf{R}^{k}) \right),
    \label{eq:solveZR}
\end{equation}
with 
\begin{equation}
    \begin{aligned}
        &\nabla f(\mathbf{z}_\mathbf{R}^{k}) 
        = 
        \underbrace{- \mathbf{D}_\mathbf{R}^{k} \otimes^{T} \left( \mathbf{I} - \sum_{i \in \{ { \mathbf{R}, \mathbf{N}} \}} \mathbf{D}_i^{k} \otimes \mathbf{z}_i^{k} - \mathbf{D}_\mathbf{T}^{k} \otimes \mathbf{z}_\mathbf{T}^{k+1} \right)}_{\nabla g(\mathbf{z}_\mathbf{R}^{k}): \; \text{gradient of $\mathbf{z}_\mathbf{R}$ \textit{w.r.t.} reconstruction term}}\\
        & \underbrace{- \tau^{k} \mathbf{M}_\mathbf{R}^{k} \otimes^{T} \left( \left( \mathbf{M}_\mathbf{T}^{k} \otimes \mathbf{z}_\mathbf{T}^{k+1} \right) \odot \left(\mathbf{z}_\mathbf{A}^{k} - \mathbf{Q}^{k}\right) \right)}_{\nabla q(\mathbf{z}_\mathbf{R}^{k}): \; \text{gradient of $\mathbf{z}_\mathbf{R}$ \textit{w.r.t.} auxiliary equality term}},\\
    \end{aligned}
    \label{eq:nabla_f}
    \nonumber
\end{equation}
where $\text{prox}_{\eta_\mathbf{R} \lambda_\mathbf{R} / \tau}(\cdot)$ is the proximal operator corresponding to the prior term $p_\mathbf{R}(\cdot)$, and
$\mathbf{Q}^{k} = \left( \mathbf{M}_\mathbf{T}^{k} \otimes \mathbf{z}_\mathbf{T}^{k+1} \right) \odot \left( \mathbf{M}_\mathbf{R}^{k} \otimes \mathbf{z}_\mathbf{R}^{k} \right)$,
$\eta_\mathbf{R}$ denotes the step-size for updating.

\textbf{Nonlinear Feature Update (NFU) unit:} 
NFU unit is used to optimize the sparse feature of the nonlinear residual image $\mathbf{N}$ with fixed $\mathbf{z}_\mathbf{T}$ and $\mathbf{z}_\mathbf{R}$ to adaptively model the nonlinear effects that cannot be well captured by $\mathbf{z}_\mathbf{T}$ and $\mathbf{z}_\mathbf{R}$.
The sub-problem of $\mathbf{z}_\mathbf{N}$ involves only the reconstruction term and its prior term: 
\begin{equation}
        \begin{aligned}
        \underset{ \mathbf{z}_\mathbf{N} }{\min} & \frac{1}{2} \Vert \mathbf{I} - \sum_{i \in \{ {{\mathbf{T}, \mathbf{R}, \mathbf{N}}} \}} \mathbf{D}_i \otimes \mathbf{z}_i \Vert_F^2  + \lambda_\mathbf{N} p_{E}(  \mathbf{z}_\mathbf{N} ).  \\
    \end{aligned}
\end{equation}

As a result, NFU unit performs a gradient descent step with respect to the reconstruction term and applies a learnable proximal operator to update the features as:
\begin{equation}
    \mathbf{z}_\mathbf{N}^{k+1} = \text{prox}_{\eta_\mathbf{N} \lambda_\mathbf{N}} \left( \mathbf{z}_\mathbf{N}^{k} - \eta_\mathbf{N} \nabla g(\mathbf{z}_\mathbf{N}^{k}) \right),
    \label{eq:solveZN}
\end{equation}
with 
\begin{equation}
    \begin{aligned}
    &\nabla g(\mathbf{z}_\mathbf{N}^{k}) = 
        \underbrace{- \mathbf{D}_\mathbf{N}^{k} \otimes^{T} ( \mathbf{I} - \sum_{i \in \{ { \mathbf{T}, \mathbf{R}} \}} \mathbf{D}_i^{k} \otimes \mathbf{z}_i^{k+1} - \mathbf{D}_\mathbf{N}^{k} \otimes \mathbf{z}_\mathbf{N}^{k} ),}_{\text{gradient of $\mathbf{z}_\mathbf{N}$ \textit{w.r.t.} reconstruction term}}
    \end{aligned}
    \nonumber
\end{equation}
where $\eta_\mathbf{N}$ denotes the step-size for updating, and $\text{prox}_{\eta_\mathbf{N} \lambda_\mathbf{N}}(\cdot)$ is the proximal operator corresponding to the prior term $p_\mathbf{N}(\cdot)$.

\textbf{Auxiliary Feature Update (AFU) unit:} 
AFU unit is used to impose the exclusion condition on the auxiliary variable ${\mathbf{z}}_\mathbf{A}$ with the given $\mathbf{z}_\mathbf{T}$ and $\mathbf{z}_\mathbf{R}$.
Its sub-problem is as follows:
\begin{equation}
    \begin{aligned}
        \underset{{\mathbf{z}}_\mathbf{A}}{\min}
        & \frac{\tau}{2} \Vert \mathbf{z}_\mathbf{A} - \left( \mathbf{M}_\mathbf{T} \otimes \mathbf{z}_\mathbf{T} \right) \odot \left( \mathbf{M}_\mathbf{R} \otimes \mathbf{z}_\mathbf{R} \right) \Vert_F^2+ \kappa p_{E}(  \mathbf{z}_\mathbf{A}).
    \end{aligned}
    \label{eq:Usub}
\end{equation}

The solution to Eqn. (\ref{eq:Usub}) is used to construct AFU unit and can then be expressed as:
\begin{equation}
    \mathbf{z}_\mathbf{A}^{k+1} = \text{prox}_{\eta_\mathbf{A} \kappa / \tau} \left( \mathbf{z}_\mathbf{A}^{k} - \eta_\mathbf{A} \nabla h(\mathbf{z}_\mathbf{A}^{k}) \right),
    \label{eq:solveZA}
\end{equation}
with
\begin{equation}
    \nabla h(\mathbf{z}_\mathbf{A}^{k}) = \underbrace{\mathbf{z}_\mathbf{A}^{k} - ( \mathbf{M}_\mathbf{T}^{k} \otimes \mathbf{z}_\mathbf{T}^{k+1}) \odot ( \mathbf{M}_\mathbf{R}^{k} \otimes \mathbf{z}_\mathbf{R}^{k+1} ) }_{\text{gradient of $\mathbf{z}_\mathbf{A}$ \textit{w.r.t.} auxiliary equality term}}
    \nonumber
\end{equation}
where 
$\eta_\mathbf{A}$ denotes the step-size for updating, $\text{prox}_{\eta_\mathbf{A} \kappa/\tau}(\cdot)$ is the proximal operator corresponding to the prior term $p_{E}(\cdot)$.

%-----------------------------------------------------------
\begin{figure*}[t]
\begin{centering}
\center\includegraphics[width=2\columnwidth]{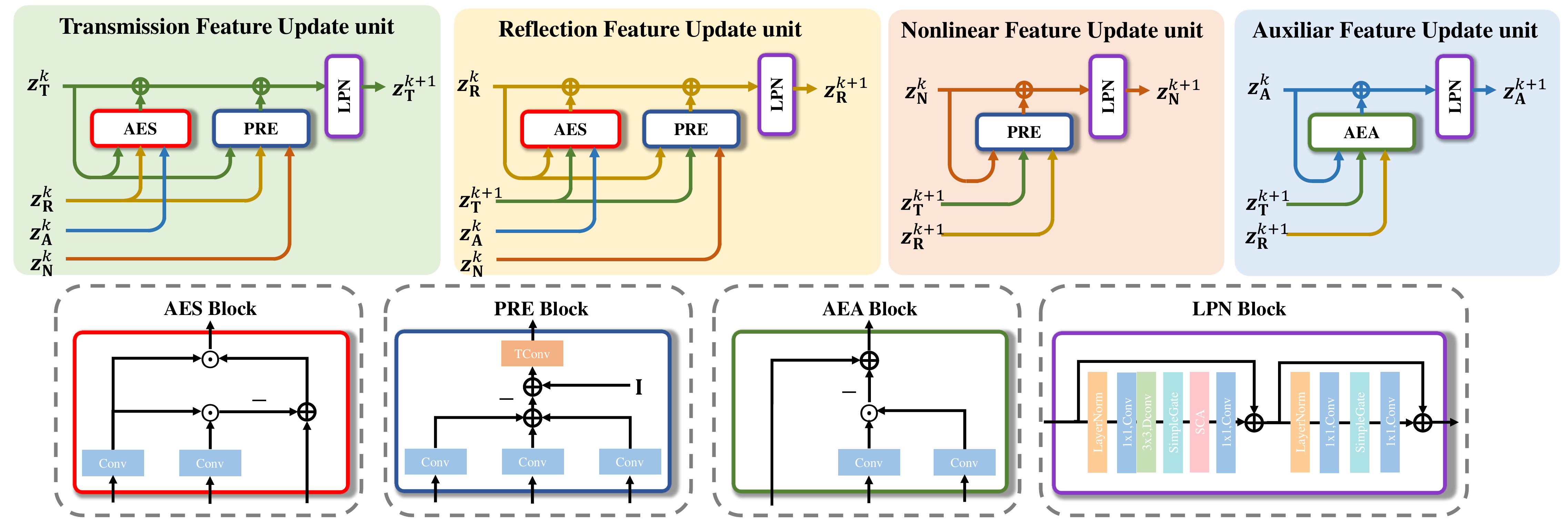}
\par\end{centering}
\caption{The network architecture of the proposed Sparse and Auxiliary Feature Update (SAFU) Module, \textit{i.e.}, Transmission Feature Update (TFU) unit, Reflection Feature Update (RFU) unit, Nonlinear Feature Update (NFU) unit, and Auxiliary Feature Update (AFU) unit. The basic network components includes AES block, AEA block, PRE block, and LPN block.}
\label{fig:FUM_nets}
\end{figure*}
%-----------------------------------------------------------

\textbf{Unfolded SAFU Architecture:}
The design of the SAFU module's network topology is firmly grounded in the principles derived from
the model-based formulations in Eqn. (\ref{eq:solveZT}), (\ref{eq:solveZR}),  (\ref{eq:solveZN}) and (\ref{eq:solveZA}), ensuring a systematic and model-driven approach to the network's structure. 
For each feature among $\mathbf{z}_\mathbf{T}$, $\mathbf{z}_\mathbf{R}$, ${\mathbf{z}}_\mathbf{N}$, and ${\mathbf{z}}_\mathbf{A}$, we have designed a corresponding feature update unit 
with 
a Learnable Proximal Network (LPN)
to learn from the training dataset and capture the prior information. 
The convolution kernels $\{ \mathbf{D}_i^{k} \}_{i=\{\mathbf{T},\mathbf{R},\mathbf{N}, \mathbf{A} \}}$ and $\{ \mathbf{M}_i^{k} \}_{i=\{\mathbf{T},\mathbf{R}\}}$ are set to be learnable.
Therefore, the $k$-th SAFU module of DExNet can be represented as follows:
\begin{equation}
    \begin{cases}
        \mathbf{z}_\mathbf{T}^{k+1} = \text{LPN}_{\mathbf{\theta_\mathbf{T}}^{k}} \left( \mathbf{z}_\mathbf{T}^{k} - \eta_\mathbf{T} \nabla f(\mathbf{z}_\mathbf{T}^{k}) \right),\\
        \mathbf{z}_\mathbf{R}^{k+1} = \text{LPN}_{\mathbf{\theta_\mathbf{R}}^{k}} \left( \mathbf{z}_\mathbf{R}^{k} - \eta_\mathbf{R} \nabla f(\mathbf{z}_\mathbf{R}^{k}) \right),\\
        \mathbf{z}_\mathbf{N}^{k+1} = \text{LPN}_{\mathbf{\theta_\mathbf{N}}^{k}} \left( \mathbf{z}_\mathbf{N}^{k} - \eta_\mathbf{N} \nabla g(\mathbf{z}_\mathbf{N}^{k}) \right),\\
        \mathbf{z}_\mathbf{A}^{k+1} = \text{LPN}_{\mathbf{\theta}_\mathbf{A}^{k}} \left( \mathbf{z}_\mathbf{A}^{k} - \eta_\mathbf{A} \nabla h(\mathbf{z}_\mathbf{A}^{k}) \right),
    \end{cases}
\end{equation}
where $\{ \mathbf{\theta}_i^{k} \}_{i=\{{\mathbf{T}, \mathbf{R}, \mathbf{N}},\mathbf{A} \}}$ denotes the learnable parameters of LPN, and $\{ \mathbf{\eta}_i^{k} \}_{i=\{{\mathbf{T}, \mathbf{R}, \mathbf{N}},\mathbf{A} \}}$ denotes the learnable step size.

Fig.~\ref{fig:FUM_nets} illustrates the network architecture of TFU, RFU, NFU and AFU. 
These units share a set of basic network components, including Auxiliary Equality for Sparse features (AES) blocks, Auxiliary Equality for Auxiliary feature (AEA) blocks, Projected Reconstruction Error (PRE) blocks, and LPN blocks. Notably, TFU and RFU have identical network structure, comprising AES, PRE and LPN. However, it is important to note that, despite having the same set of network blocks, different feature update units do not share model parameters.
The mathematical interpretation of these basic network blocks are as follows:
\begin{itemize}
    \item An AES block is constructed \textit{w.r.t.} $\nabla q(\cdot)$ as in Eqn. (\ref{eq:solveZT}) and Eqn. (\ref{eq:solveZR})
    and works for evaluating the gradient of $\mathbf{z}_{\mathbf{T}}, \mathbf{z}_{\mathbf{R}}$ with respect to auxiliary equality term. It consists of two element-wise multiplication operations between features, and facilitates the interactions between the common image contents.

    \item  An AEA block is constructed \textit{w.r.t.} $\nabla h(\cdot)$ as in Eqn. (\ref{eq:solveZA})
    and updates the auxiliary variable with respect to the common feature between two image layers. It consists of one element-wise multiplication operation between features and then measures the difference with the input auxiliary variable. 

    \item A PRE block is constructed \textit{w.r.t.} $\nabla g(\cdot)$ as in Eqn. (\ref{eq:solveZT}), Eqn. (\ref{eq:solveZR}) and Eqn. (\ref{eq:solveZN}) and imposes the reconstruction condition. It measures the reconstruction residual between the input reflection-contaminated image and the estimated reconstruction image and then projects the residual into the feature domain. 
    
    \item The LPN block is used to learn prior from data and can be an arbitrary network, here we adopt the Nonlinear Activation Free (NAF) block~\cite{chen2022simple} to learn the proximal operators $\text{prox}_{\lambda}(\cdot)$ due to the fact that it has proven to be highly effective on image restoration tasks.
\end{itemize}

In Section~\ref{sec:ablation}, we demonstrate through ablation study that 
AES, AEA and PRE blocks are 
able to substantially enhance the
separation performance with a marginal parameter footprint, and LNP block enables effective prior learning.

%-----------------------------------------------------------
\begin{figure}[t]
    \centering
    \includegraphics[width=0.45\textwidth]{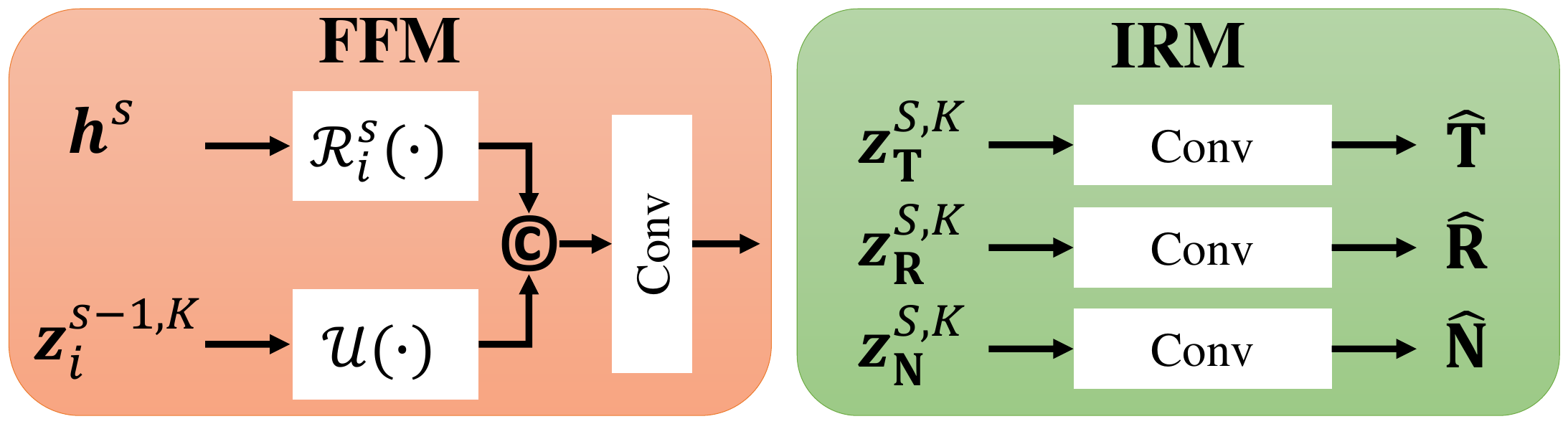}
    \caption{The network architecture of Feature Fusion Module (FFM) and Image Restoration module (IRM). }
    \label{fig:FFM_IRM}
\end{figure}
%-----------------------------------------------------------

\subsubsection{Feature Fusion Module (FFM) }

The FFM works for fusing the upscaled features $\{ \mathbf{z}_\mathbf{T}, \mathbf{z}_\mathbf{R}, {\mathbf{z}}_\mathbf{N}, \mathbf{z}_\mathbf{A} \}$ from the lower scale with the hypercolumn features~\cite{hariharan2015hypercolumns}.
Let us take the Feature Fusion Module at the $s$-th scale (with $s > 1$) as an example for illustration. As depicted in Fig. ~\ref{fig:FFM_IRM}, the input feature $\mathbf{z}_i^{s,0}$ with $i \in \{ \mathbf{T},\mathbf{R},\mathbf{N}, \mathbf{A} \}$ is obtained by:
\begin{equation}
    \mathbf{z}_i^{s,0} = \mathbf{C}_{1 \times 1, i}^s \otimes \left( \mathcal{U}(\mathbf{z}_i^{s-1,K}) \copyright \mathcal{R}_i^s(\mathbf{h}^{s}) \right),
\end{equation}
where $\mathbf{C}_{1 \times 1, i}^{s}$ denotes a $1 \times 1$ convolution operator, $\copyright$ denotes the channel-wise concatenation operator, $\mathcal{U}(\cdot)$ represents the bilinear upsampling operator, $\mathcal{R}_i^s(\cdot)$ is a residual block that maps the downscaled hypercolumn feature $\mathbf{h}^s$ to $n$ channels.

\subsubsection{Image Reconstruction Module (IRM)}

Based on the Convolutional Sparse Coding (CSC) model, each image is parameterized as a sparse representation over a dictionary. Therefore, as shown in Fig. ~\ref{fig:FFM_IRM}, the reconstructed transmission image $\hat{\mathbf{T}}$, reflection image $\hat{\mathbf{R}}$ and nonlinear residual image $\hat{\mathbf{N}}$ are mapped from the final updated features $\mathbf{z}_i^{S,K}$ ($i \in \{ \mathbf{T}, \mathbf{R}, \mathbf{N}\}$) to image domain via convolution with the corresponding convolutional dictionaries:
\begin{equation}
    \begin{cases}
        \hat{\mathbf{T}} = \mathbf{D}_\mathbf{T} \otimes \mathbf{z}_\mathbf{T}^{S,K},\\
        \hat{\mathbf{R}} = \mathbf{D}_\mathbf{R} \otimes \mathbf{z}_\mathbf{R}^{S,K},\\
        \hat{\mathbf{N}} = \mathbf{D}_\mathbf{N} \otimes \mathbf{z}_\mathbf{N}^{S,K},\\
    \end{cases}
\end{equation}
where $\mathbf{D}_i \in \mathbb{R}^{w \times w \times 3 \times n}$ with $i \in \{ \mathbf{T}, \mathbf{R}, \mathbf{N}\}$ denote convolution dictionaries.

%-----------------------------------------------------------
\begin{table*}[t]
\caption{Quantitative results of the estimated \textit{Transmission Image} of different methods evaluated on \textit{Real20} dataset~\cite{perceptual_loss_2018} and three subset of $SIR^{2}$ dataset~\cite{wan2017benchmarking}. 
(The best scores and the second best scores are in \textbf{bold} and with \underline{underline}, respectively. $^\dagger$ denotes the data directly from the original paper and - denotes the unavailable data.)}
    \centering
    \begin{tabular}{c|c|C{1cm}C{0.9cm}C{0.9cm}C{0.9cm}C{0.9cm}C{0.9cm}C{0.9cm}C{0.9cm}|C{0.9cm}C{0.9cm}} \hline 
    \toprule[1pt]
    \rowcolor[HTML]{EFEFEF}  \cellcolor[HTML]{EFEFEF} & &\multicolumn{2}{c}{\textbf{\textit{Real20}} (20)}&\multicolumn{2}{c}{\textbf{\textit{Objects}} (200)}&     \multicolumn{2}{c}{\textbf{\textit{Postcard}} (199)}&\multicolumn{2}{c}{\textbf{\textit{Wild}} (55)}&\multicolumn{2}{|c}{\textbf{Average} (474)}\\ 
\cline{3-12} \rowcolor[HTML]{EFEFEF} \multirow{-2}{*}{\textbf{Methods}} &  \multirow{-2}{*}{\textbf{Model Size}} &  \textbf{PSNR }& \textbf{SSIM  }& \textbf{PSNR }& \textbf{SSIM }&     \textbf{PSNR }&\textbf{SSIM }&\textbf{PSNR }&\textbf{SSIM }&\textbf{PSNR }&\textbf{SSIM }\\ \hline 
Zhang\textit{ et al.}~\cite{perceptual_loss_2018}  & 3.16M & 22.55&       0.788           & 22.68  &0.879&16.81&0.797&21.52&0.832&20.08 &0.835\\ 
 BDN~\cite{yang2018seeing}    &   24.37M   &18.41&0.726&22.72&0.856      &20.71&0.859&22.36&0.830 &21.65&0.849\\
  Zheng \textit{et al.}~\cite{zheng2021single} & 36.66M
 &20.17 &0.755 &25.20 &0.880 &23.26 &0.905 &25.39 &0.878 &24.19 &0.885\\
 CoRRN\cite{wan2019corrn}     &   59.50M   &21.57&0.807&25.13&0.912&20.87&0.866&24.34&0.891&23.10&0.866\\
 ERRNet\cite{wei2019single}   &   19.30M   &22.89&0.803&24.87&0.896&22.04&0.876&24.25&0.853&23.53&0.879\\
 IBCLN~\cite{cascaded_Refine_2020} & 21.61M &21.86&0.762&24.84&0.893&23.39&0.875&24.71&0.886&24.09&0.879\\
 YTMT ~\cite{hu2021trash}          & 38.40M &23.26&0.806&24.87&0.896&22.91&0.884&25.48&0.890&24.05&0.886\\
 LASIRR~\cite{dong2021location}
& 10.42M&23.34 &0.812 &24.36 &0.898 &23.72 &0.903 &25.73 &0.902 &24.21 &0.897\\
 DMGN~\cite{FengTIP2021}
& 62.48M &20.71 &0.770 &24.98 &0.899 &22.92 &0.877 &23.81 &0.835 &23.80 &0.877\\

RAGNet~\cite{LiAI2023} & 139.92M &22.95&0.793&26.15&0.903&23.67&0.879&25.53&0.880&24.90&0.886\\
 DSRNet~\cite{Hu_2023_ICCV}  & 124.60M
 &\textbf{23.91} &\textbf{0.818} &\textbf{26.74} &\textbf{0.920} &\underline{24.83} &\underline{0.911} &26.11 &0.906 &\underline{25.75} &\underline{0.910}\\
 MoG-SIRR~\cite{Shao2021MoGSIRR}$^\dagger$ &- &21.63 &0.814& 24.57 &0.911 &22.78 &0.892 &24.13 &0.890 &23.64	&0.866\\
CGDNet~\cite{zhang2022content}$^\dagger$ &- &- &-&25.29 &0.901 &24.03 &0.912 &26.22 &\textbf{0.917} & -	&-\\
DURRNet~\cite{Huang24DURRNet} & 12.10M &\underline{23.80}&0.810 &24.52&0.891 &22.26 &0.866 &25.62 &0.899 &24.00 & 0.854	\\
 \textbf{DExNet$_S$ (Proposed) }           &  4.52M              
 &23.13 	&0.813 	&26.46 	&0.915 	&\underline{24.83} 	&0.908 	&\underline{26.64} 	&\underline{0.909} 	&25.66 	&0.907 \\
 \textbf{DExNet$_L$ (Proposed)}            &   9.66M             &{23.50}&\underline{0.817}&\underline{26.38}&\underline{0.916}&\textbf{25.52}&\textbf{0.918}&\textbf{26.95}&{0.908}&\textbf{25.96}&\textbf{0.912}\\\hline
 \toprule[1pt]
 \end{tabular}
    \label{tab:real20_SIR2}
\end{table*}
%-----------------------------------------------------------

%-----------------------------------------------------------
\begin{table*}[t]
\caption{Quantitative results of the estimated \textit{Reflection Image} of different methods evaluated on \textit{Real20} dataset~\cite{perceptual_loss_2018} and three subset of $SIR^{2}$ dataset~\cite{wan2017benchmarking}. 
(The best scores and the second best scores are in \textbf{bold} and with \underline{underline}, respectively.)}
    \centering
    \begin{tabular}{c|c|C{1cm}C{0.9cm}C{0.9cm}C{0.9cm}C{0.9cm}C{0.9cm}C{0.9cm}C{0.9cm}|C{0.9cm}C{0.9cm}} \hline 
    \toprule[1pt]
    \rowcolor[HTML]{EFEFEF}  \cellcolor[HTML]{EFEFEF} & &\multicolumn{2}{c}{\textbf{\textit{Real20}} (20)}&\multicolumn{2}{c}{\textbf{\textit{Objects}} (200)}&     \multicolumn{2}{c}{\textbf{\textit{Postcard}} (199)}&\multicolumn{2}{c}{\textbf{\textit{Wild}} (55)}&\multicolumn{2}{|c}{\textbf{Average} (474)}\\ 
\cline{3-12} \rowcolor[HTML]{EFEFEF} \multirow{-2}{*}{\textbf{Methods}} &  \multirow{-2}{*}{\textbf{Model Size}} &  \textbf{PSNR }& \textbf{SSIM  }& \textbf{PSNR }& \textbf{SSIM }&     \textbf{PSNR }&\textbf{SSIM }&\textbf{PSNR }&\textbf{SSIM }&\textbf{PSNR }&\textbf{SSIM }\\ \hline 
Zhang\textit{ et al.}~\cite{perceptual_loss_2018}  & 3.16M & 19.20 & 0.316 & 23.82 & 0.418 &21.93 & 0.374& 25.89 & 0.500 &23.07 &0.405\\
 BDN~\cite{yang2018seeing}    &   24.37M   & 10.26 & 0.249 & 10.68 & 0.286 & 8.75 & 0.256 & 10.84 & 0.401 & 9.87& 0.285\\
  Zheng \textit{et al.}~\cite{zheng2021single} & 36.66M & 20.60& 0.411& 27.05& 0.549& 22.55 & 0.459& 27.39 & 0.516 & 24.92&0.502\\
 CoRRN\cite{wan2019corrn}     &   59.50M & 21.16 & 0.387&24.55 &0.449 & 18.83& 0.283& 23.68& 0.356& 21.90 &0.366\\
 ERRNet\cite{wei2019single}   &   19.30M   &23.55&0.446&26.02&0.446&22.47&0.419&25.52&0.460& 24.54 & 0.443 \\
 IBCLN~\cite{cascaded_Refine_2020} & 21.61M &22.36&0.469&19.97&0.226&13.16&0.230&20.83&0.298& 20.22 & 0.332 \\
 YTMT ~\cite{hu2021trash}          & 38.40M & 22.95 & 0.454 & 25.47 & 0.419 & 22.61 & 0.440 & 27.90 & 0.611 & 24.18 & 0.444 \\
 LASIRR~\cite{dong2021location}
&10.42M & 17.78 & 0.366 & 22.71 & 0.540 & 13.00 & 0.211 & 21.80  & 0.440 & 18.32 &0.383\\
 DMGN~\cite{FengTIP2021} & 62.48M & 20.39 & 0.314 & 23.79 & 0.422 & 15.12 & 0.269 & 24.16 & 0.440& 20.05&0.355\\
 RAGNet~\cite{LiAI2023} & 139.92M & 22.11 & 0.284 & 25.61 & 0.327 & 22.43 & 0.245 & 24.58 & 0.252 & 24.01 & 0.282\\
  DSRNet~\cite{Hu_2023_ICCV}  & 124.60M
  & 23.91 & 0.520 & 27.03 & 0.532 & 21.51 & 0.380 & 25.31 & 0.485 & 25.00 & 0.507 \\
DURRNet~\cite{Huang24DURRNet} &12.10M &\textbf{24.64}	&\underline{0.547} 	&25.81	&0.468 	&22.67	&0.435 	&26.76	&0.574 	&24.99	&0.502\\
  \textbf{DExNet$_S$ (Proposed) }           &  4.52M  & {23.81} & {0.538} & \textbf{27.91} & \textbf{0.592} & \underline{23.00} & \underline{0.450} & \underline{27.91} & \underline{0.560} & \underline{25.61} & \underline{0.551}\\
 \textbf{DExNet$_L$ (Proposed)}            &   9.66M &\underline{24.20}& \textbf{0.579} & \underline{27.62} & \underline{0.591} &\textbf{23.28}&\textbf{0.491}&\textbf{28.21}&\textbf{0.602}& \textbf{25.70} & \textbf{0.574}\\\hline
 \toprule[1pt]
 \end{tabular}
    \label{tab:real20_Reflection}
\end{table*}
%-----------------------------------------------------------

%-----------------------------------------------------------
\begin{figure*}[t]
    \centering
    \subfigure[Input]{
    \begin{minipage}[b]{0.09\textwidth}
        \includegraphics[width=1.0\linewidth]{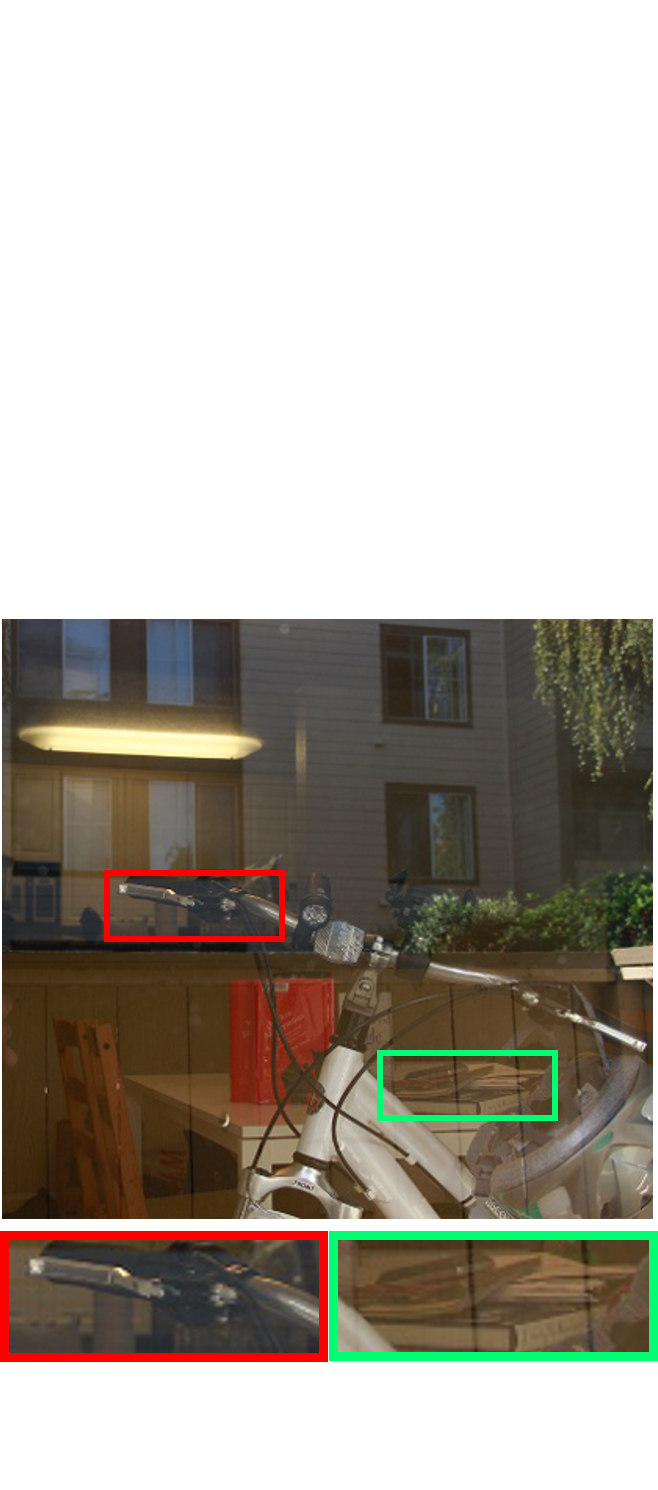}\vspace{1pt}
        \includegraphics[width=1.0\linewidth]{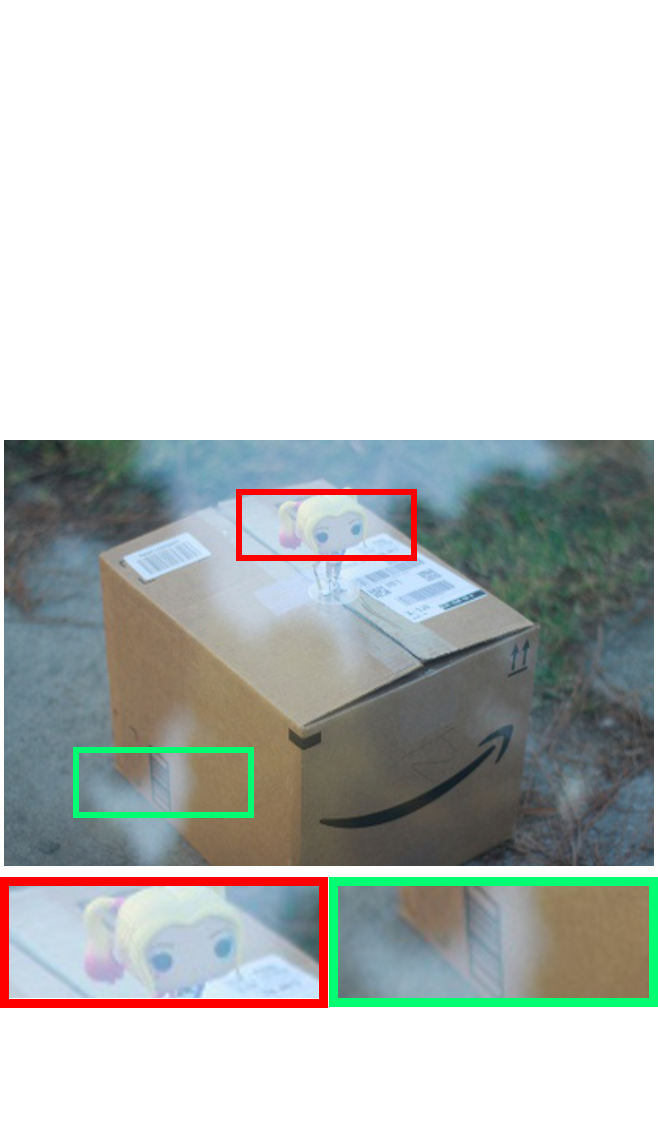}\vspace{1pt}
        \includegraphics[width=1.0\linewidth]{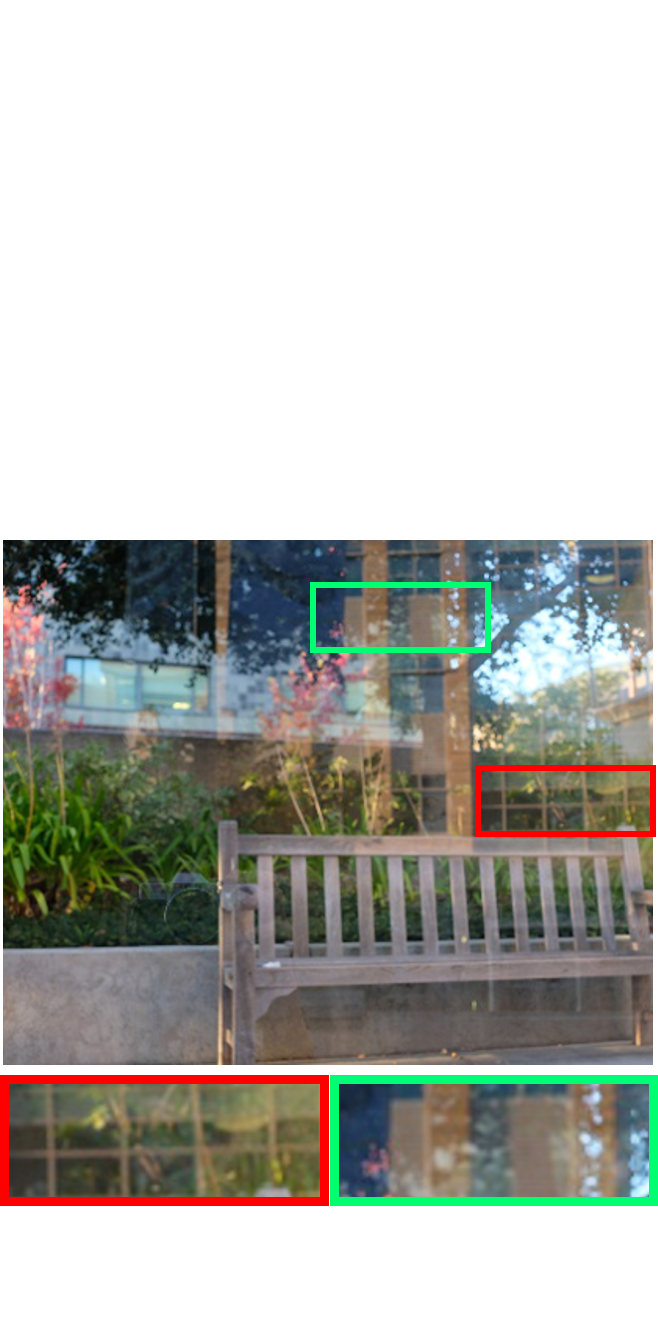}\vspace{1pt}
    \end{minipage}
    }
    \subfigure[ {Zhang}]{
    \begin{minipage}[b]{0.09\textwidth}
    \includegraphics[width=1.0\linewidth]{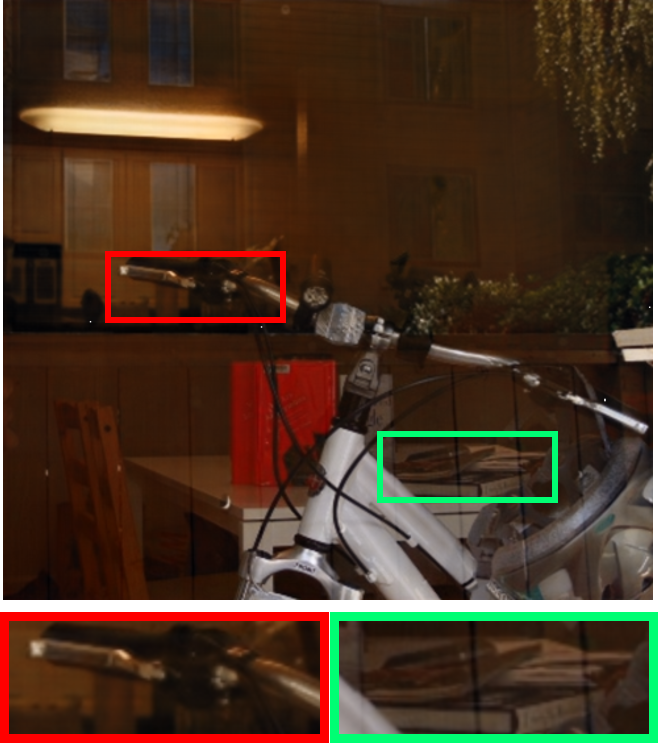}\vspace{0.5pt}
    \includegraphics[width=1.0\linewidth]{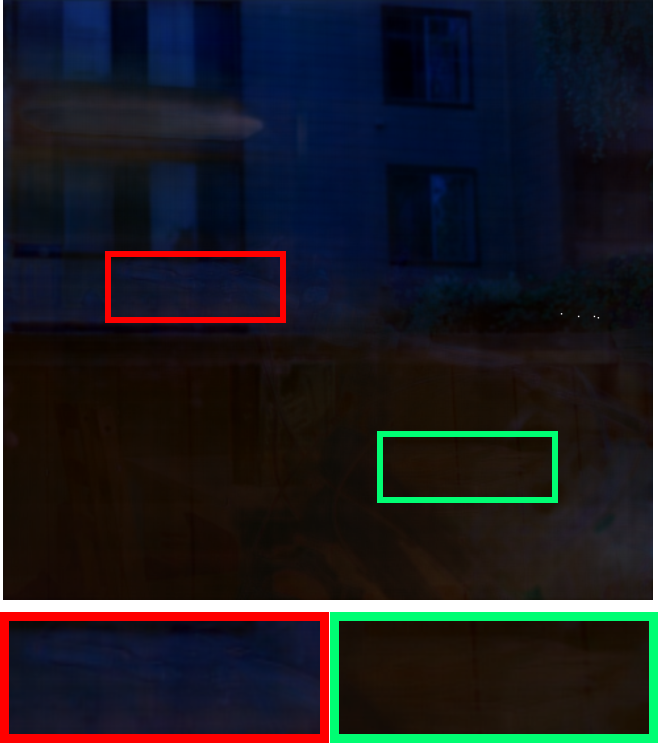}\vspace{1pt}
    \includegraphics[width=1.0\linewidth]{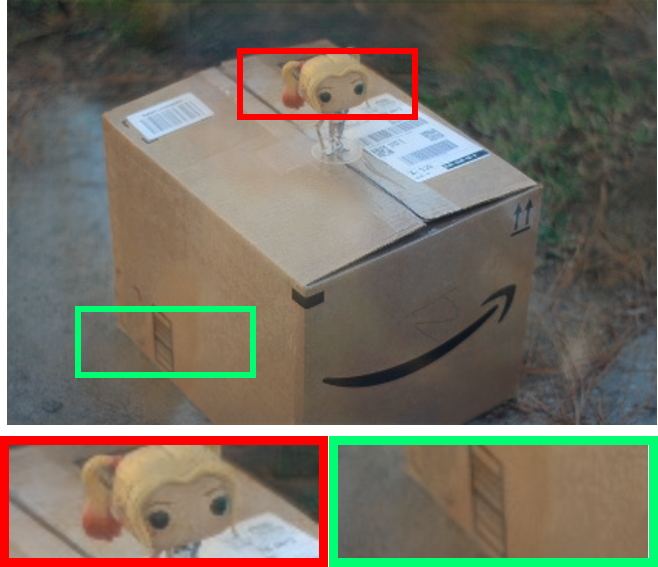}\vspace{0.5pt}
    \includegraphics[width=1.0\linewidth]{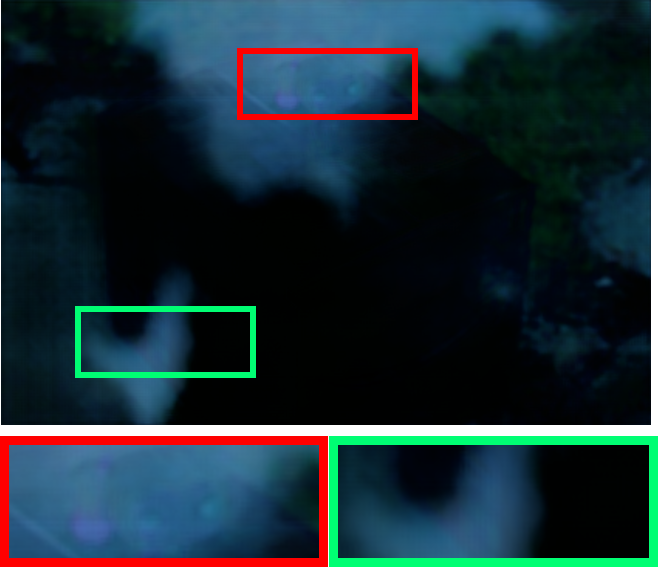}\vspace{1pt}
    \includegraphics[width=1.0\linewidth]{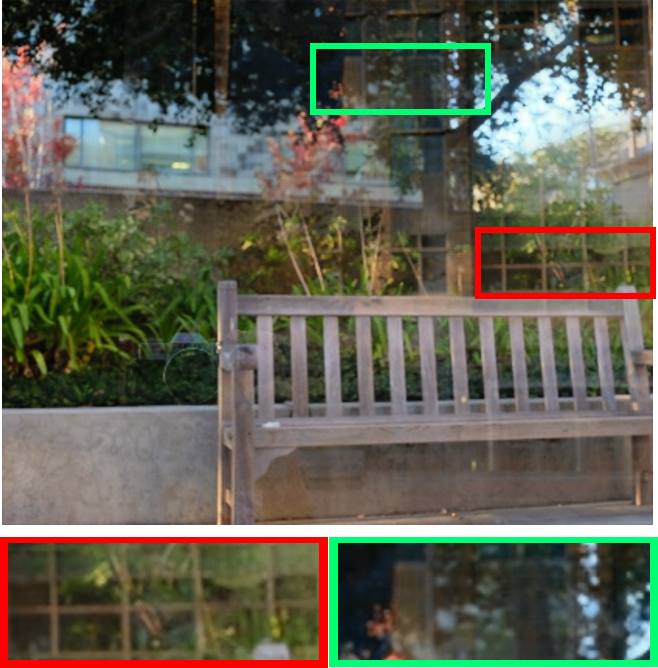}\vspace{0.5pt}
    \includegraphics[width=1.0\linewidth]{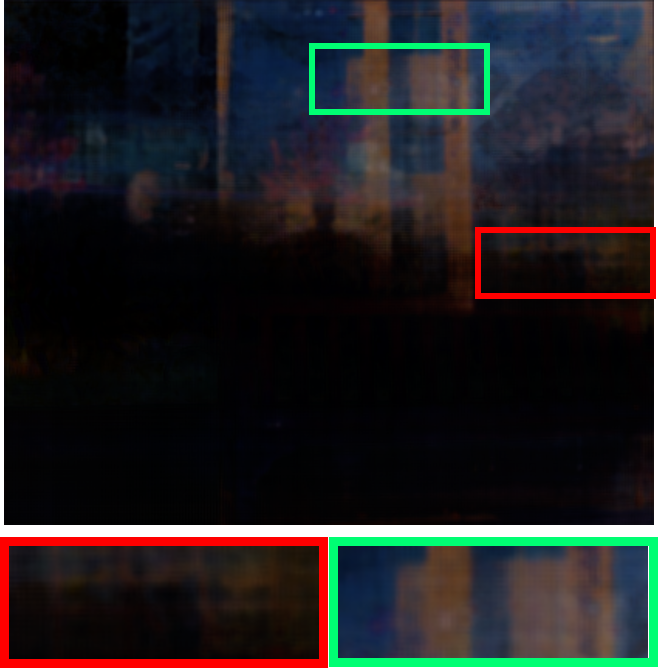}\vspace{1pt}
    \end{minipage}
    }
    \subfigure[ {IBCLN}]{
    \begin{minipage}[b]{0.09\textwidth}
    \includegraphics[width=1.0\linewidth]{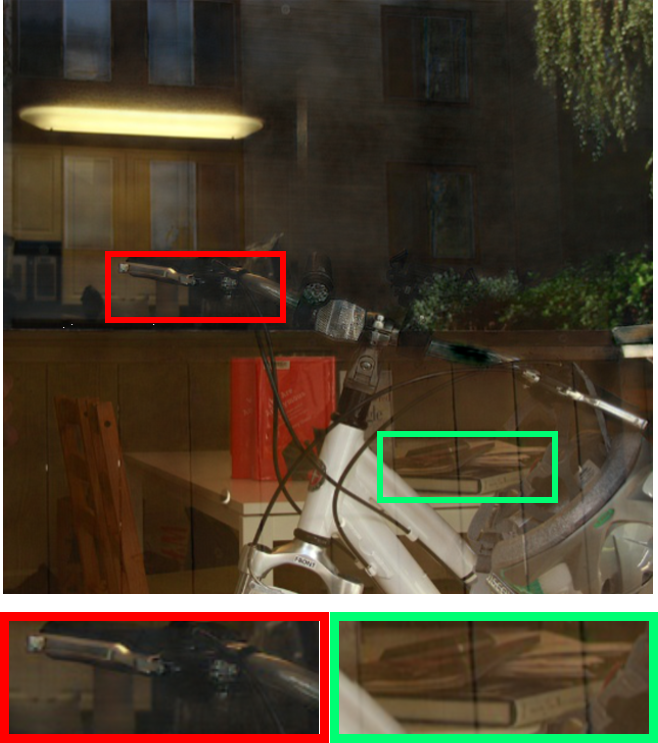}\vspace{0.5pt}
    \includegraphics[width=1.0\linewidth]{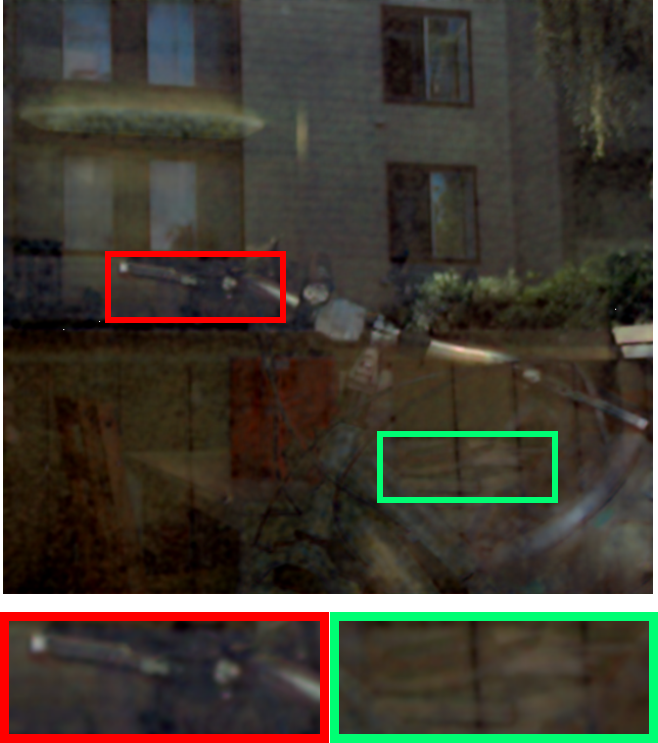}\vspace{1pt}
    \includegraphics[width=1\linewidth]{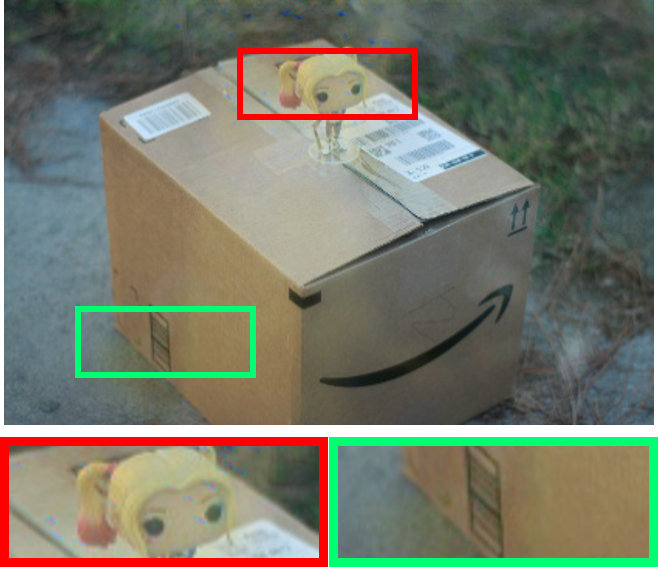}\vspace{0.5pt}
    \includegraphics[width=1\linewidth]{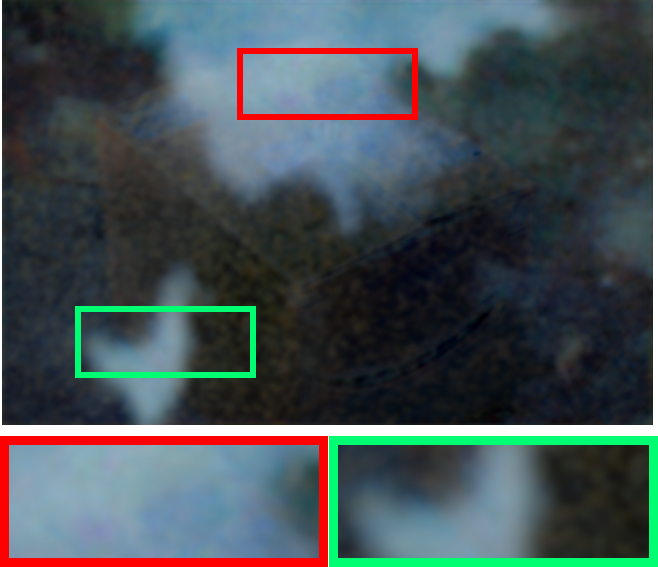}\vspace{1pt}
    \includegraphics[width=1.0\linewidth]{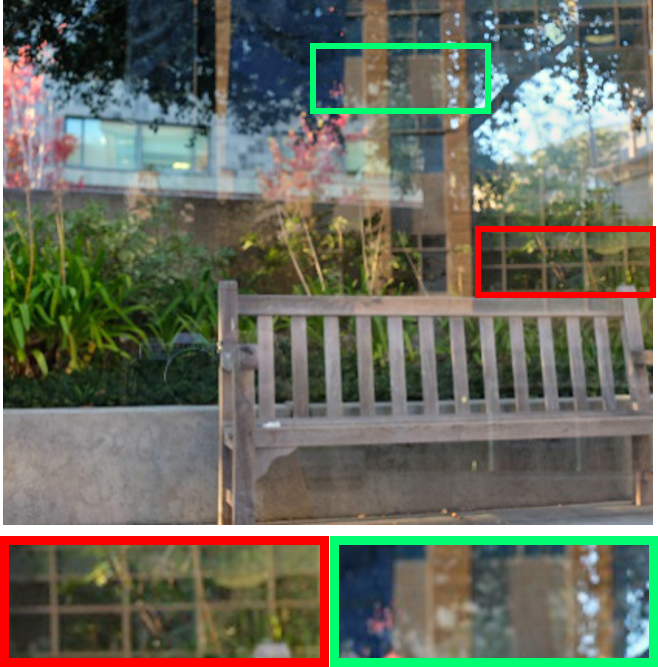}\vspace{0.5pt}
    \includegraphics[width=1.0\linewidth]{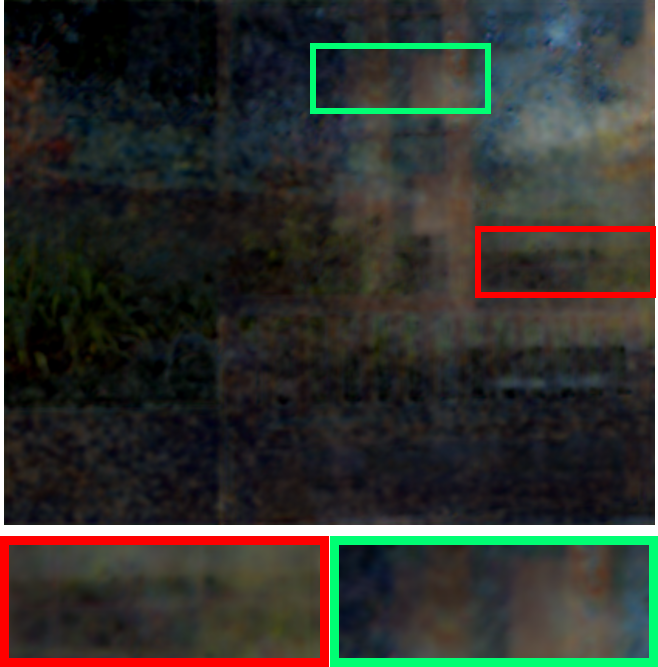}\vspace{1pt}
    \end{minipage}
    }
    \subfigure[ERRNet]{
    \begin{minipage}[b]{0.09\textwidth}
    \includegraphics[width=1.0\linewidth]{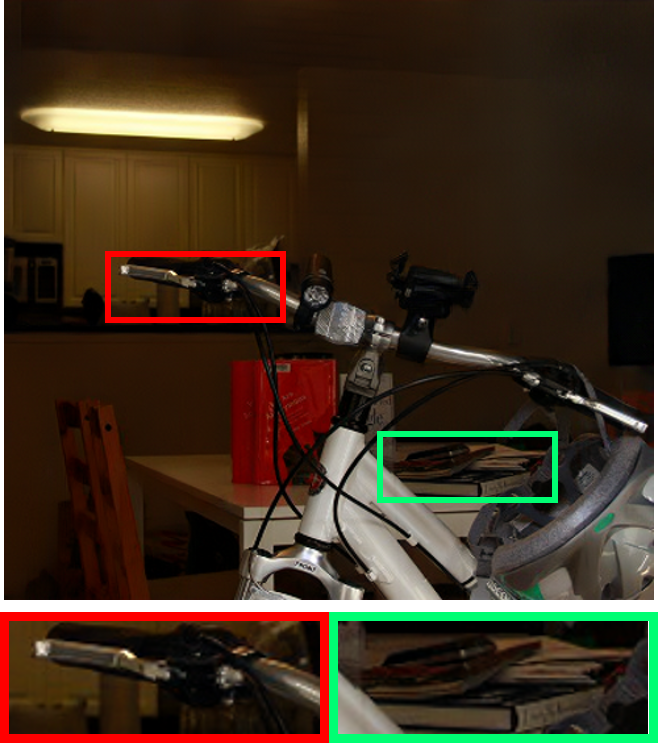}\vspace{0.5pt}
    \includegraphics[width=1.0\linewidth]{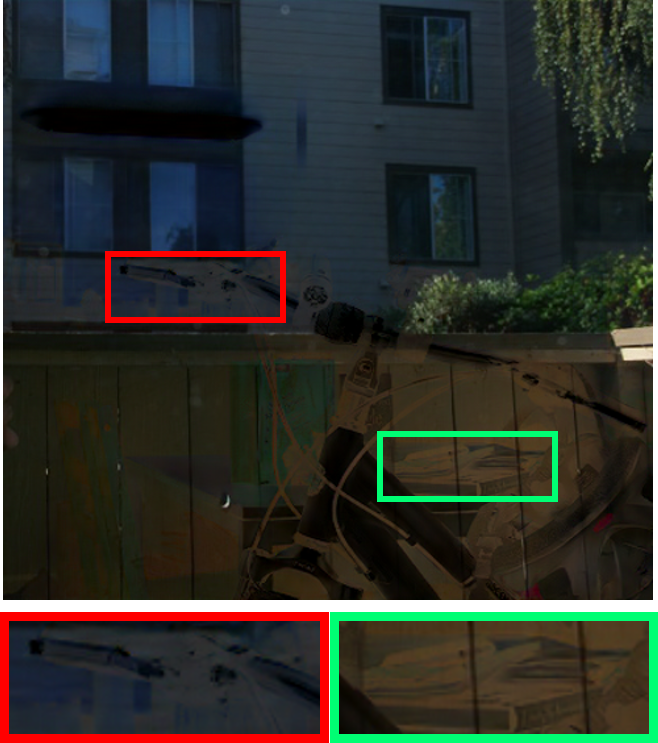}\vspace{1pt}
    \includegraphics[width=1.0\linewidth]{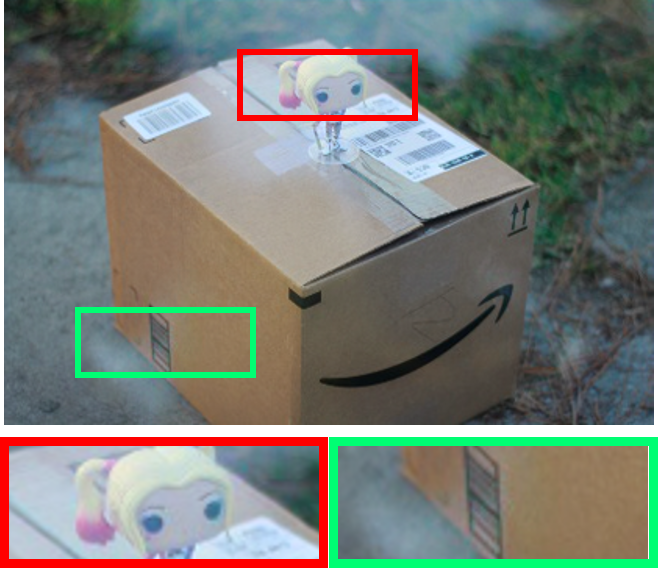}\vspace{0.5pt}
    \includegraphics[width=1.0\linewidth]{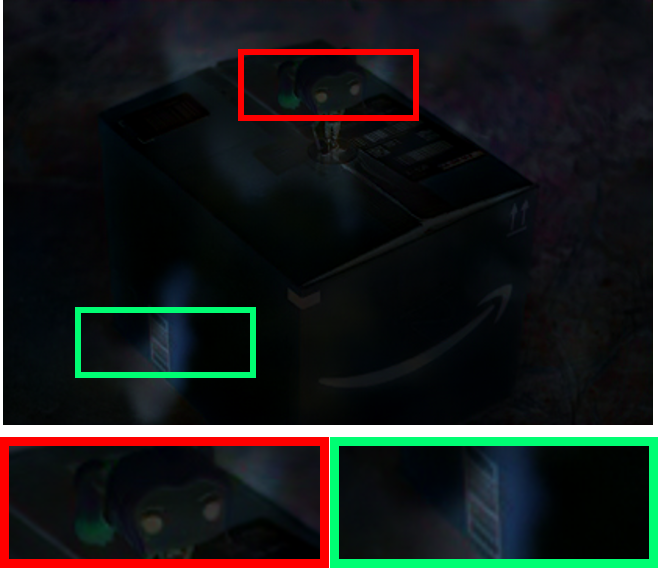}\vspace{1pt}
    \includegraphics[width=1.0\linewidth]{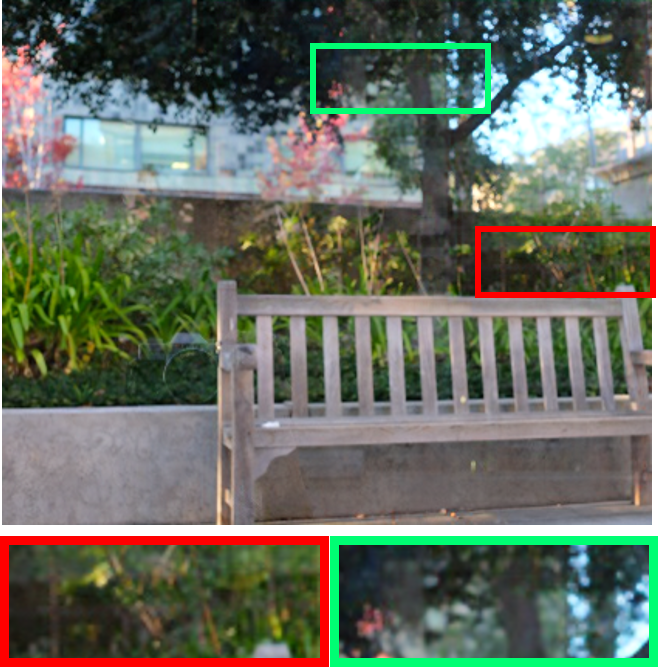}\vspace{0.5pt}
    \includegraphics[width=1.0\linewidth]{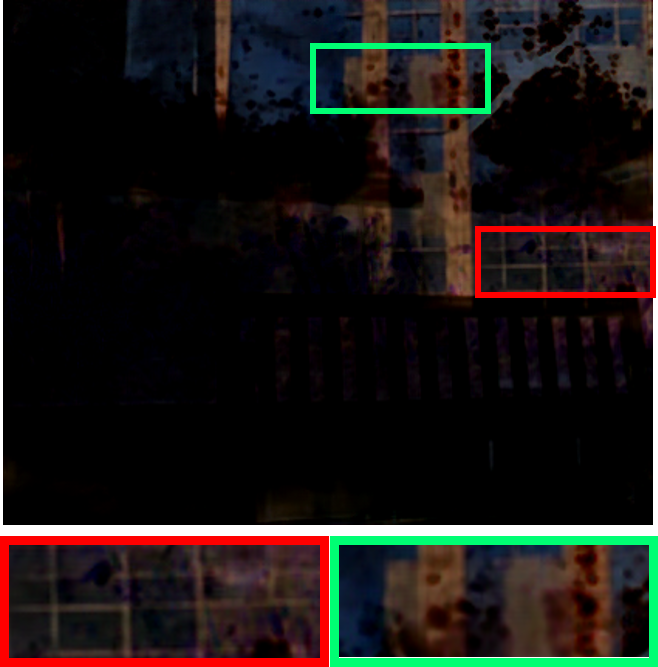}\vspace{1pt}
    \end{minipage}
    }
    \subfigure[YTMT]{
    \begin{minipage}[b]{0.09\textwidth}
    \includegraphics[width=1.0\linewidth]{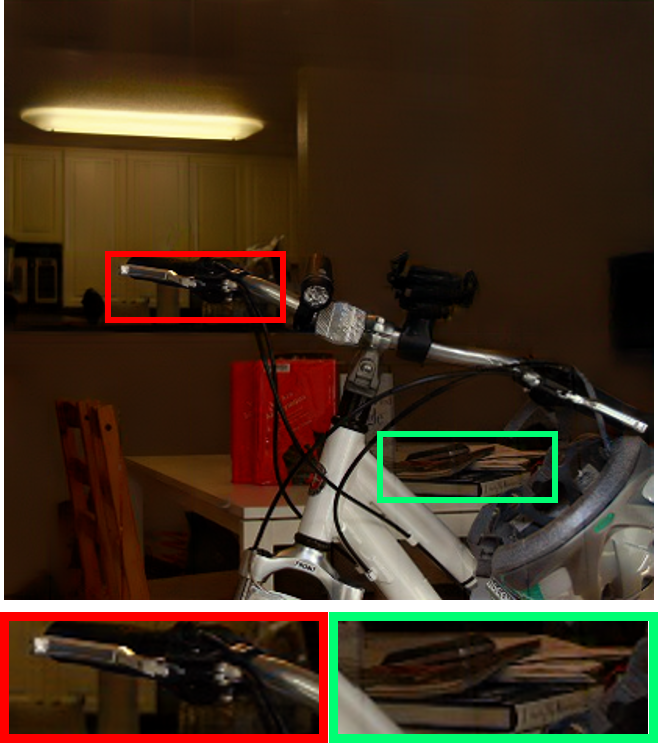}\vspace{0.5pt}
    \includegraphics[width=1.0\linewidth]{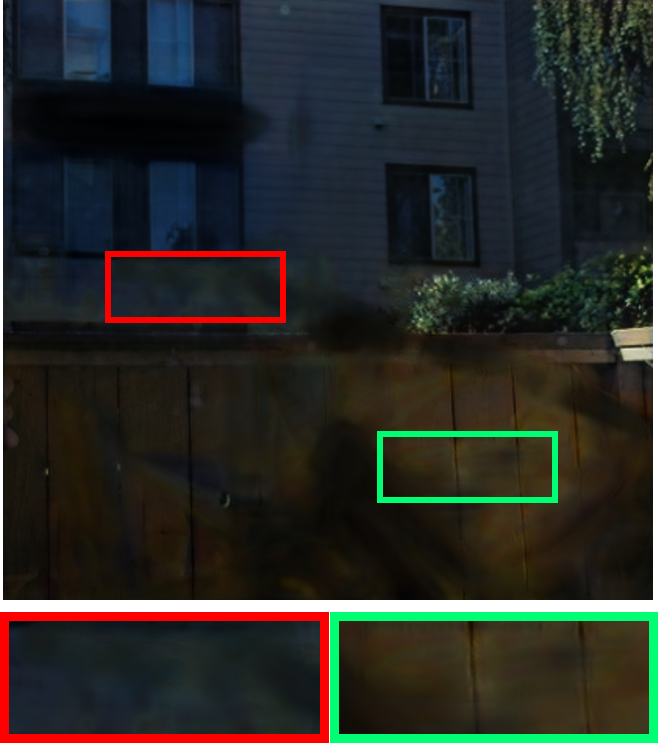}\vspace{1pt}
    \includegraphics[width=1.0\linewidth]{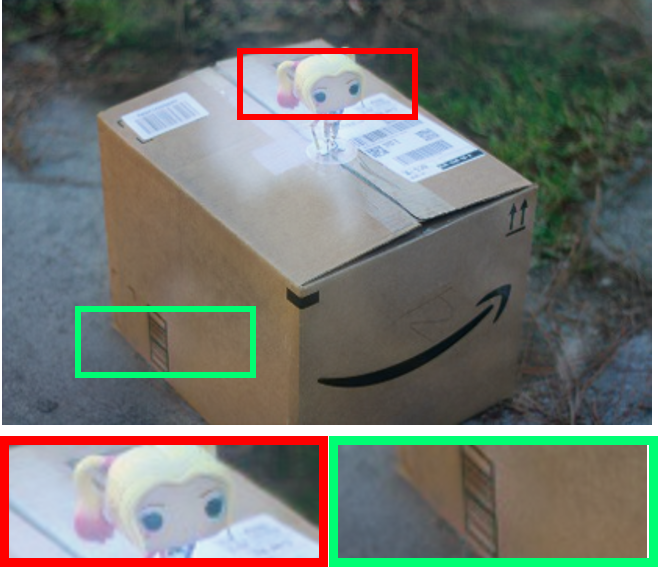}\vspace{0.5pt}
    \includegraphics[width=1.0\linewidth]{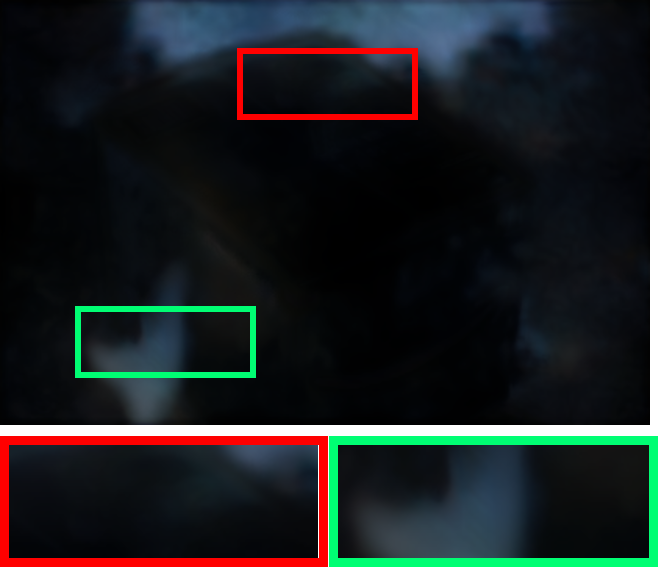}\vspace{1pt}
    \includegraphics[width=1.0\linewidth]{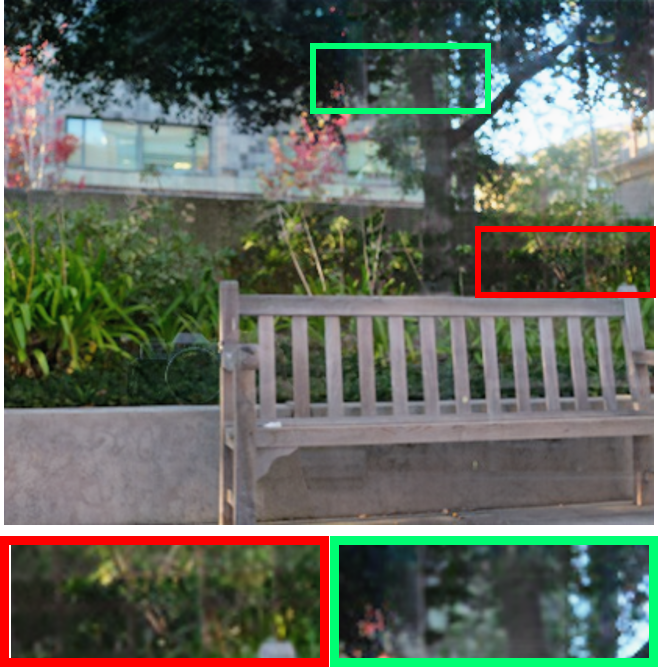}\vspace{0.5pt}
    \includegraphics[width=1.0\linewidth]{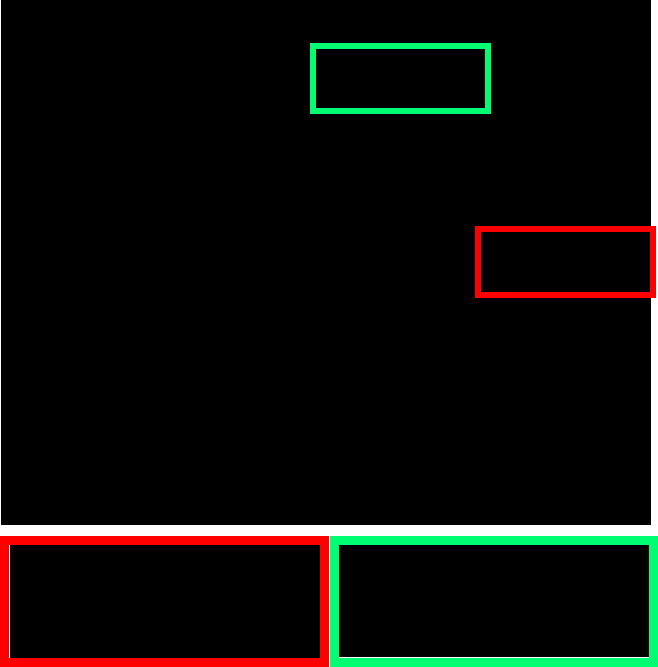}\vspace{1pt}
    \end{minipage}
    }
    \subfigure[DSRNet]{
    \begin{minipage}[b]{0.09\textwidth}
    \includegraphics[width=1.0\linewidth]{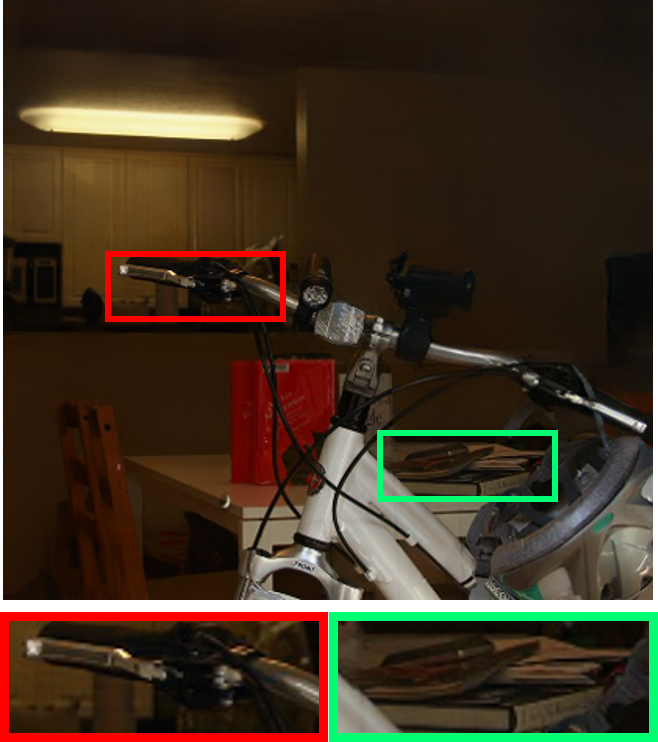}\vspace{0.5pt}
    \includegraphics[width=1.0\linewidth]{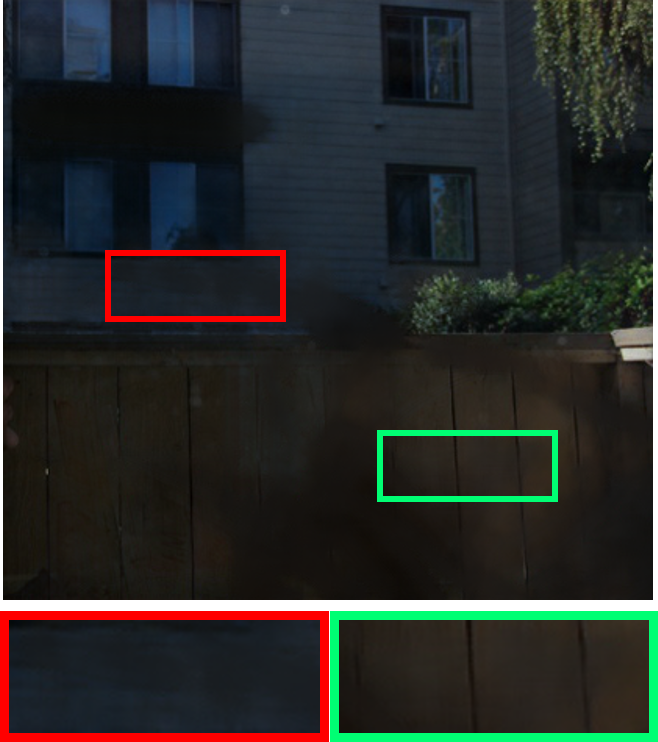}\vspace{1pt}
    \includegraphics[width=1.0\linewidth]{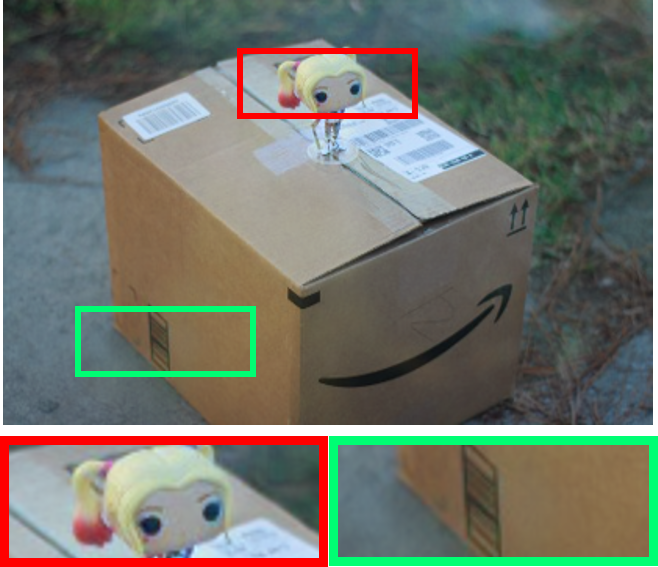}\vspace{0.5pt}
    \includegraphics[width=1.0\linewidth]{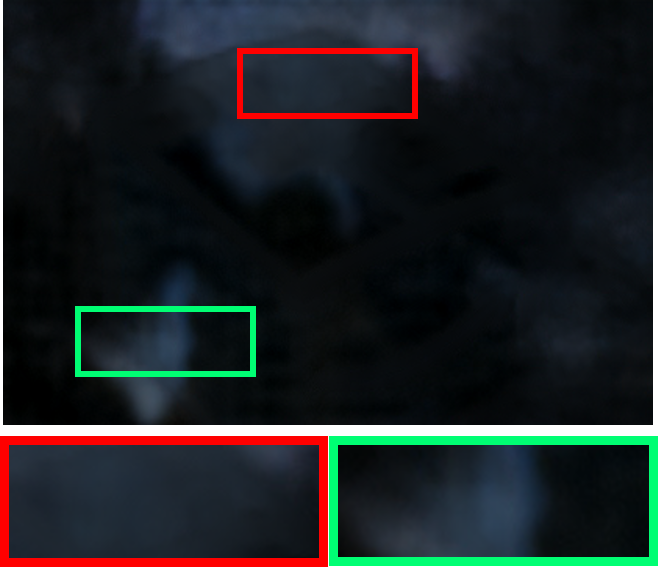}\vspace{1pt}
    \includegraphics[width=1.0\linewidth]{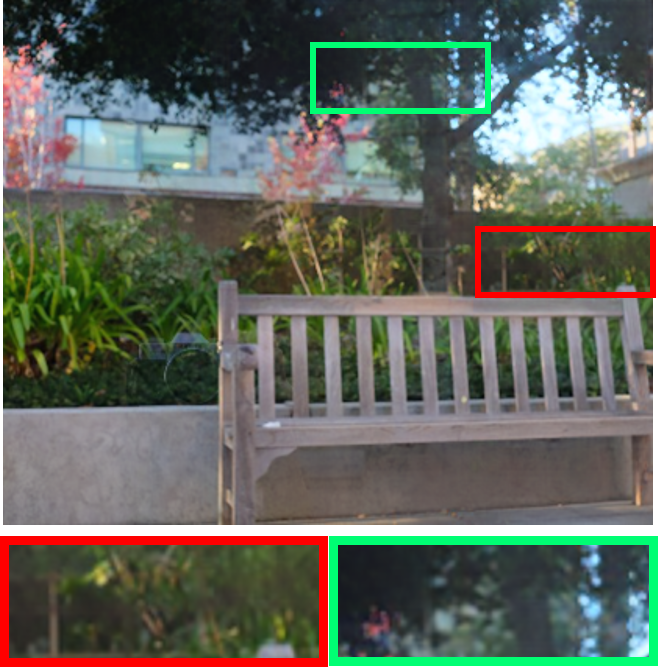}\vspace{0.5pt}
    \includegraphics[width=1.0\linewidth]{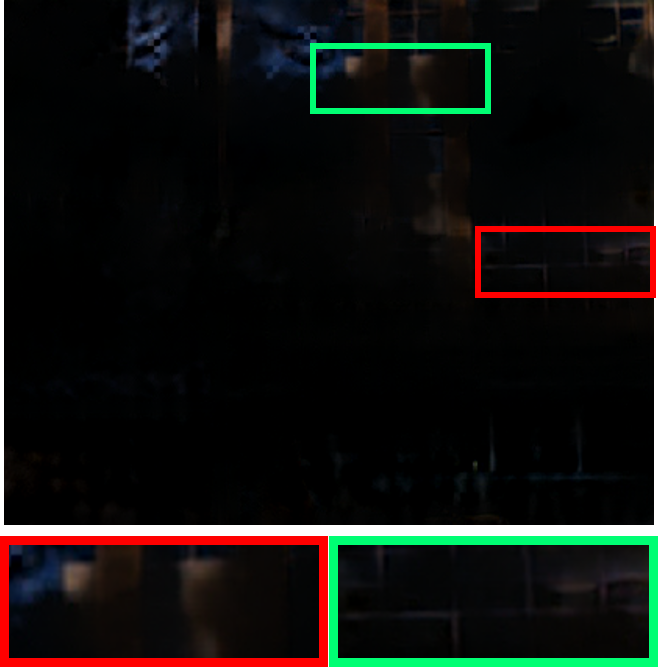}\vspace{1pt}
    \end{minipage}
    }
    \subfigure[DURRNet]{
        \begin{minipage}[b]{0.09\textwidth}
        \includegraphics[width=1.0\linewidth]{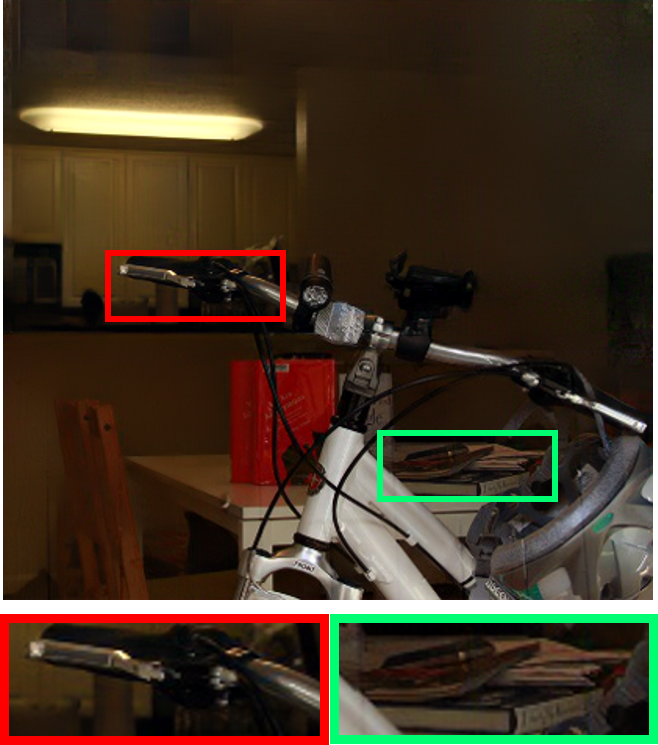}\vspace{0.5pt}
        \includegraphics[width=1.0\linewidth]{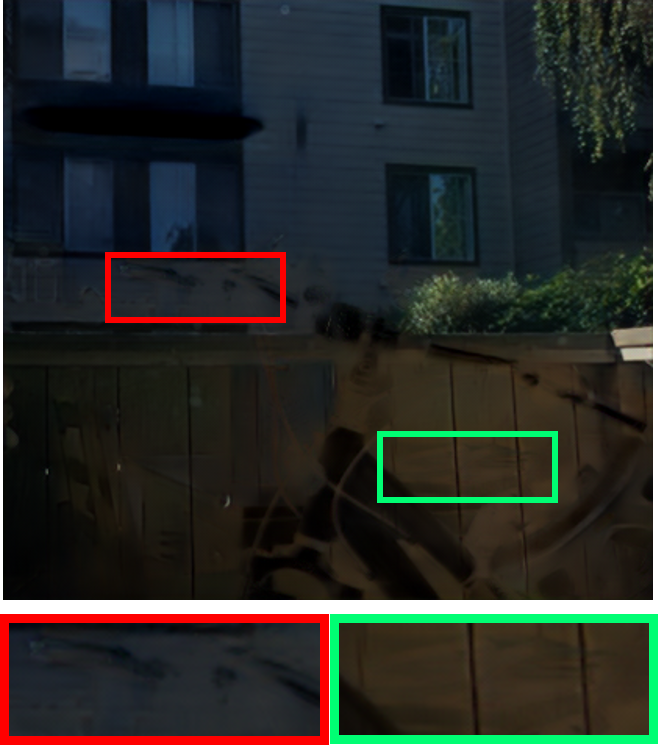}\vspace{1pt}
        \includegraphics[width=1.0\linewidth]{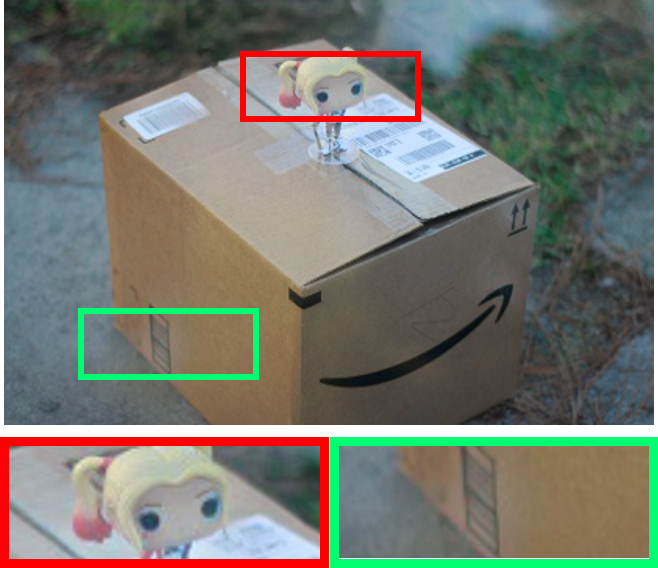}\vspace{0.5pt}
        \includegraphics[width=1.0\linewidth]{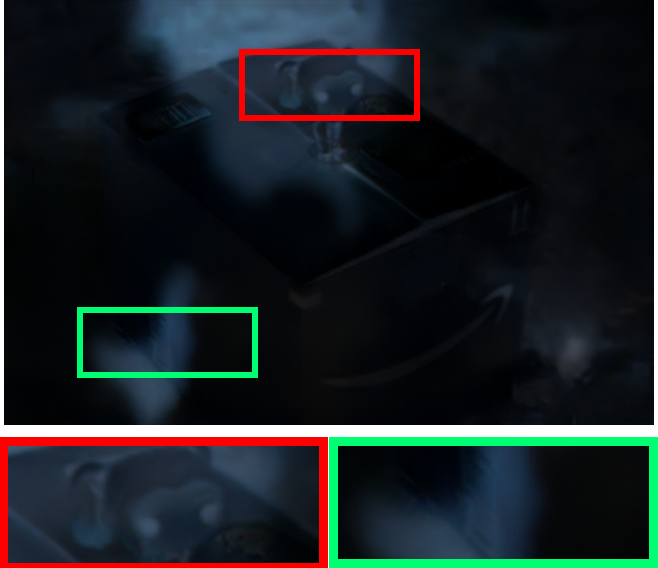}\vspace{1pt}
        \includegraphics[width=1.0\linewidth]{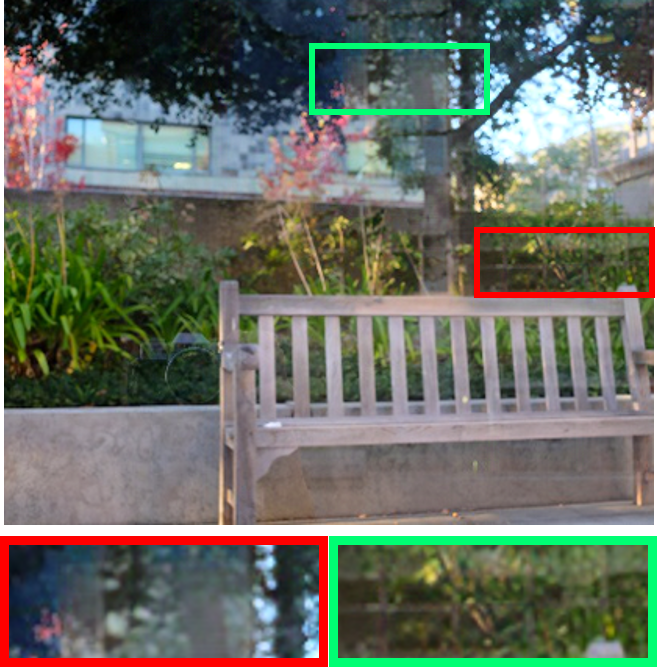}\vspace{0.5pt}
        \includegraphics[width=1.0\linewidth]{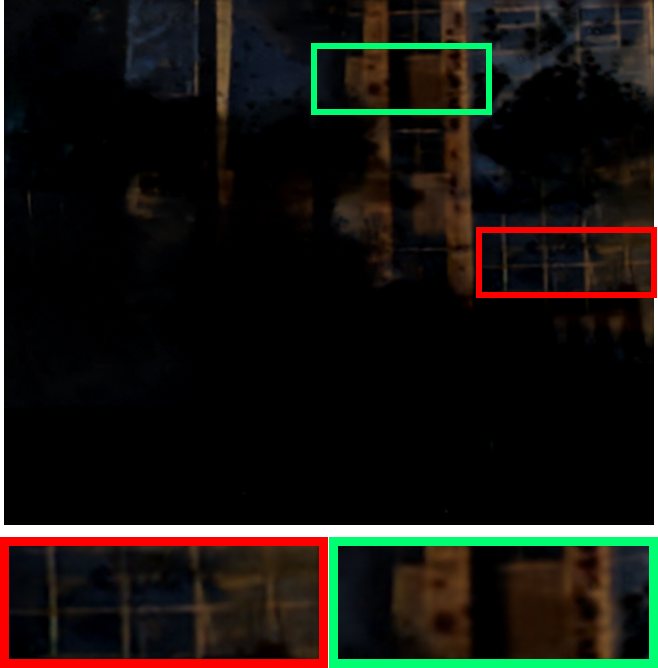}\vspace{1pt}
        \end{minipage}
    }
    \subfigure[DExNet]{
    \begin{minipage}[b]{0.09\textwidth}
    \includegraphics[width=1.0\linewidth]{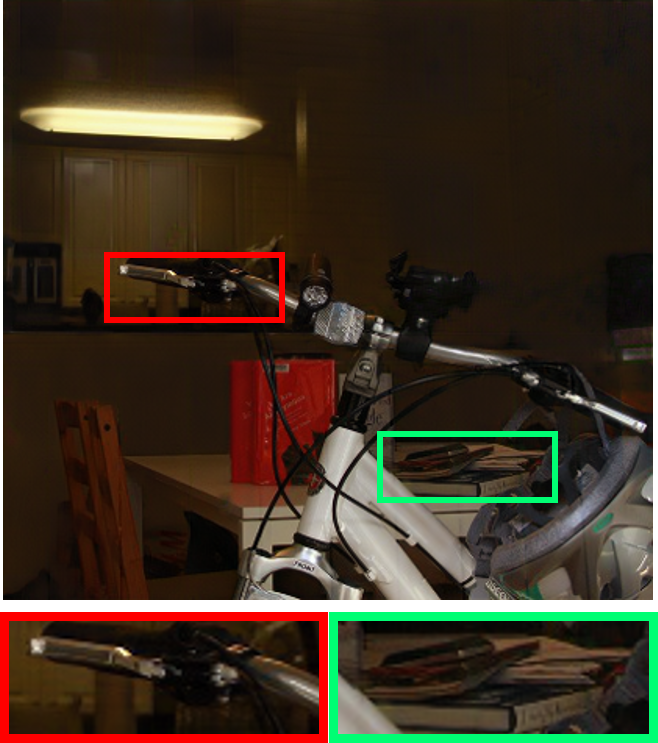}\vspace{0.5pt}
    \includegraphics[width=1.0\linewidth]{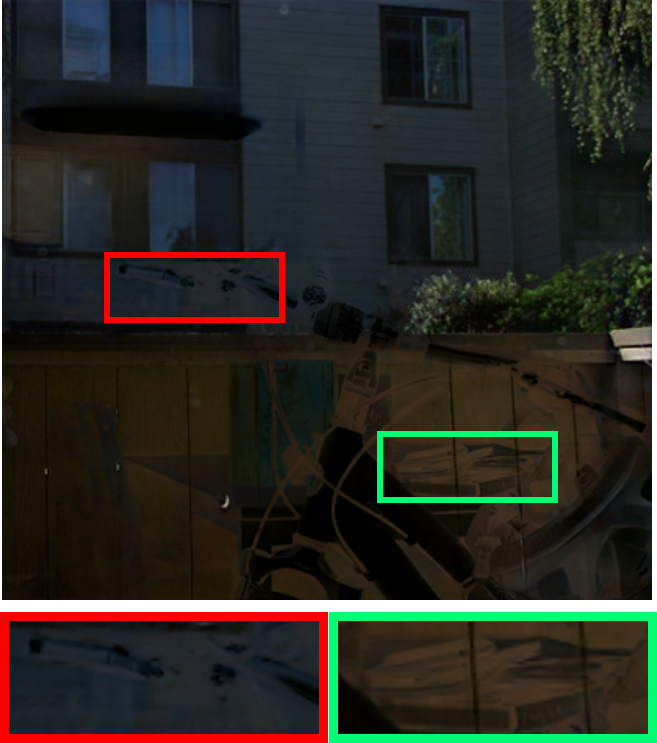}\vspace{1pt}
    \includegraphics[width=1.0\linewidth]{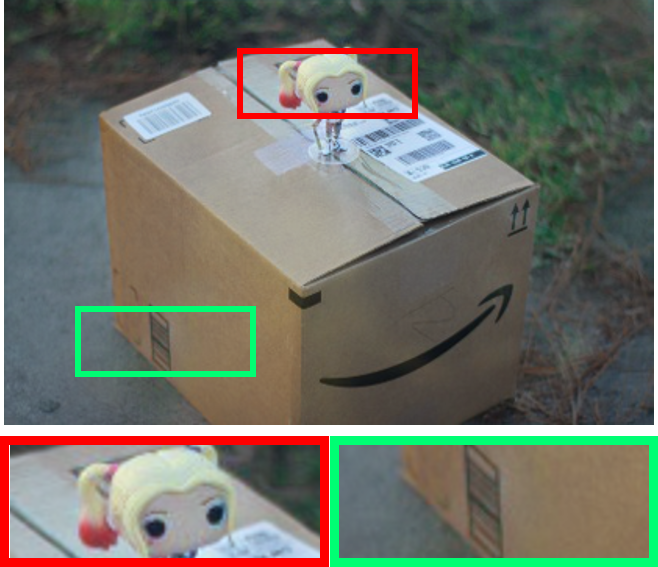}\vspace{0.5pt}
    \includegraphics[width=1.0\linewidth]{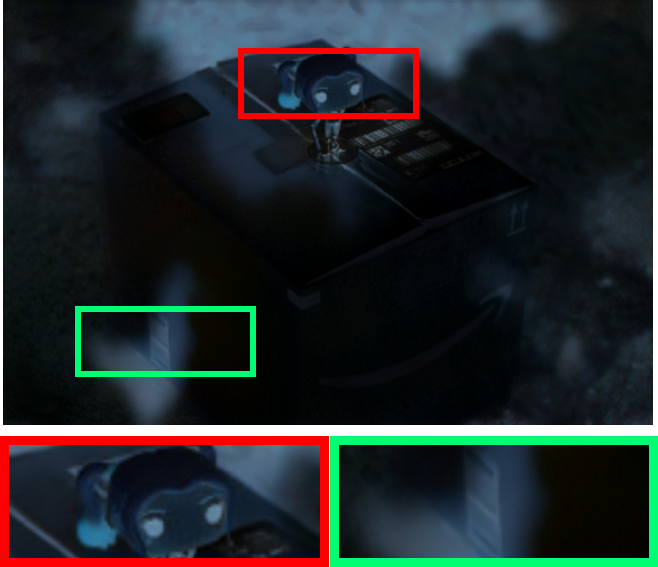}\vspace{1pt}
    \includegraphics[width=1.0\linewidth]{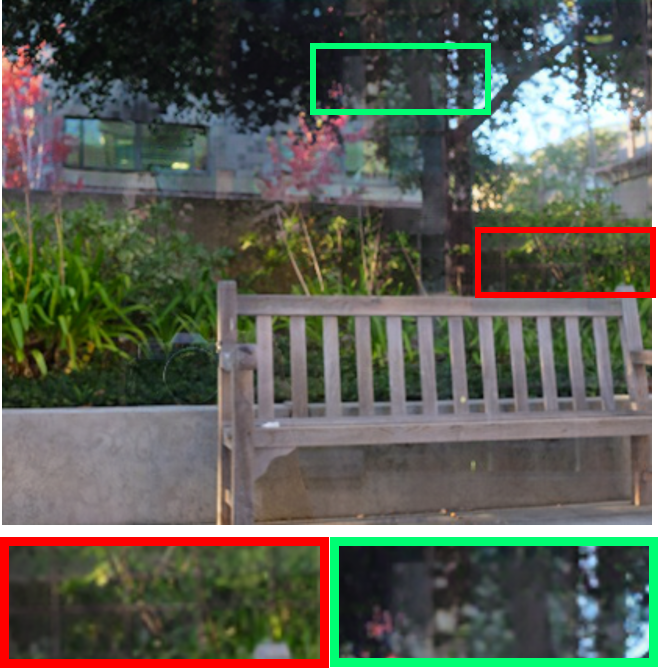}\vspace{0.5pt}
    \includegraphics[width=1.0\linewidth]{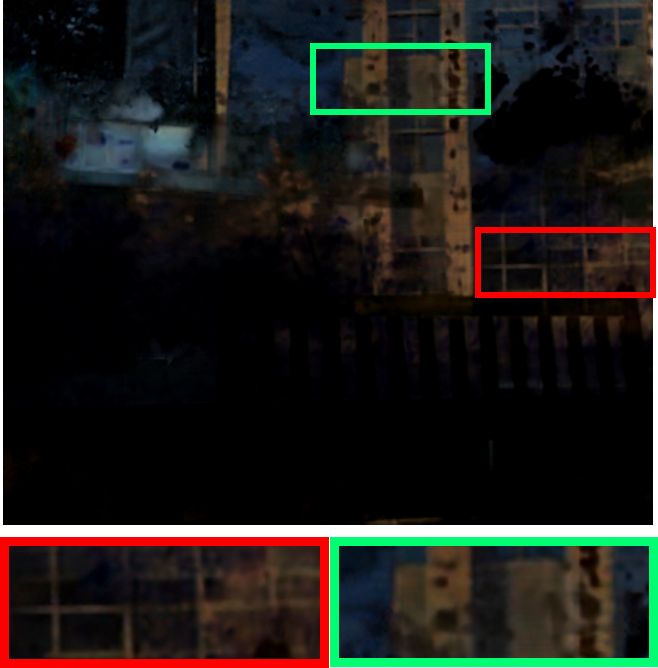}\vspace{1pt}
    \end{minipage}
    }
    \subfigure[GT]{
    \begin{minipage}[b]{0.09\textwidth}
    \includegraphics[width=1.0\linewidth]{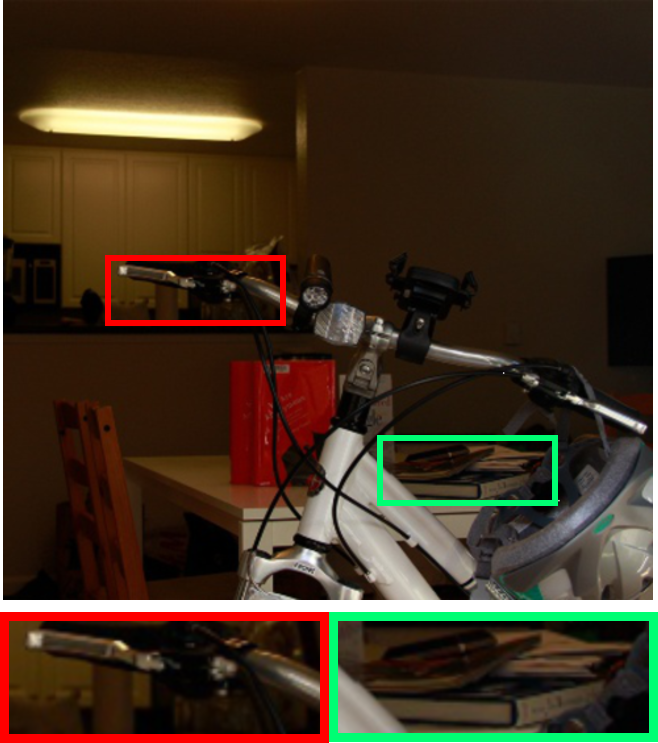}\vspace{0.5pt}
    \includegraphics[width=1.0\linewidth]{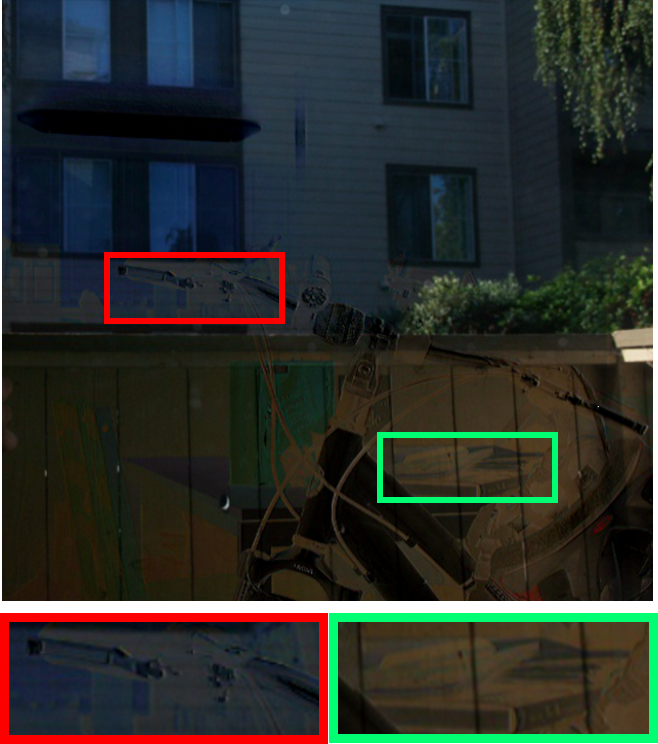}\vspace{1pt}
    \includegraphics[width=1.0\linewidth]{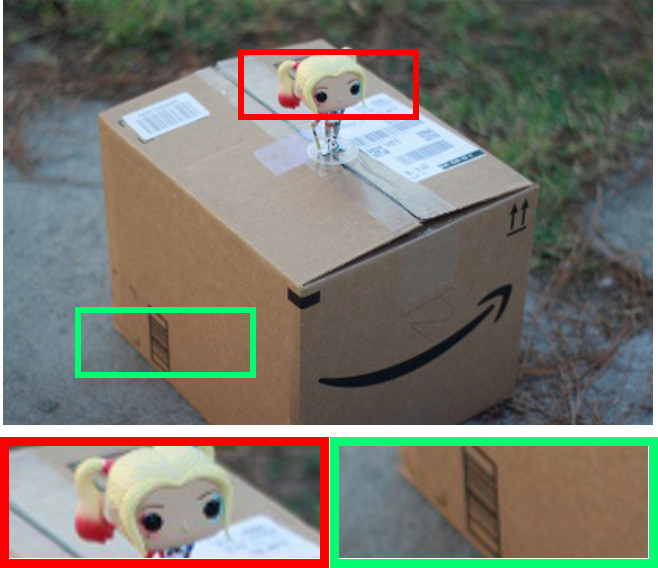}\vspace{0.5pt}
    \includegraphics[width=1.0\linewidth]{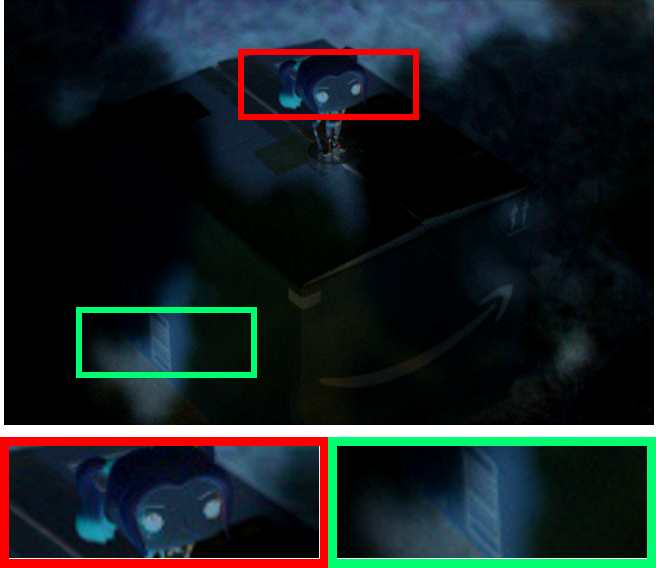}\vspace{1pt}
    \includegraphics[width=1.0\linewidth]{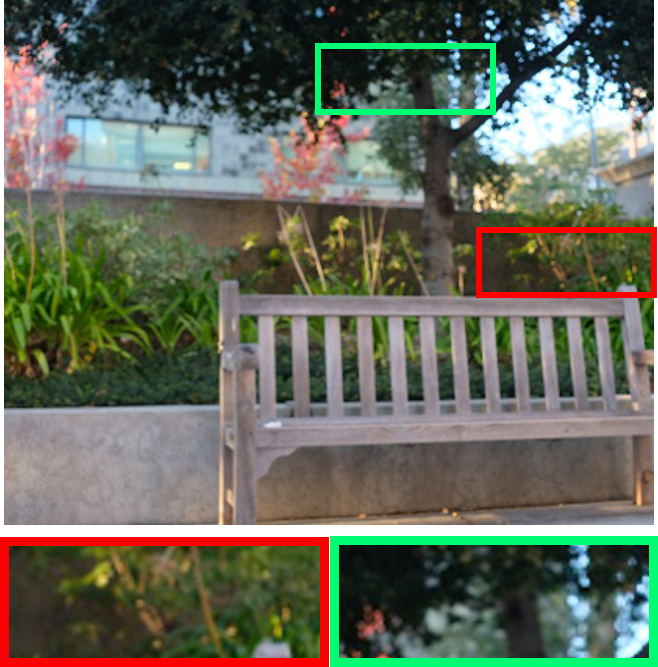}\vspace{0.5pt}
    \includegraphics[width=1.0\linewidth]{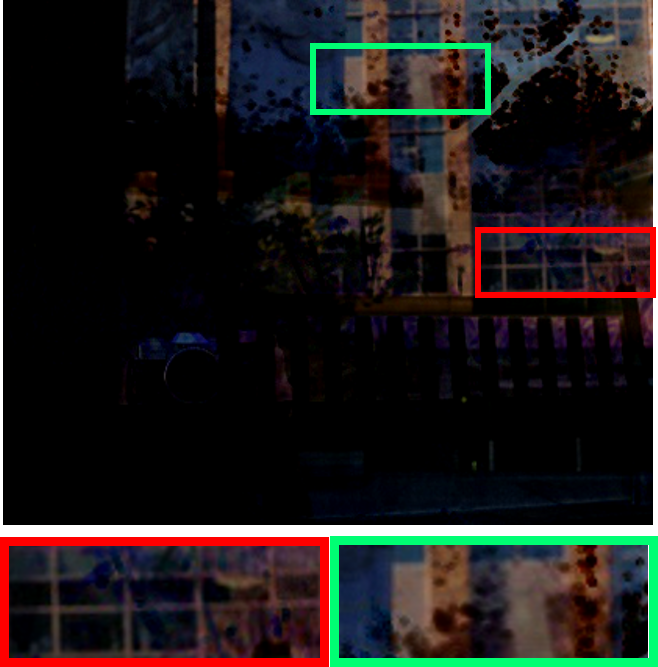}\vspace{1pt}
    \end{minipage}
    }
    
   \caption{Visual comparisons on \textit{Real20} dataset~\cite{perceptual_loss_2018} of different single image reflection methods. The estimated transmission images are shown in the odd rows and the estimated reflection images are shown in the even rows. The last column shows the ground-truth transmission image and reflection image for reference.}
    \label{fig:real20}
\end{figure*}
% -----------------------------------------------------------

%-----------------------------------------------------------
\begin{figure*}[t]
    \centering
    \subfigure[Input]{
    \begin{minipage}[b]{0.09\textwidth}
    \includegraphics[width=1\linewidth]{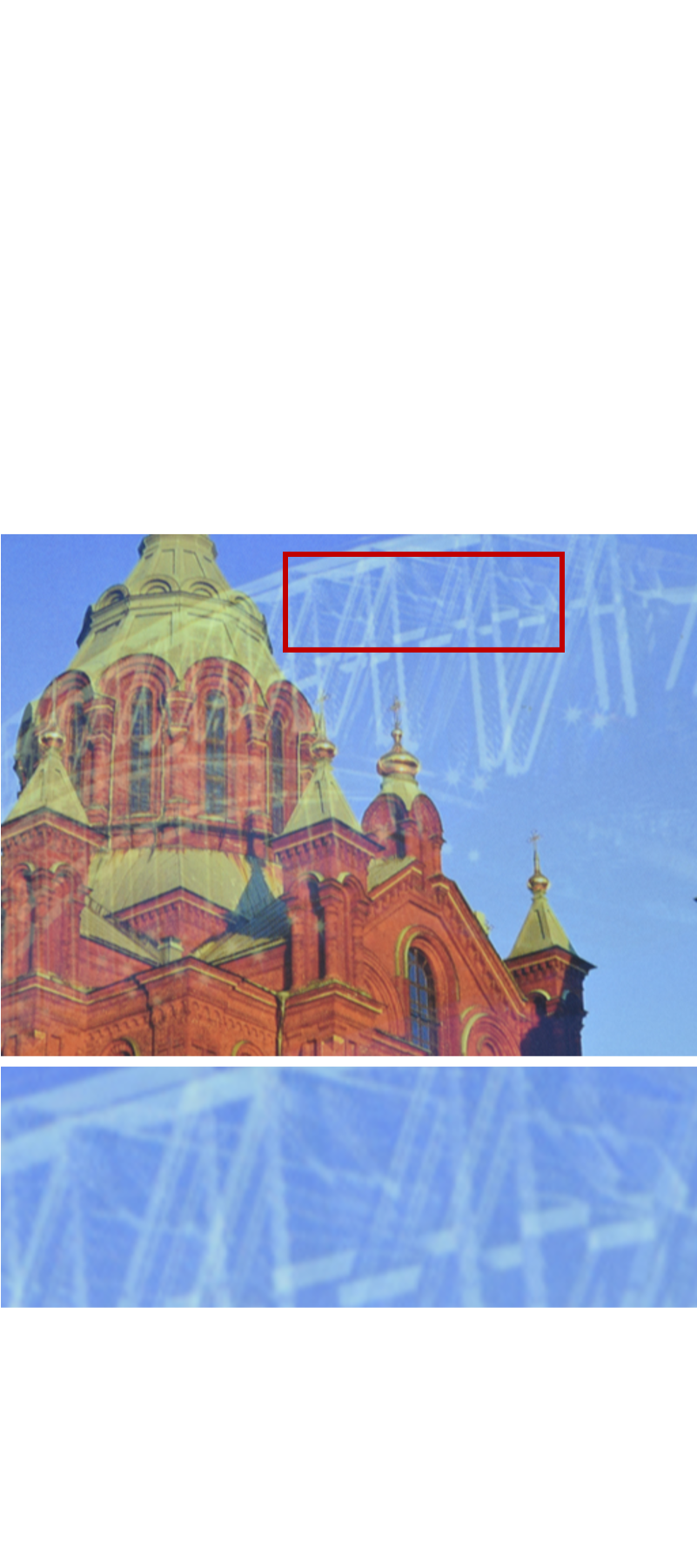}\vspace{1pt}
    \includegraphics[width=1\linewidth]{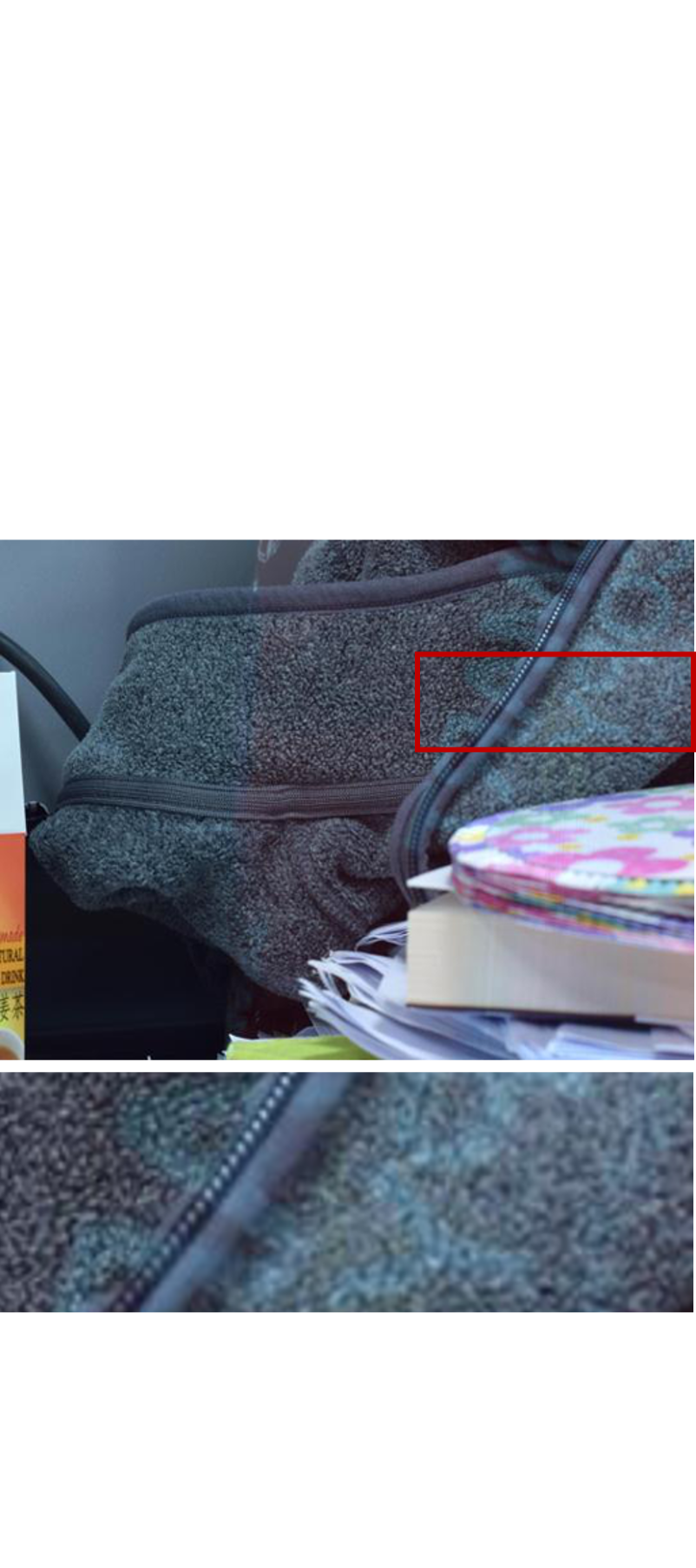}\vspace{1pt}
    \includegraphics[width=1\linewidth]{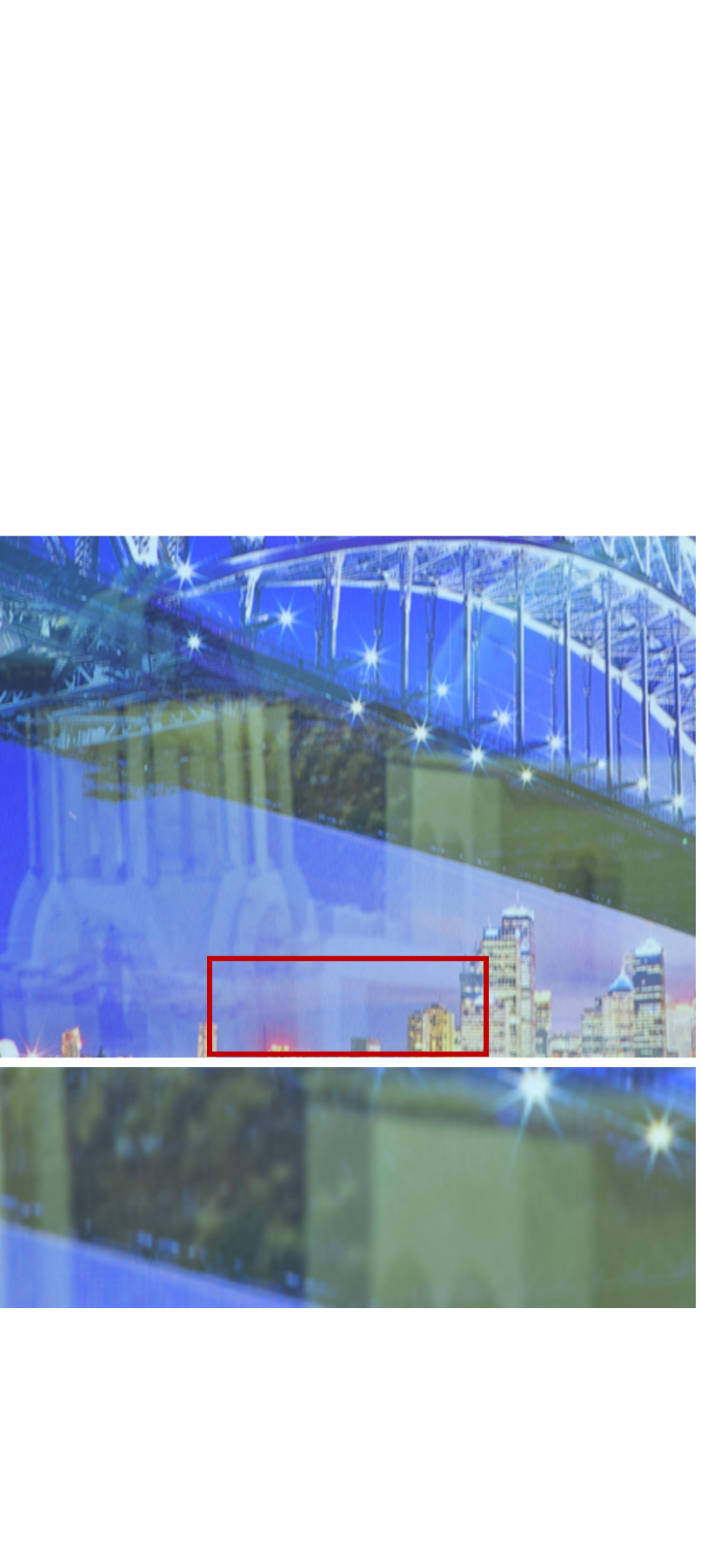}\vspace{1pt}
    \end{minipage}
    }
    \subfigure[Zhang]{
    \begin{minipage}[b]{0.09\textwidth}
    \includegraphics[width=1\linewidth]{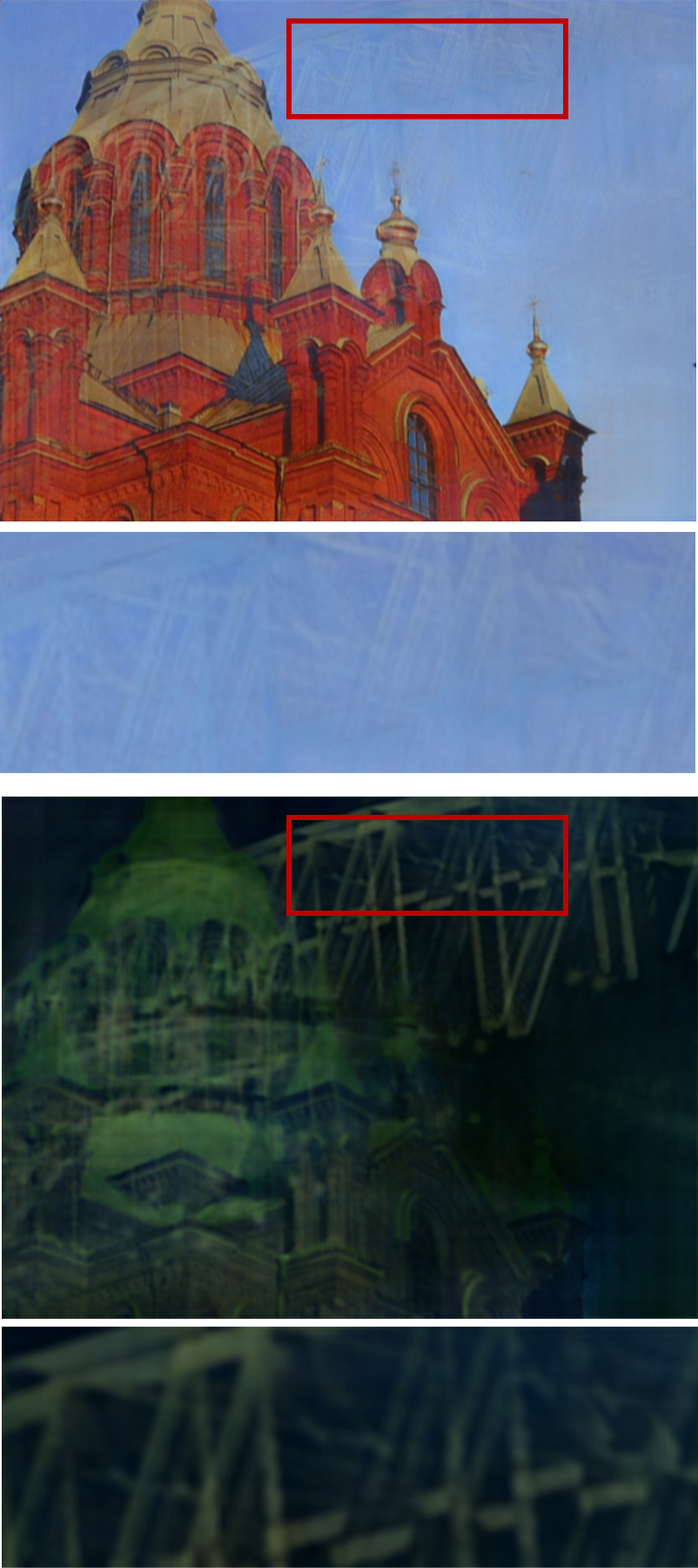}\vspace{1pt}
    \includegraphics[width=1\linewidth]{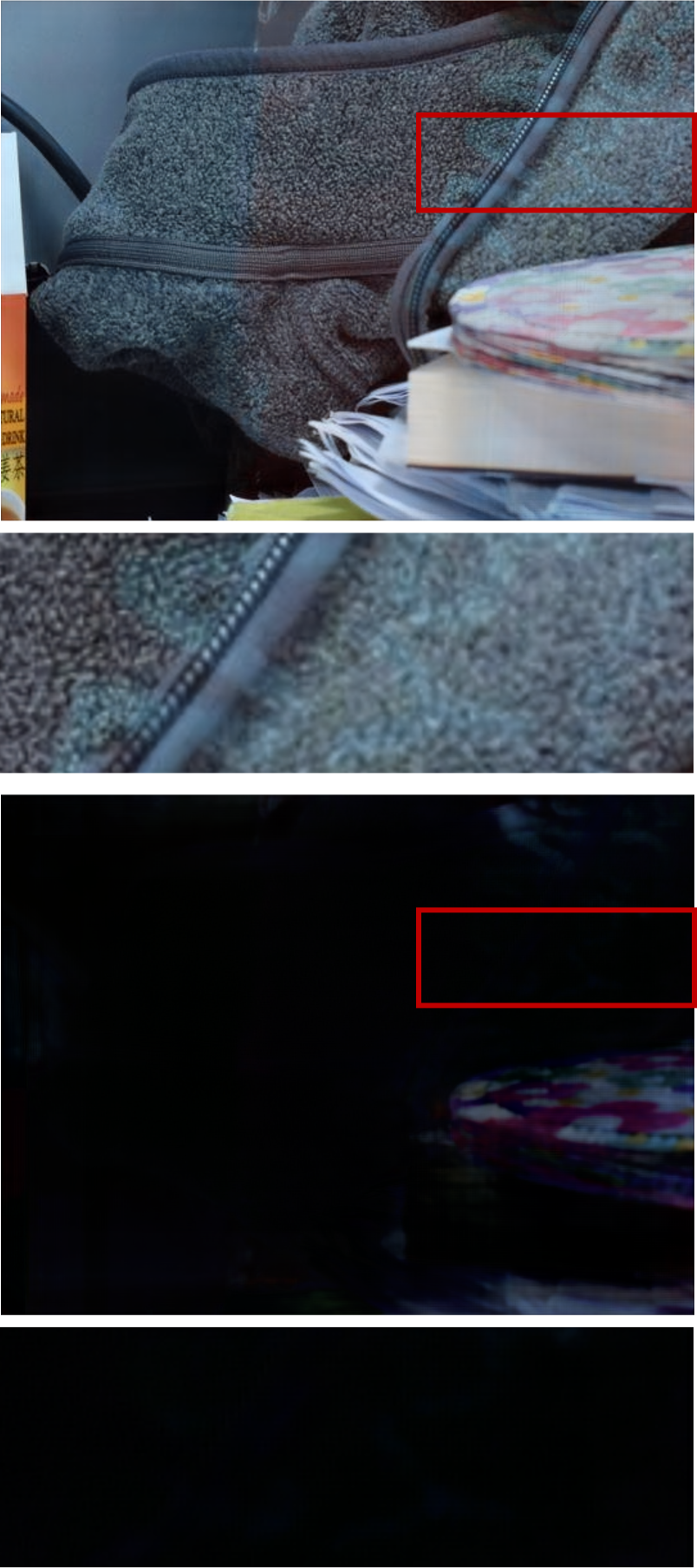}\vspace{1pt}
    \includegraphics[width=1\linewidth]{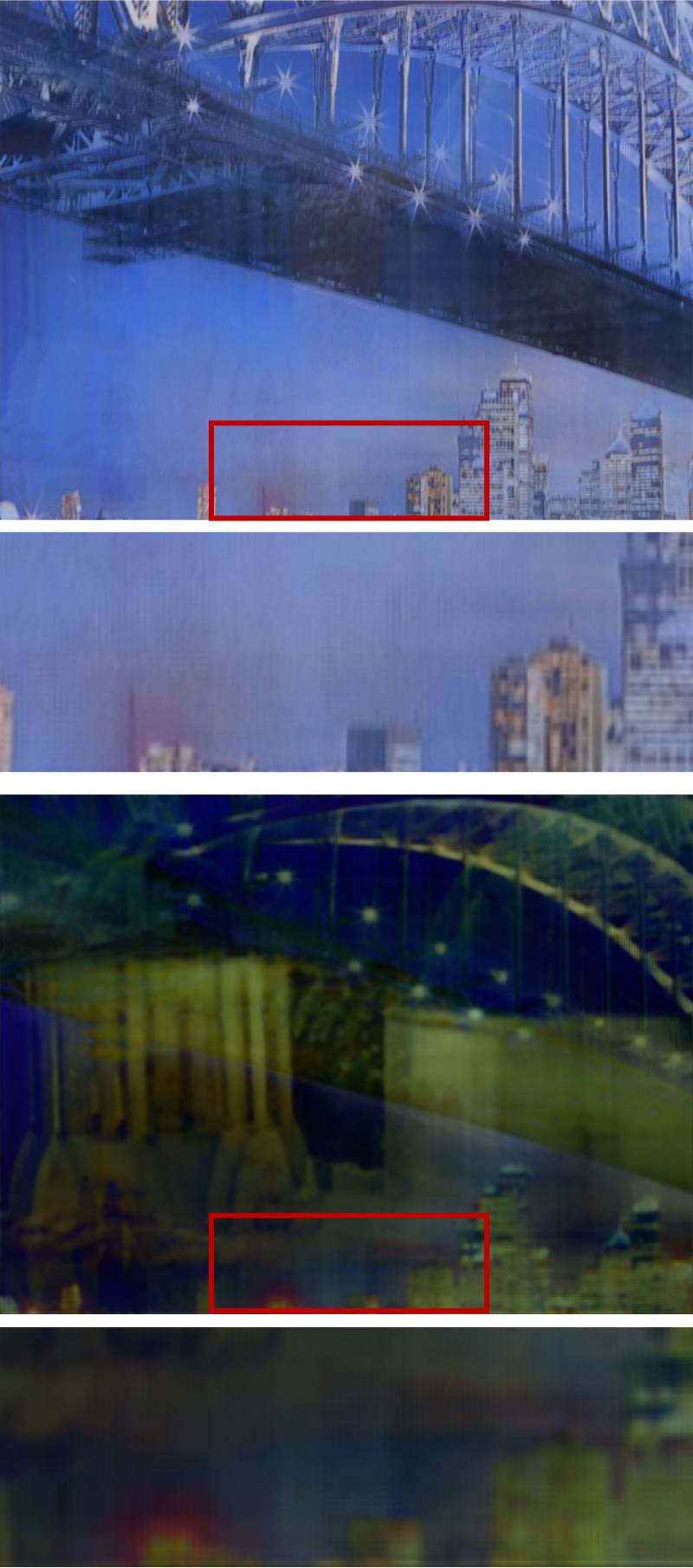}\vspace{1pt}
    \end{minipage}
    }
    \subfigure[IBCLN]{
    \begin{minipage}[b]{0.09\textwidth}
    \includegraphics[width=1\linewidth]{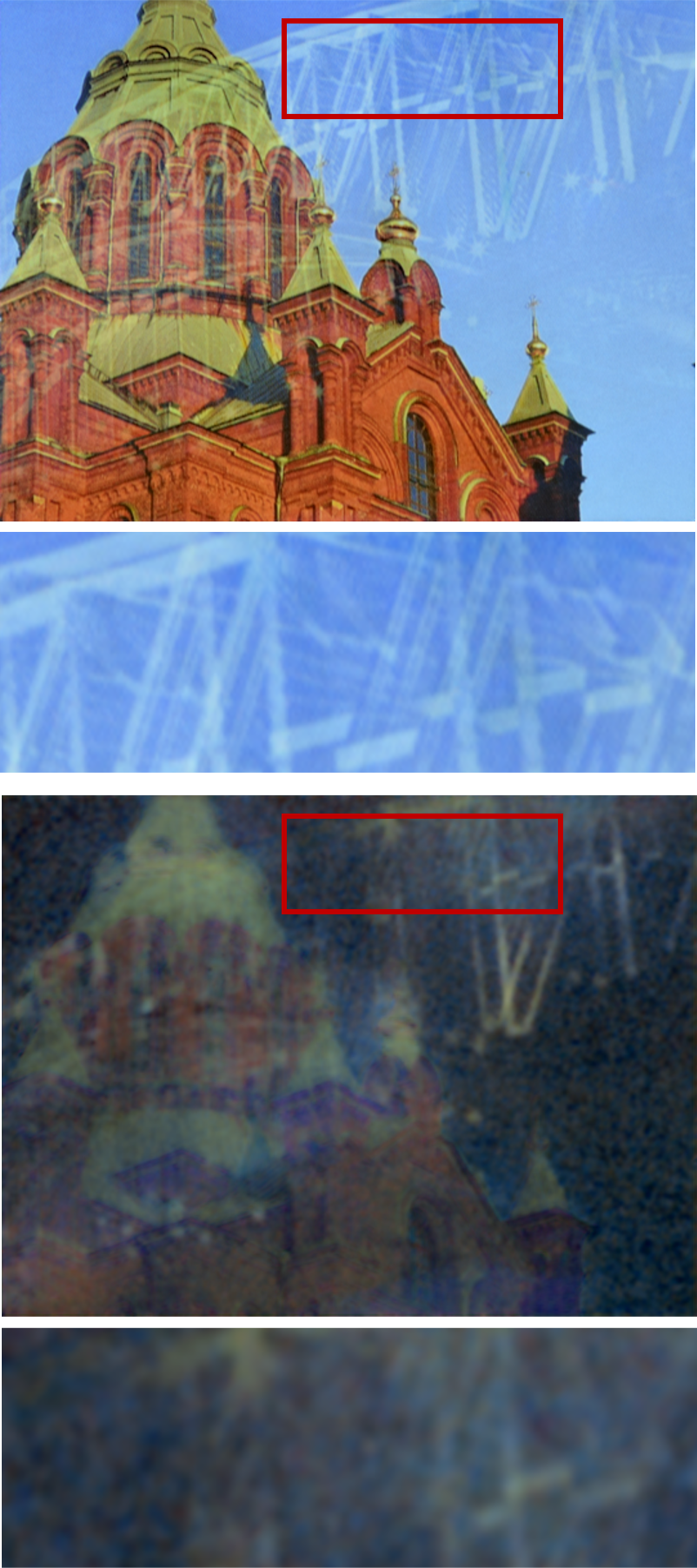}\vspace{1pt}
    \includegraphics[width=1\linewidth]{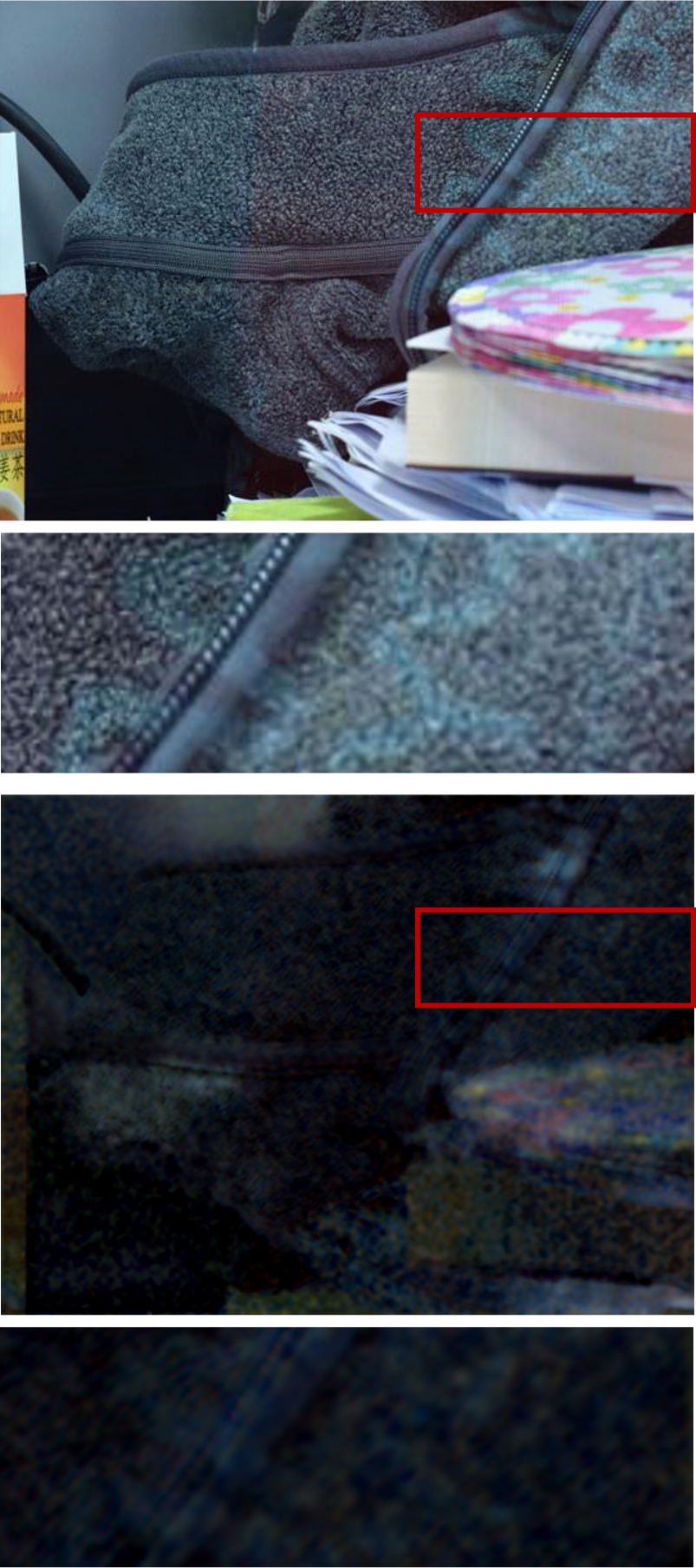}\vspace{1pt}
    \includegraphics[width=1\linewidth]{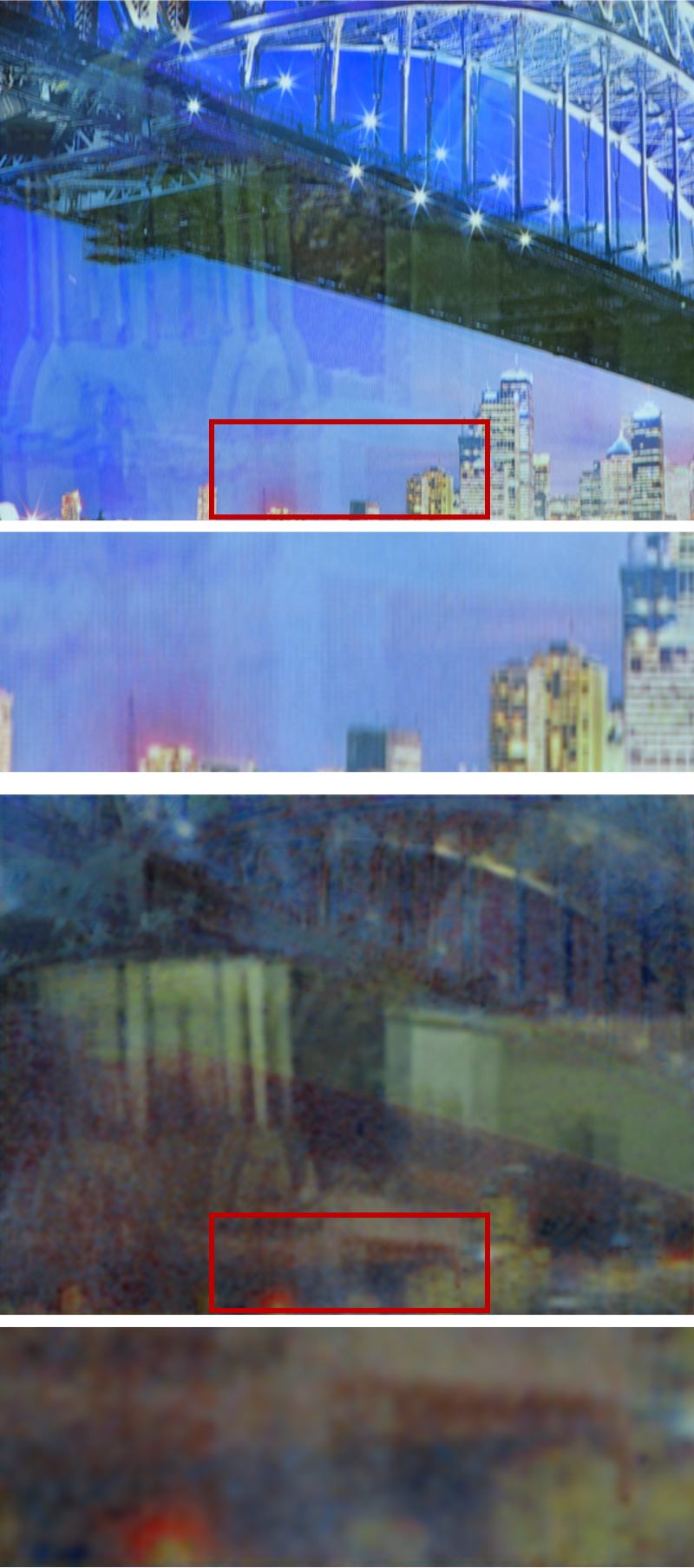}\vspace{1pt}
    \end{minipage}
    }
    \subfigure[ERRNet]{
    \begin{minipage}[b]{0.09\textwidth}
    \includegraphics[width=1\linewidth]{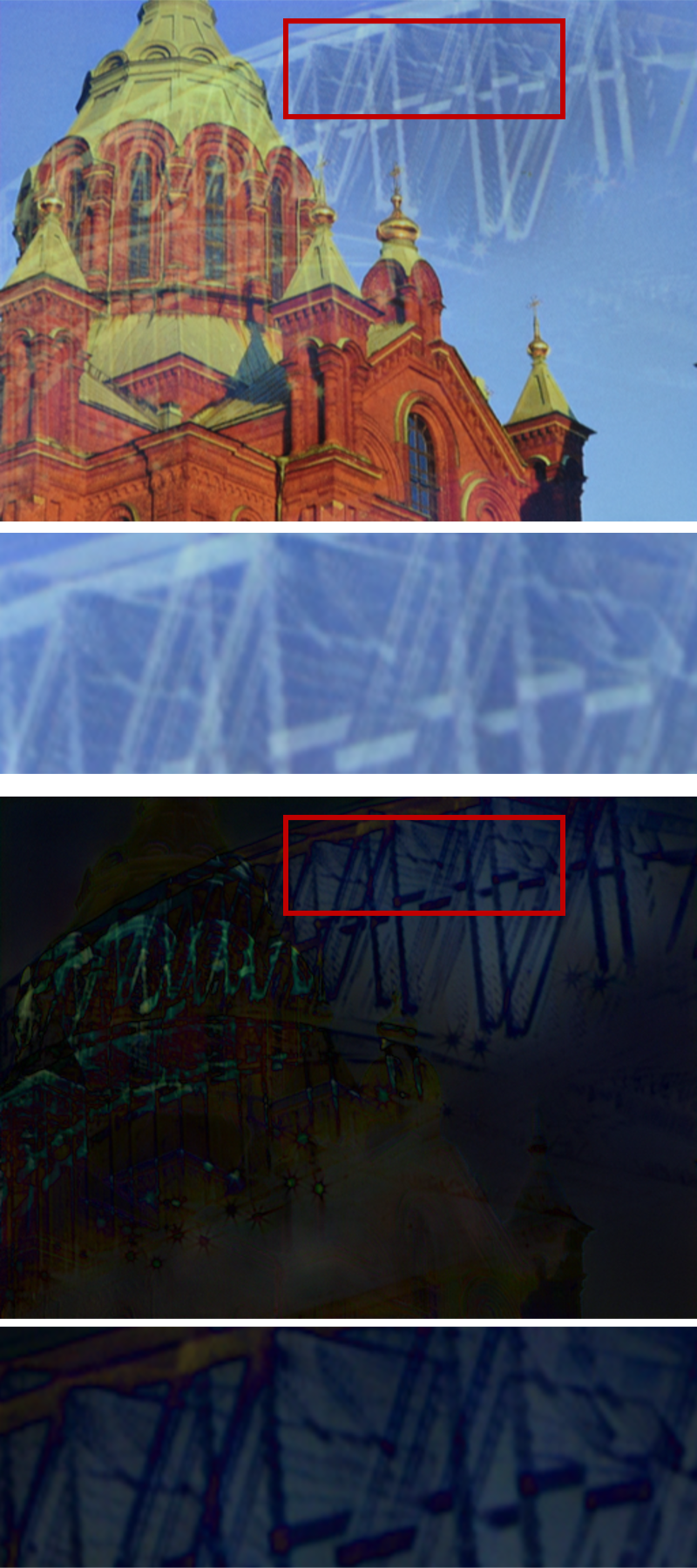}\vspace{1pt}
    \includegraphics[width=1\linewidth]{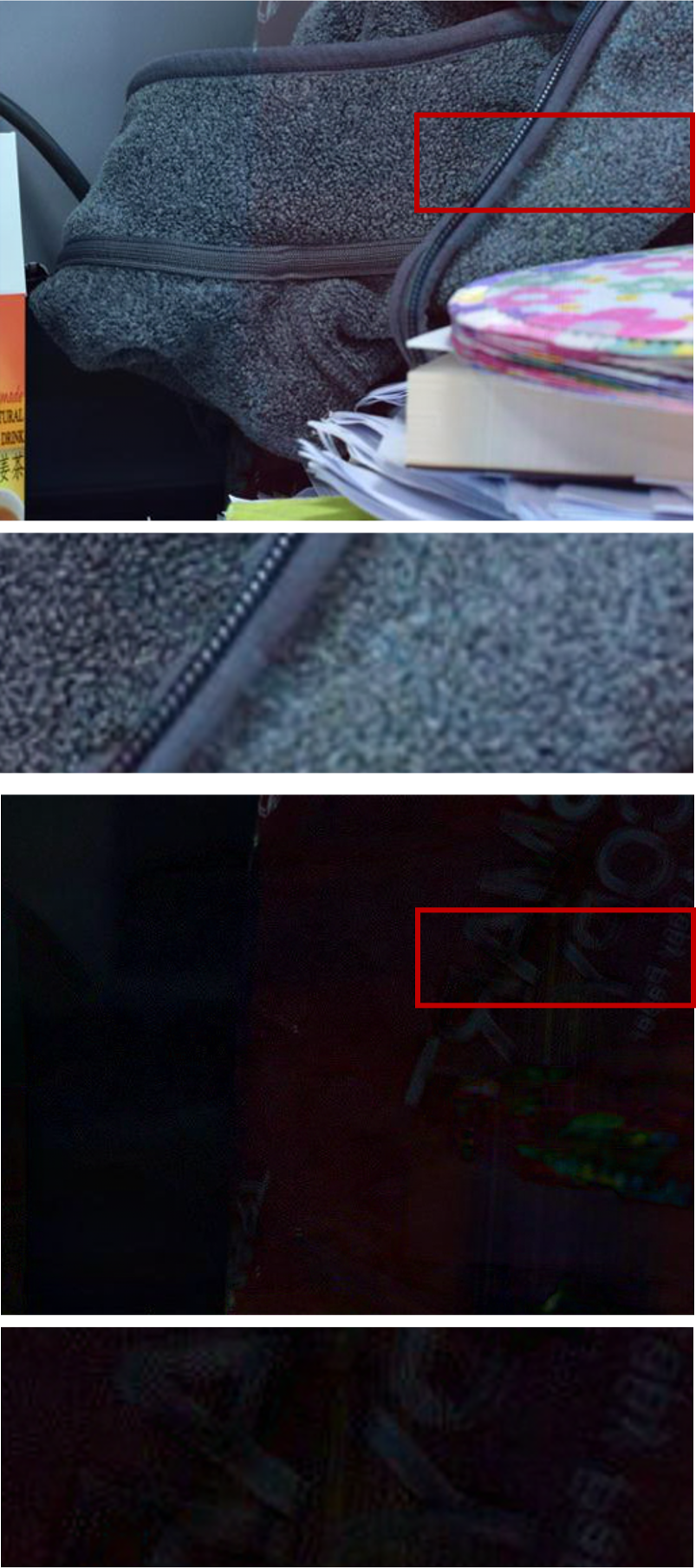}\vspace{1pt}
    \includegraphics[width=1\linewidth]{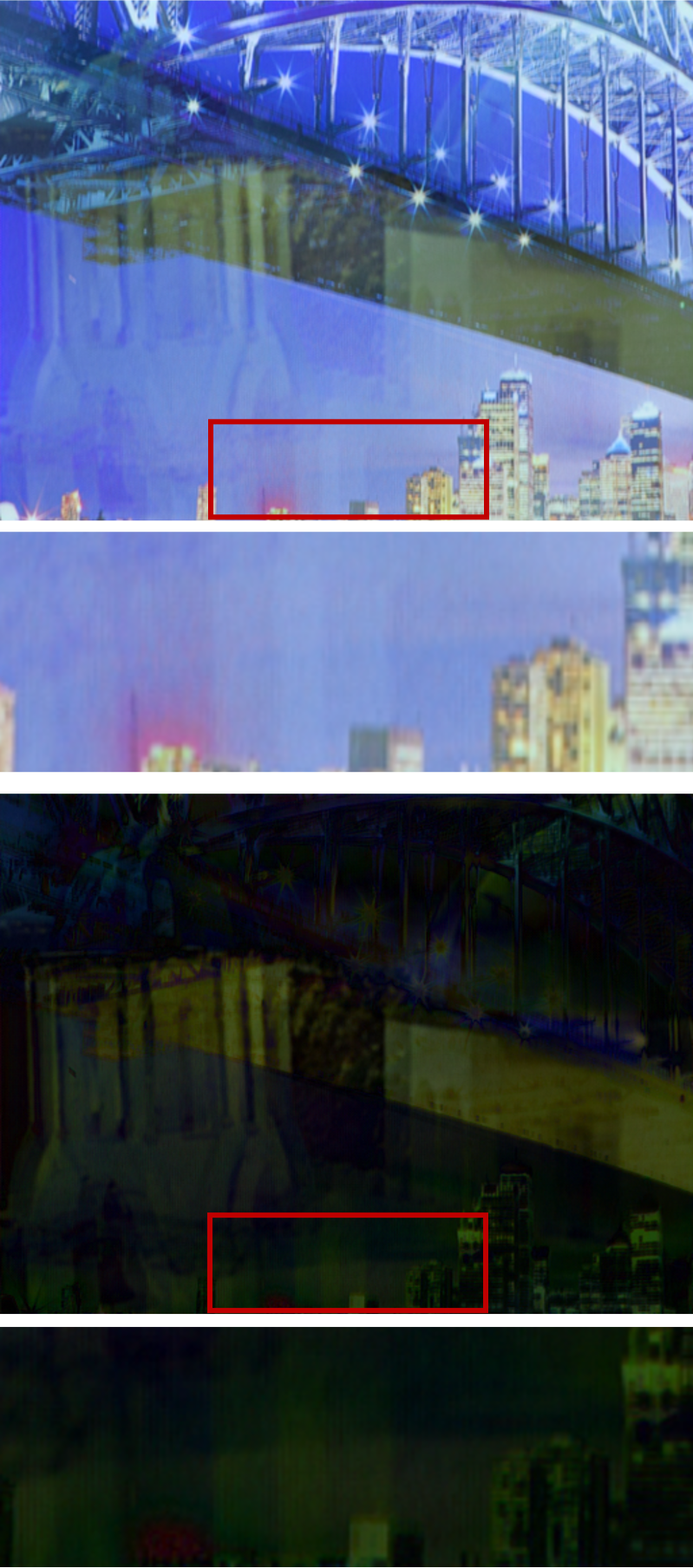}\vspace{1pt}
    \end{minipage}
    }
    \subfigure[YTMT]{
    \begin{minipage}[b]{0.09\textwidth}
    \includegraphics[width=1\linewidth]{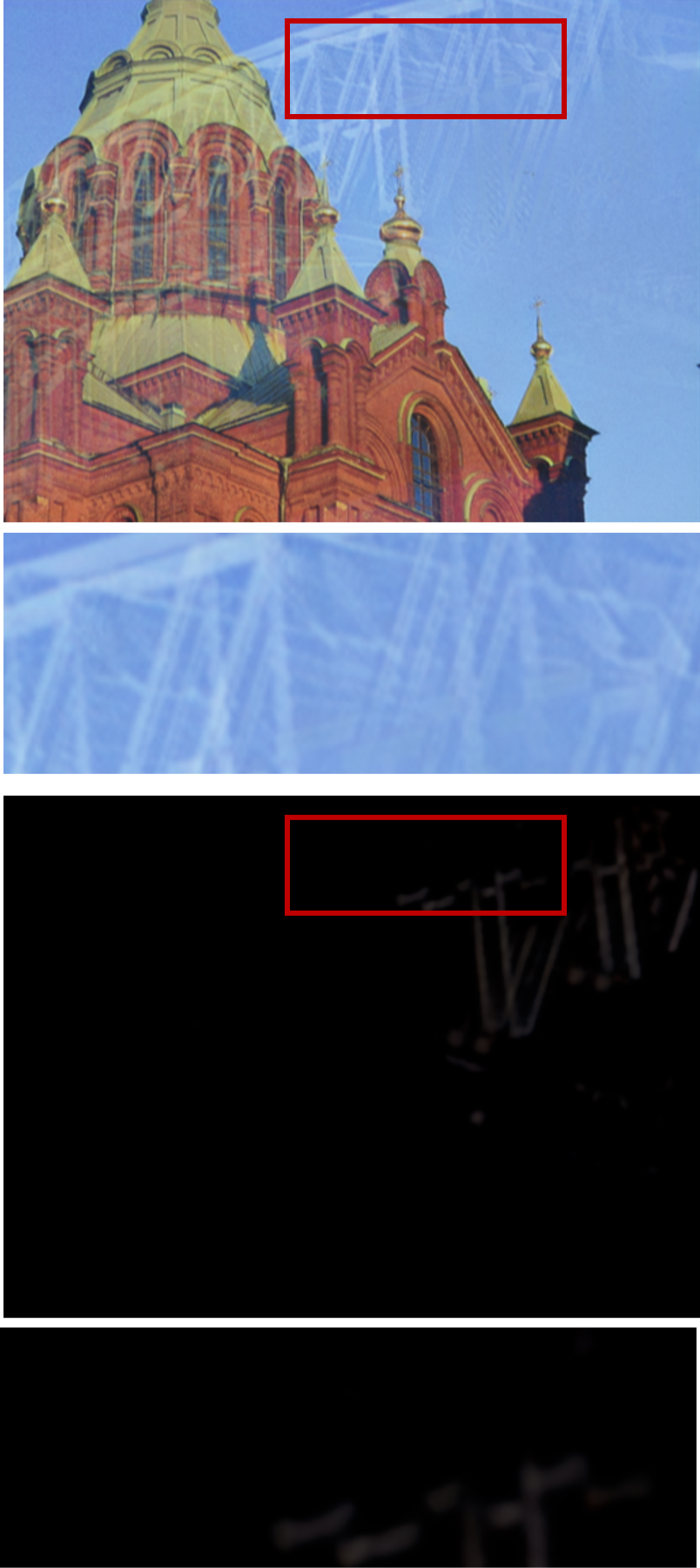}\vspace{1pt}
    \includegraphics[width=1\linewidth]{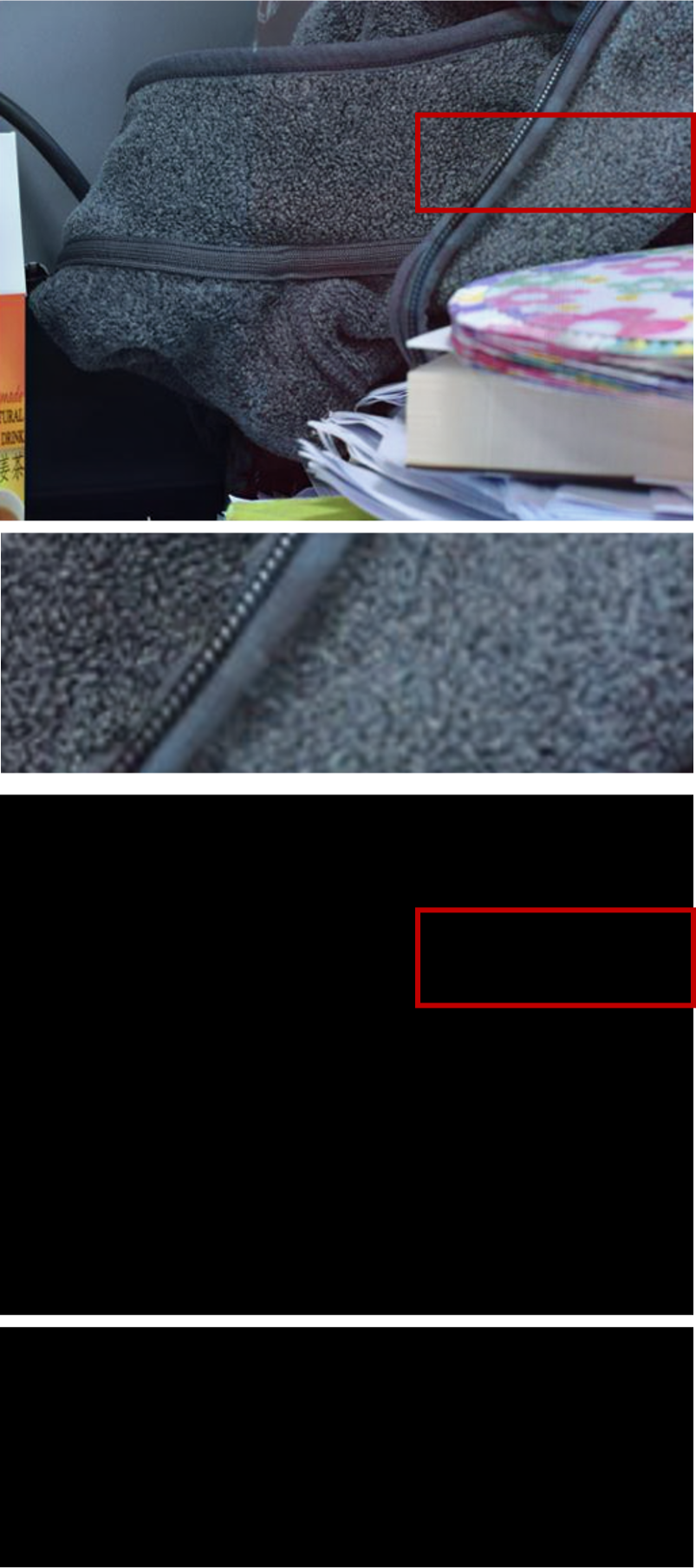}\vspace{1pt}
    \includegraphics[width=1\linewidth]{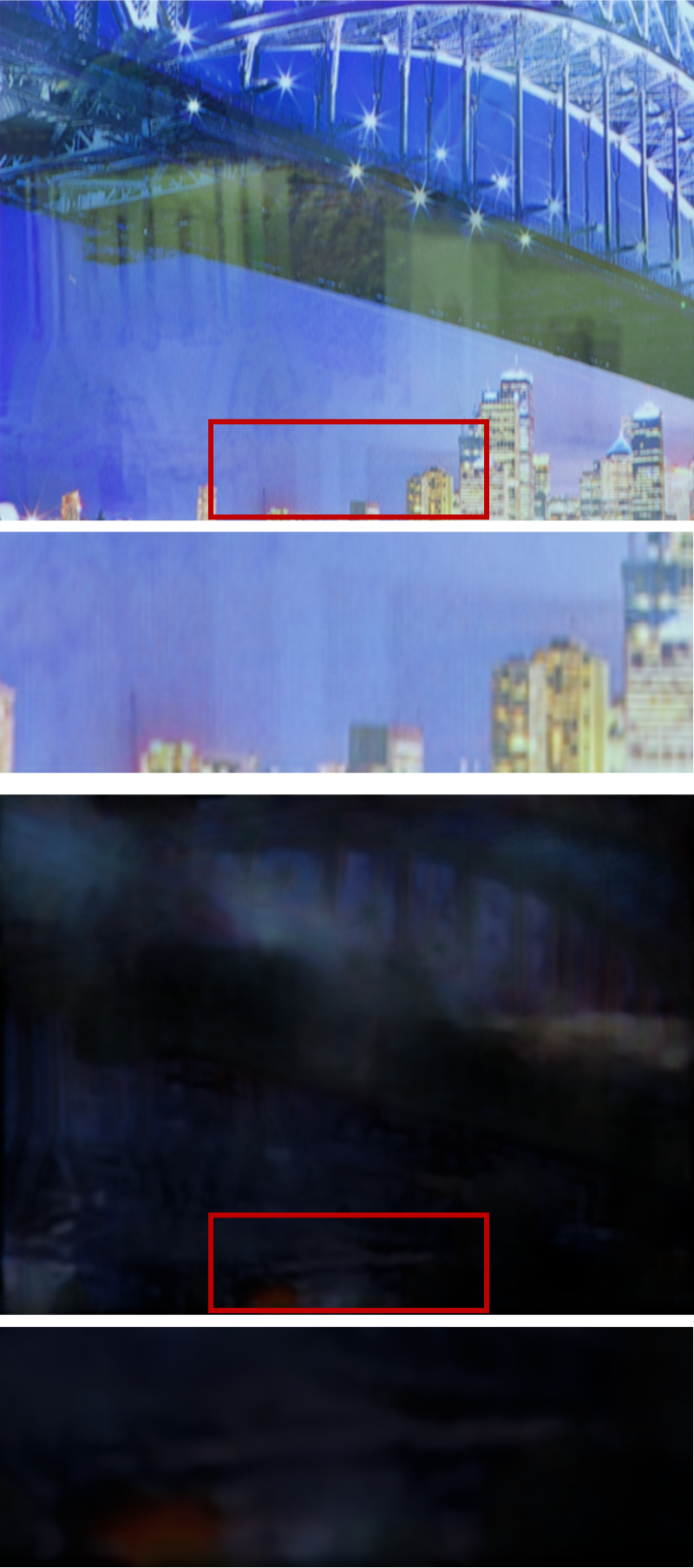}\vspace{1pt}
    \end{minipage}
    }
    \subfigure[DSRNet]{
    \begin{minipage}[b]{0.09\textwidth}
    \includegraphics[width=1\linewidth]{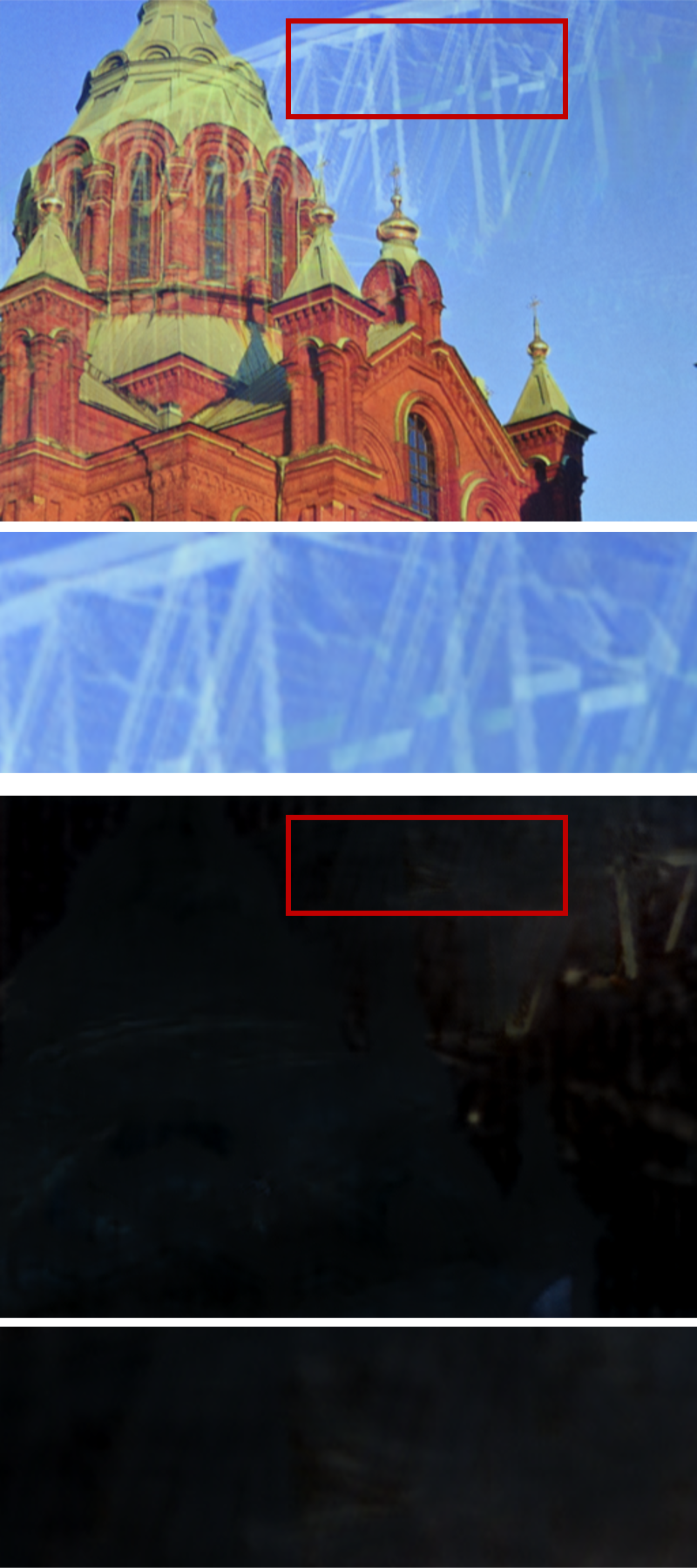}\vspace{1pt}
    \includegraphics[width=1\linewidth]{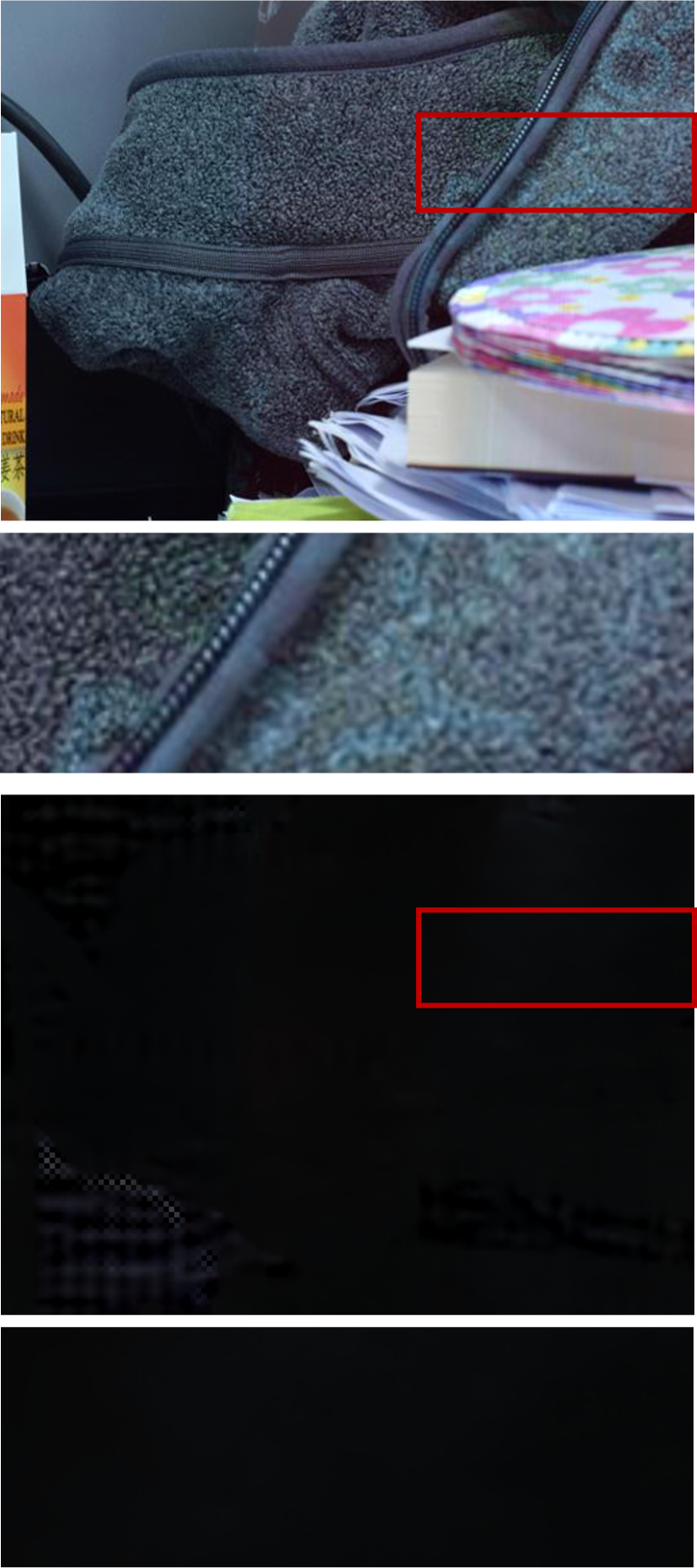}\vspace{1pt}
    \includegraphics[width=1\linewidth]{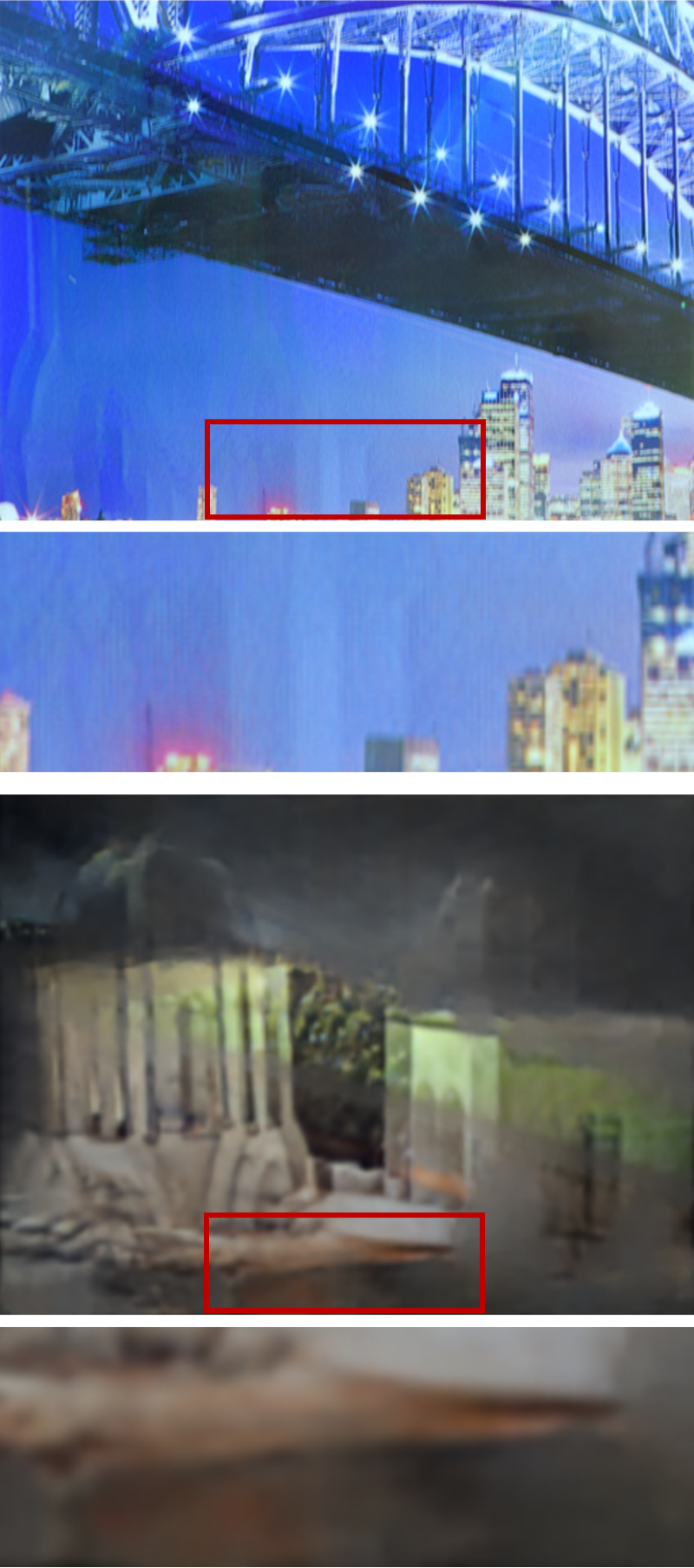}\vspace{1pt}
    \end{minipage}
    }
    \subfigure[DURRNet]{
    \begin{minipage}[b]{0.0903\textwidth}
    \includegraphics[width=1\linewidth]{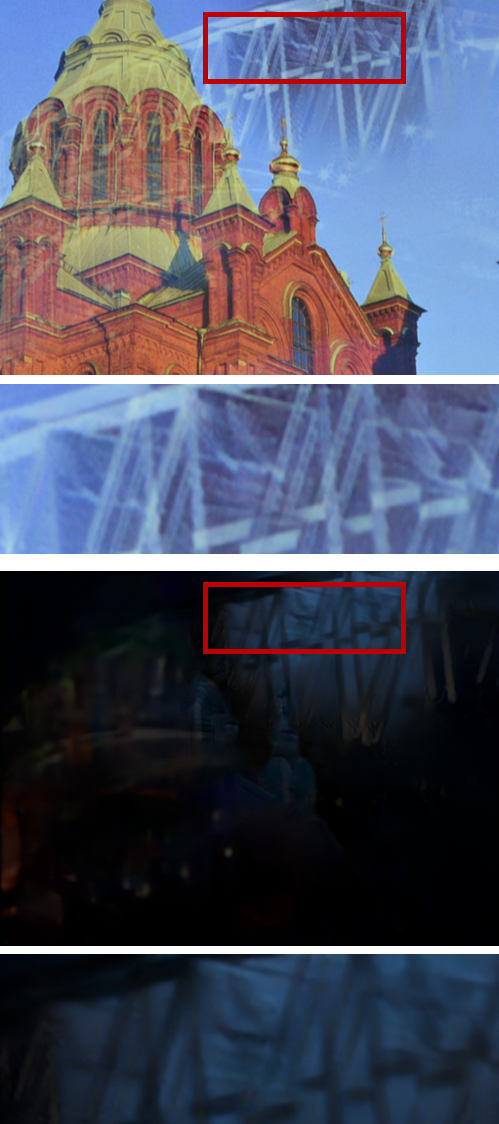}\vspace{1pt}
    \includegraphics[width=1\linewidth]{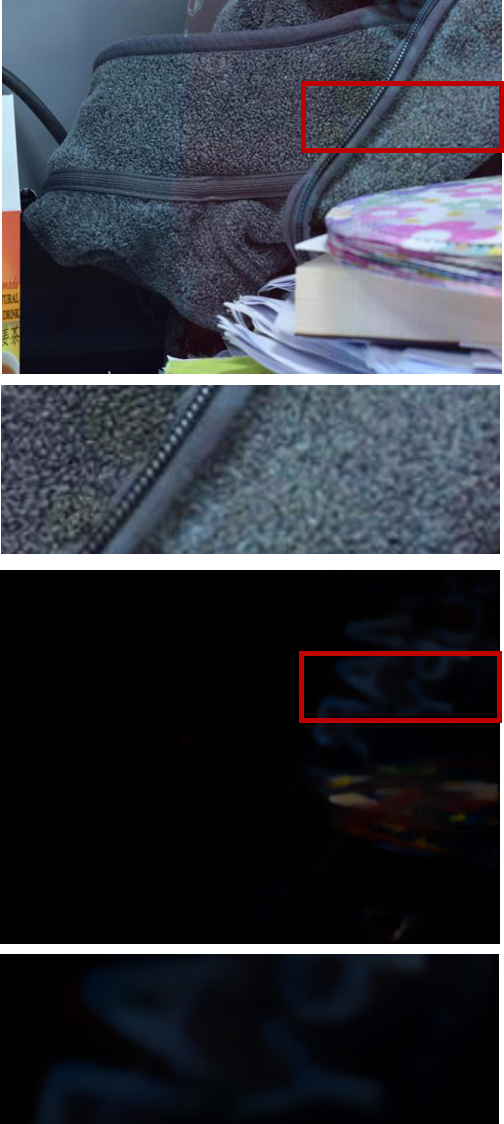}\vspace{1pt}
    \includegraphics[width=1\linewidth]{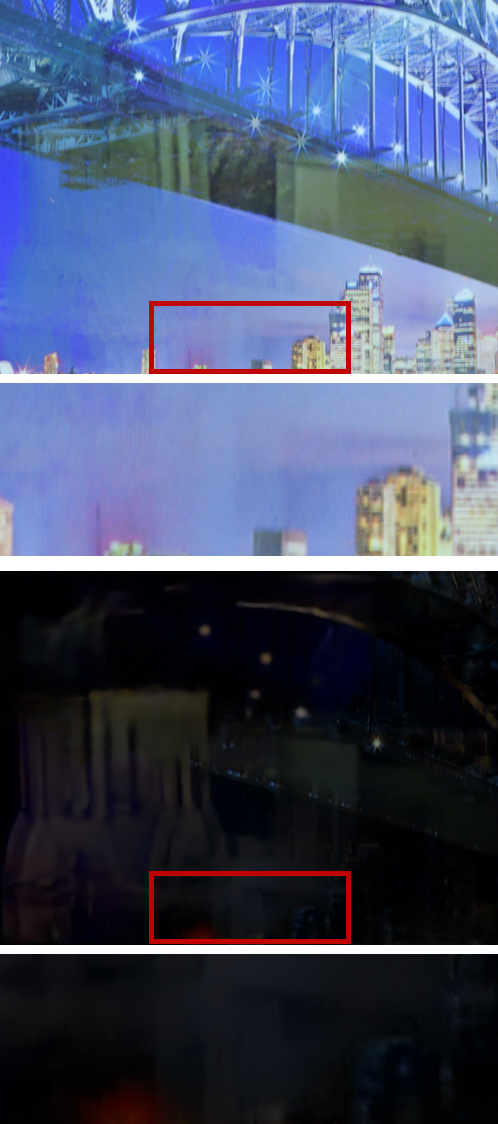}\vspace{1pt}
    \end{minipage}
    }
    \subfigure[DExNet]{
    \begin{minipage}[b]{0.09\textwidth}
    \includegraphics[width=1\linewidth]{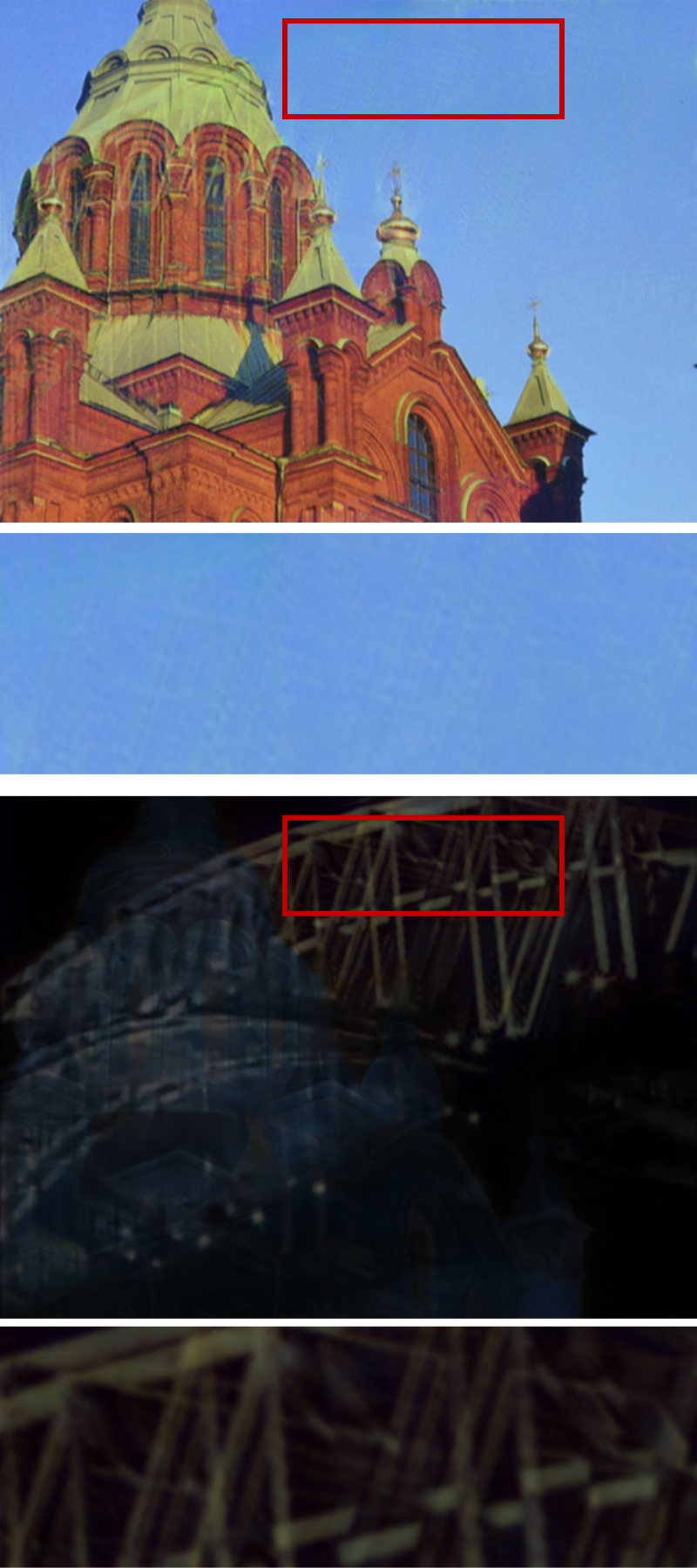}\vspace{1pt}
    \includegraphics[width=1\linewidth]{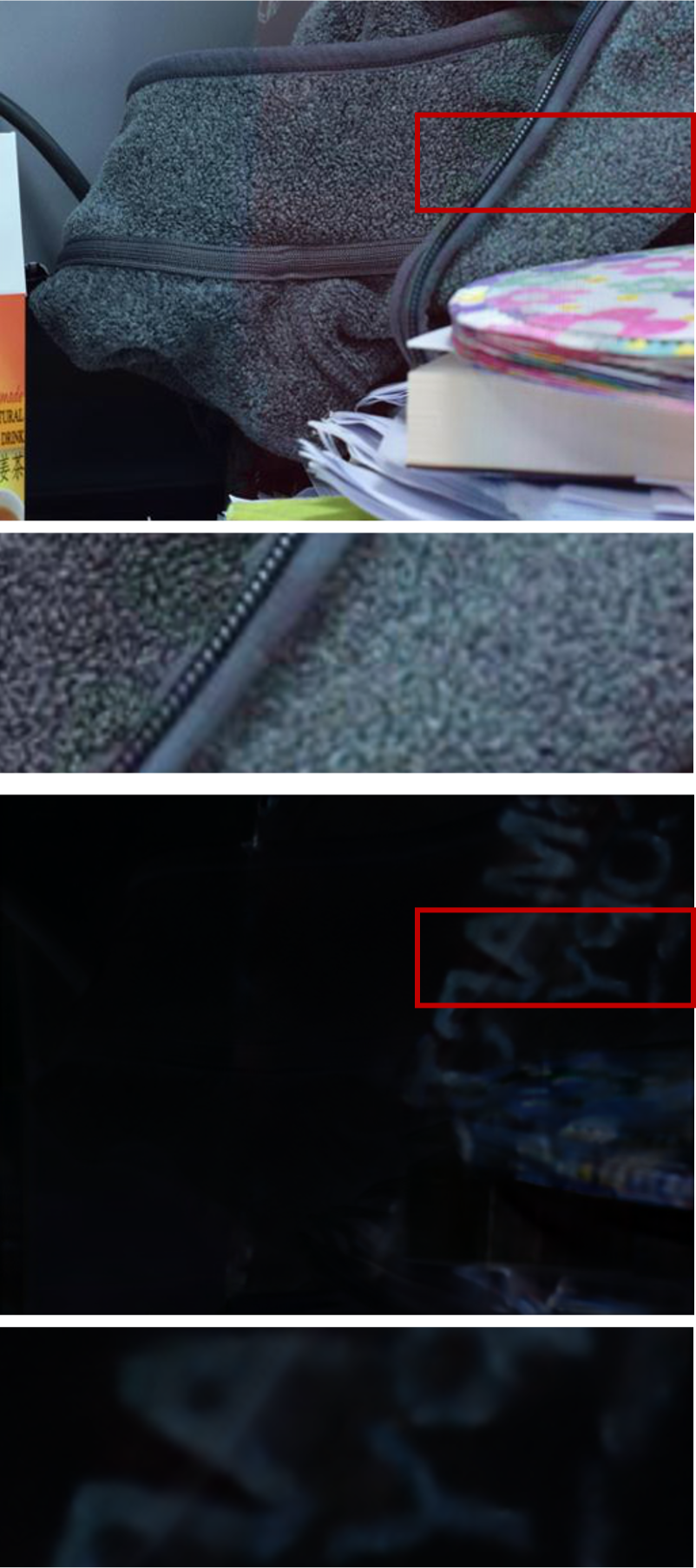}\vspace{1pt}
    \includegraphics[width=1\linewidth]{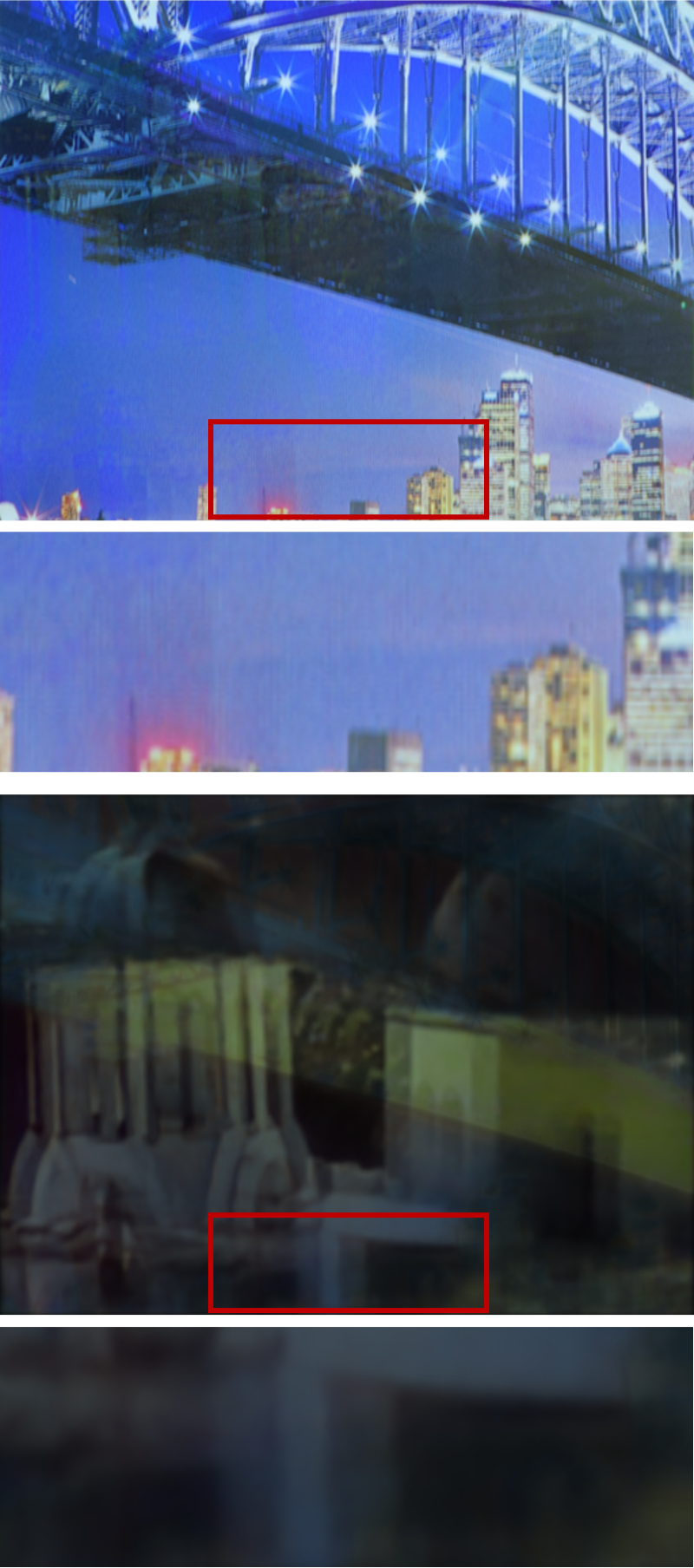}\vspace{1pt}
    \end{minipage}
    }
    \subfigure[GT]{
    \begin{minipage}[b]{0.09\textwidth}
    \includegraphics[width=1\linewidth]{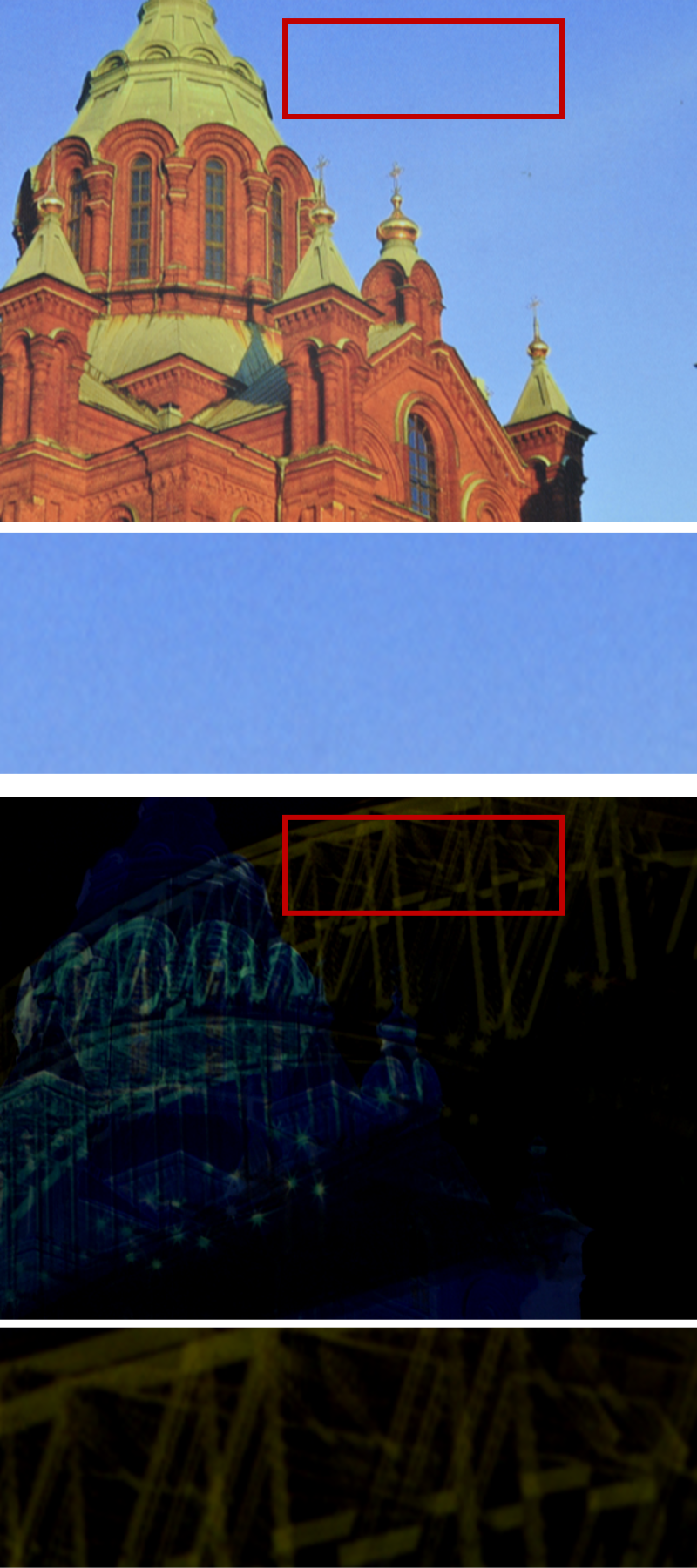}\vspace{1pt}
    \includegraphics[width=1\linewidth]{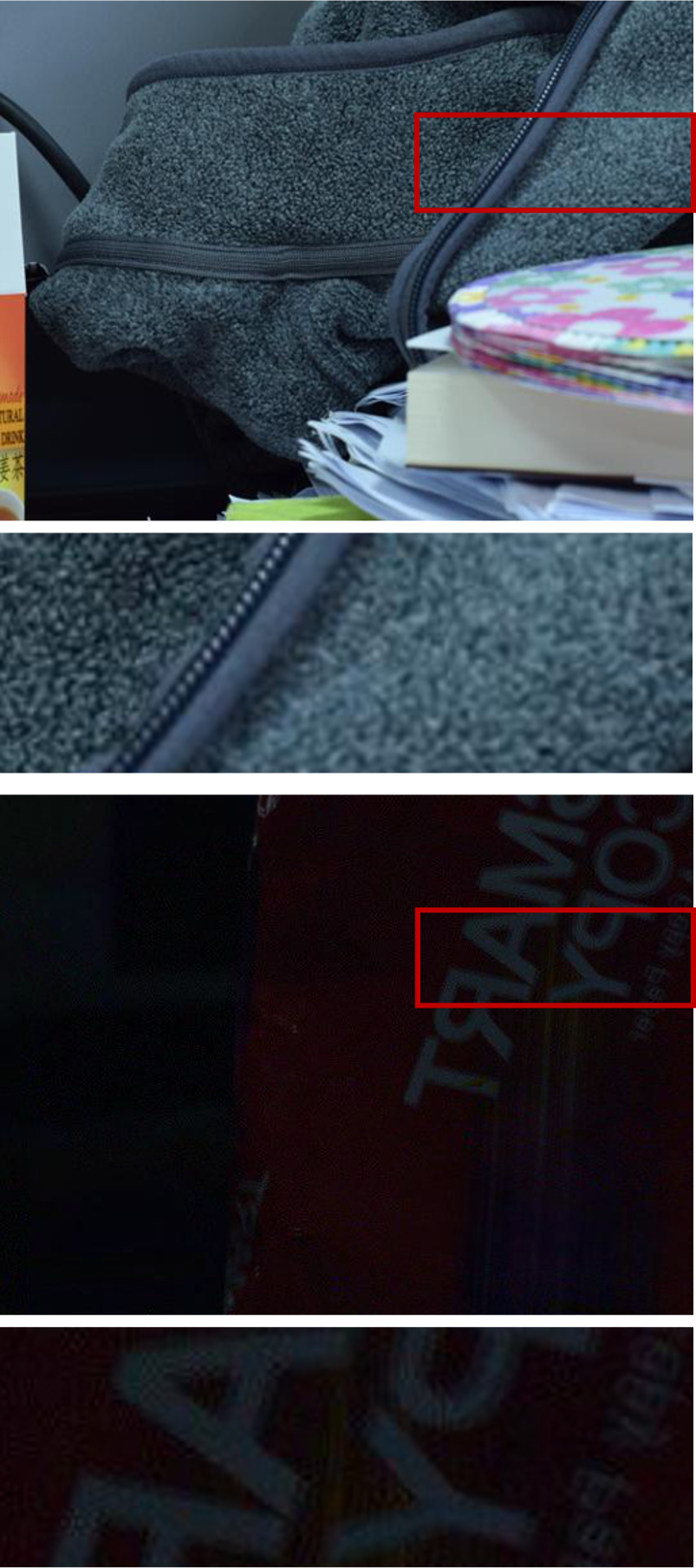}\vspace{1pt}
    \includegraphics[width=1\linewidth]{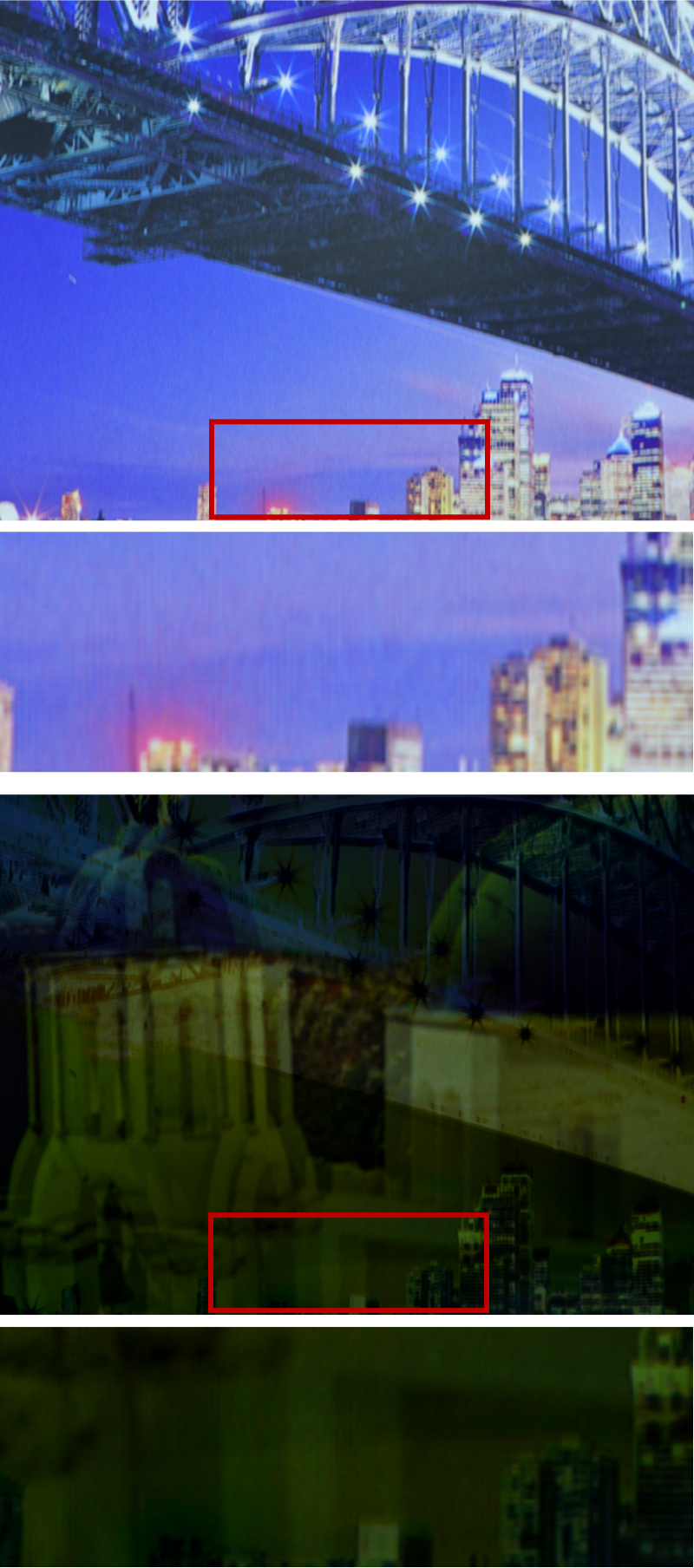}\vspace{1pt}
    \end{minipage}
    }
    \caption{Visual comparisons on \textit{SIR$^{2}$} dataset of different single image reflection methods. The first column shows the input images and the last column displays the ground-truth transmission image. }
    \label{fig:sir2}
\end{figure*}
%-----------------------------------------------------------

%-----------------------------------------------------------
\begin{figure*}[t]
    \centering
    \subfigure[ {Input}]{
    \begin{minipage}[b]{0.10\textwidth}

    \includegraphics[width=1.0\linewidth]{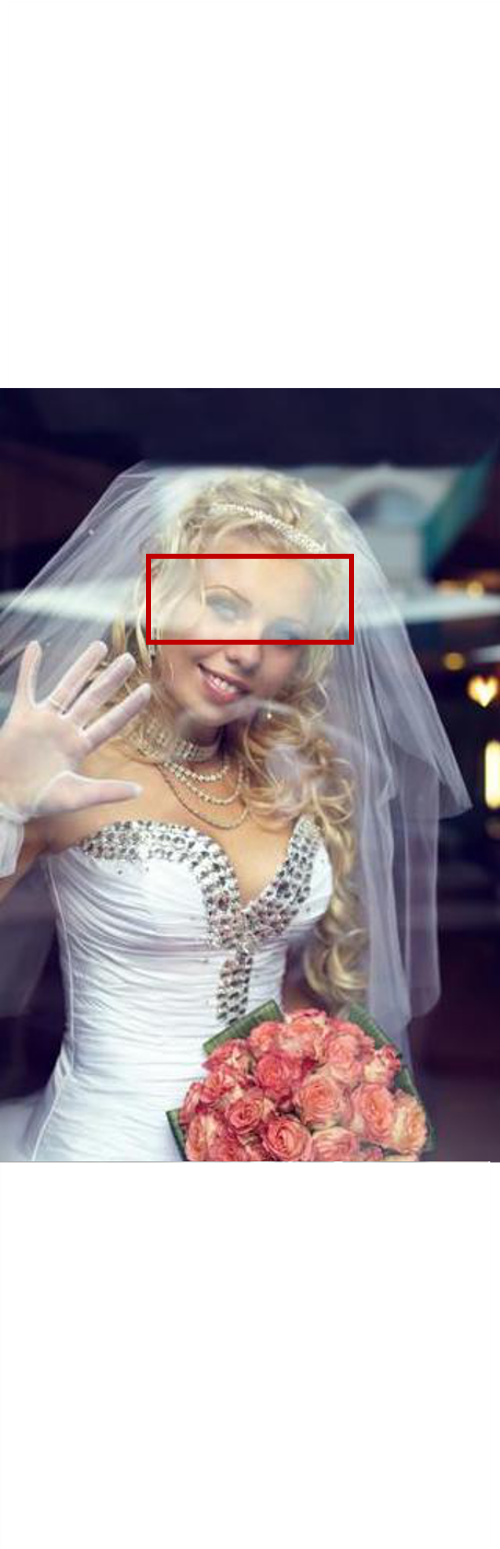}\vspace{0.5pt}
    \includegraphics[width=1.0\linewidth]{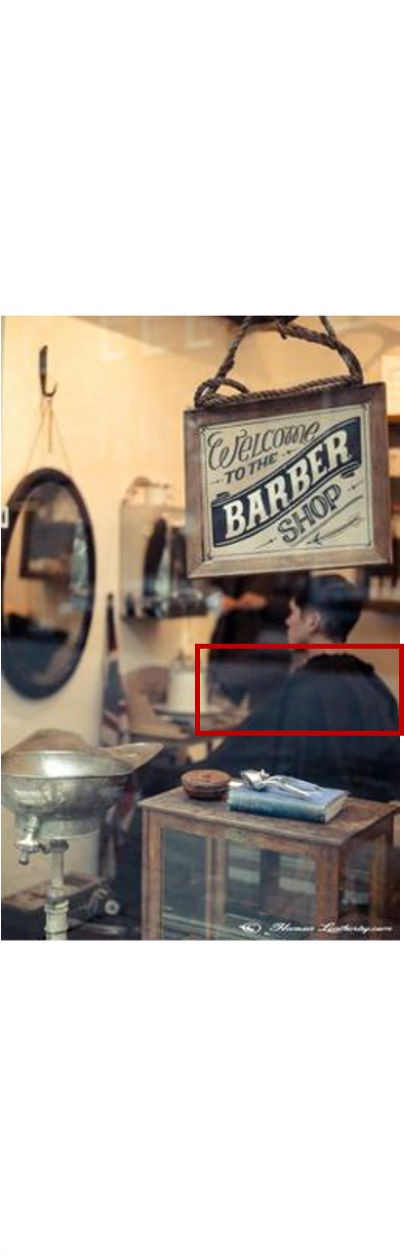}\vspace{0.5pt}
    \includegraphics[width=1.0\linewidth]{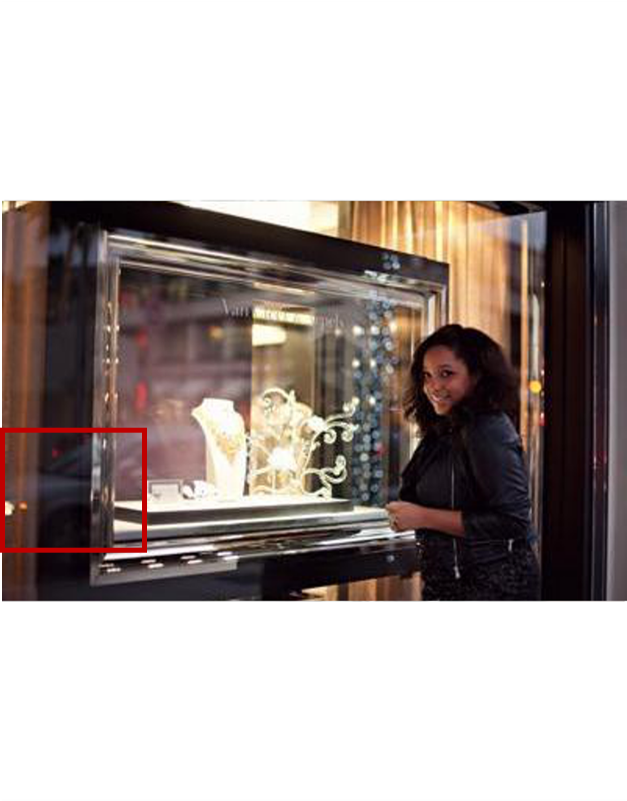}\vspace{0.5pt}
    \end{minipage}
    }
    \subfigure[Zhang]{
    \begin{minipage}[b]{0.10\textwidth}
    \includegraphics[width=1\linewidth]{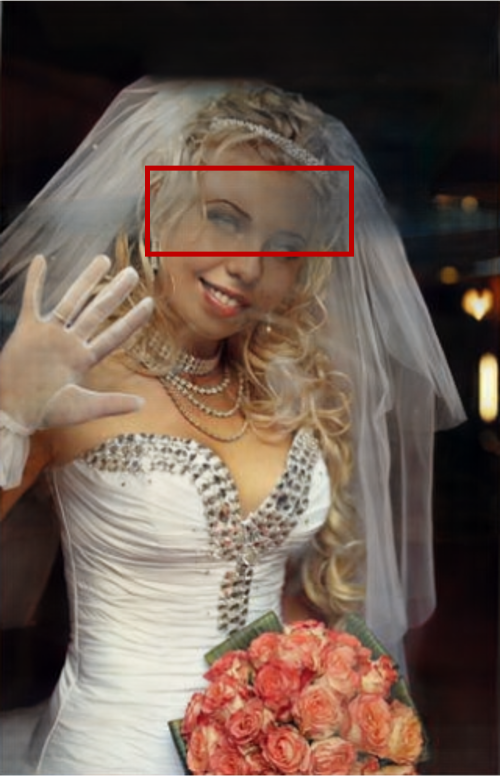}\vspace{0.5pt}
    \includegraphics[width=1\linewidth]{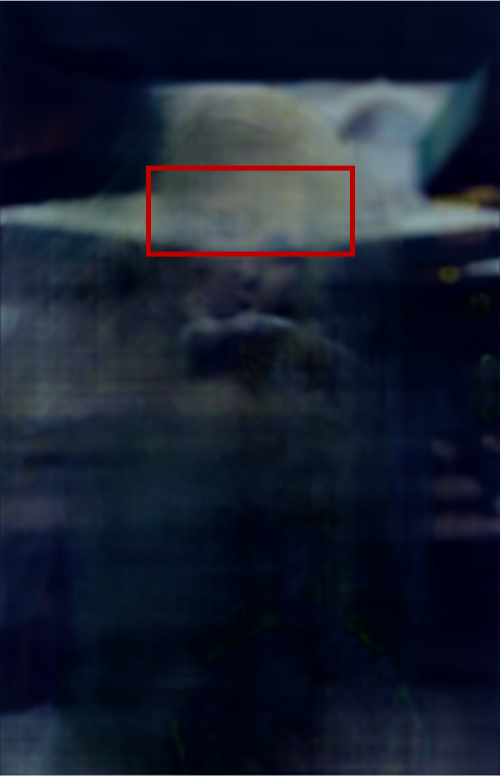}\vspace{1pt}
    \includegraphics[width=1\linewidth]{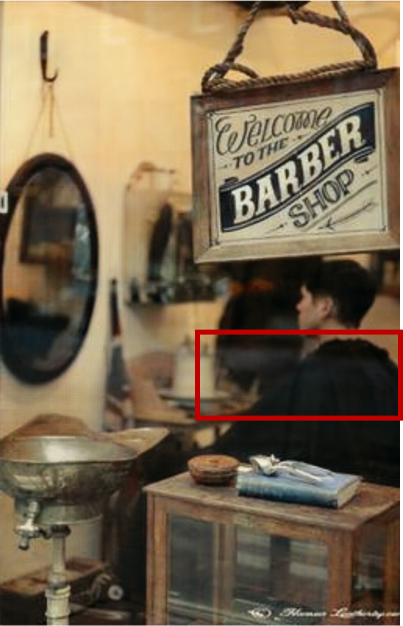}\vspace{0.5pt}
    \includegraphics[width=1\linewidth]{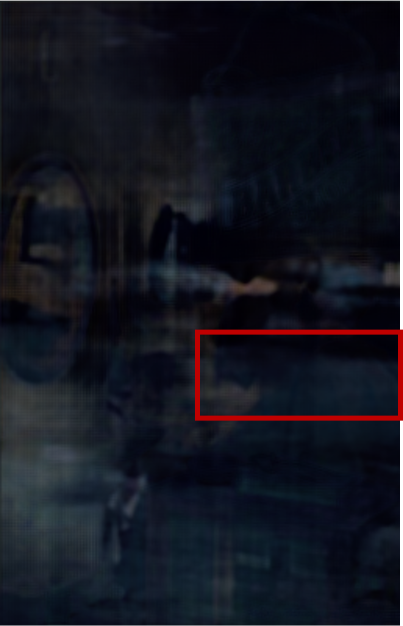}\vspace{1pt}
    \includegraphics[width=1\linewidth]{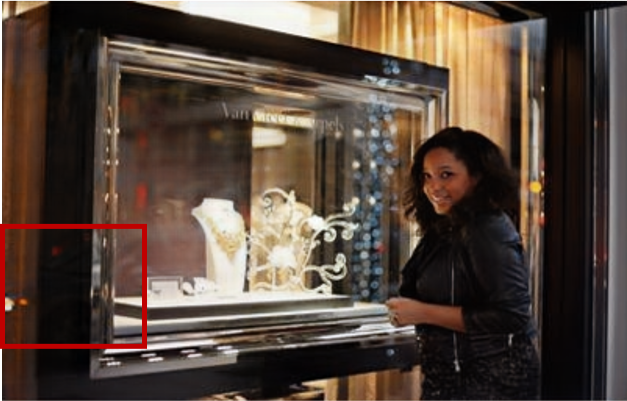}\vspace{0.5pt}
    \includegraphics[width=1\linewidth]{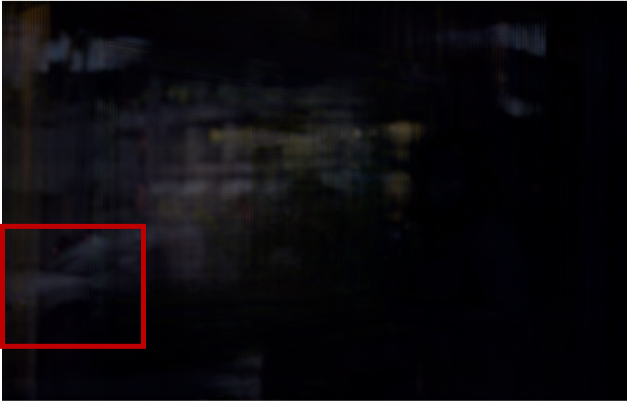}\vspace{1pt}
    \end{minipage}
    }
    \subfigure[ IBCLN]{
    \begin{minipage}[b]{0.10\textwidth}
    \includegraphics[width=1.0\linewidth]
    {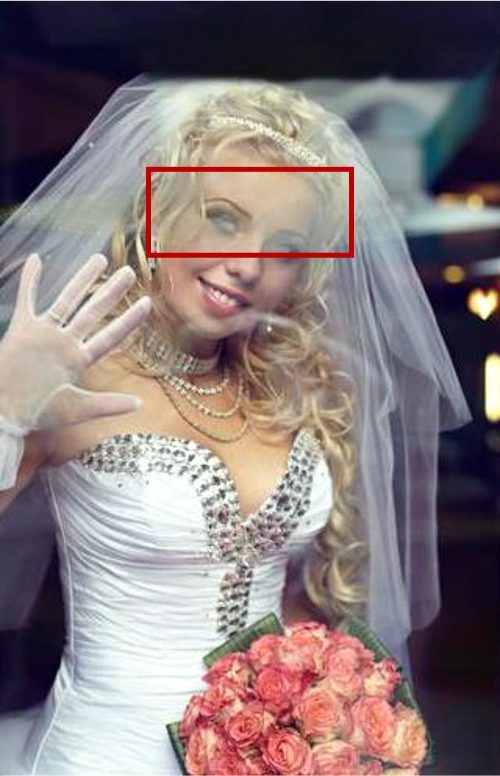}\vspace{0.5pt}
    \includegraphics[width=1.0\linewidth]{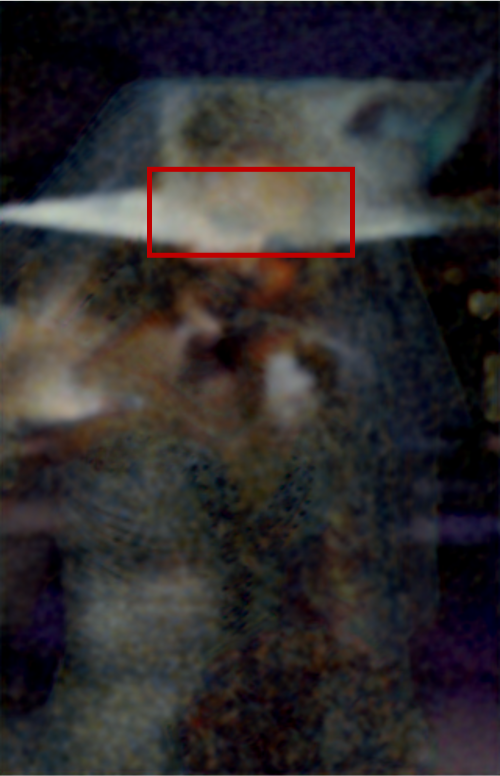}\vspace{1pt}
    \includegraphics[width=1.0\linewidth]{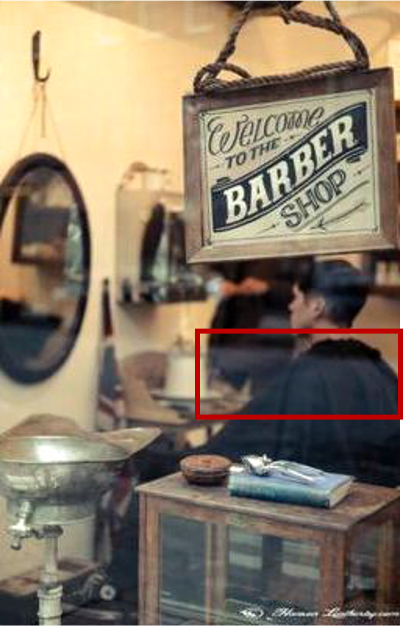}\vspace{0.5pt}
    \includegraphics[width=1.0\linewidth]{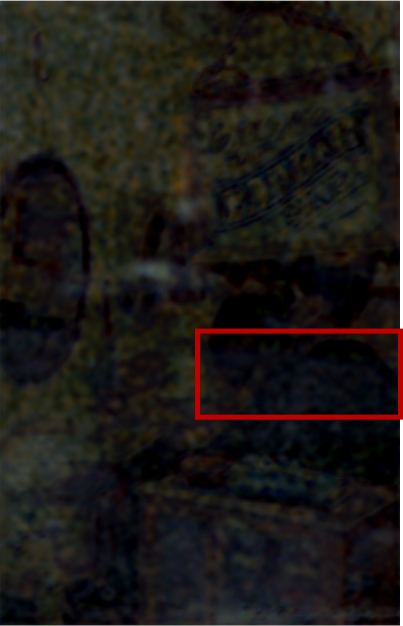}\vspace{1pt}
    \includegraphics[width=1.0\linewidth]{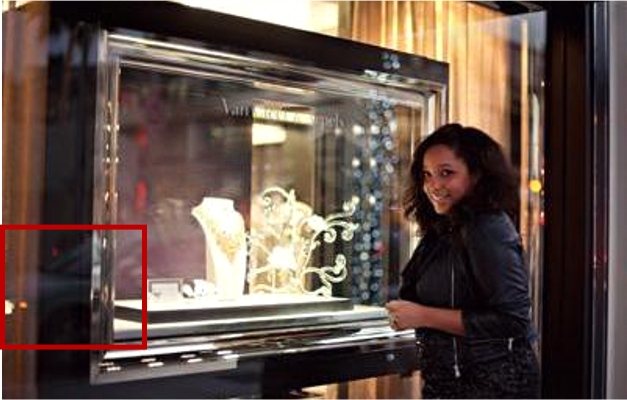}\vspace{0.5pt}
    \includegraphics[width=1.0\linewidth]{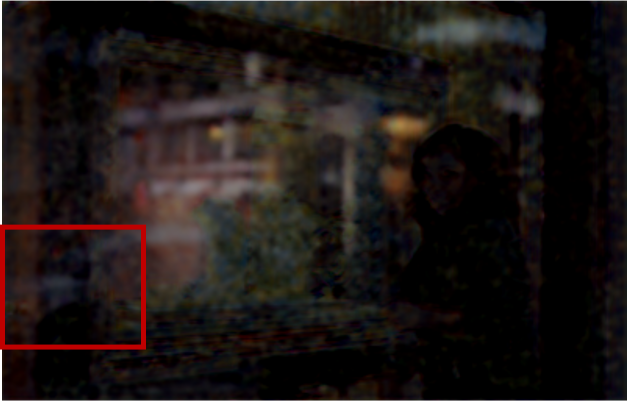}\vspace{1pt}
    \end{minipage}
    }
    \subfigure[ERRNet]{
    \begin{minipage}[b]{0.10\textwidth}

    \includegraphics[width=1.0\linewidth]{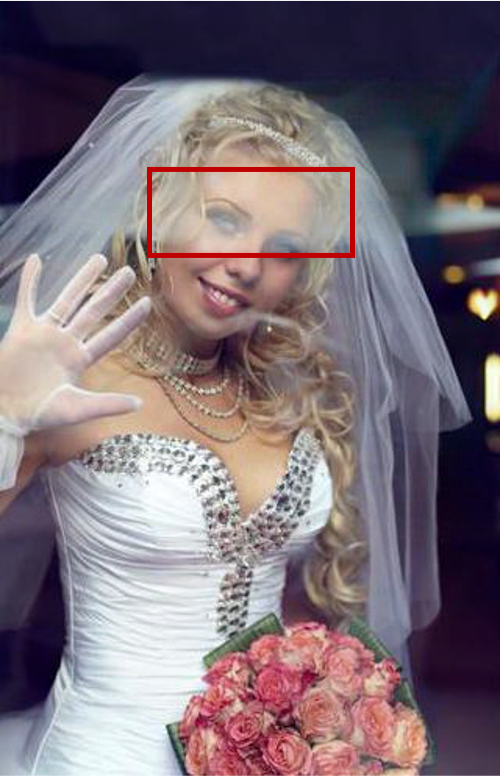}\vspace{0.5pt}
    \includegraphics[width=1.0\linewidth]{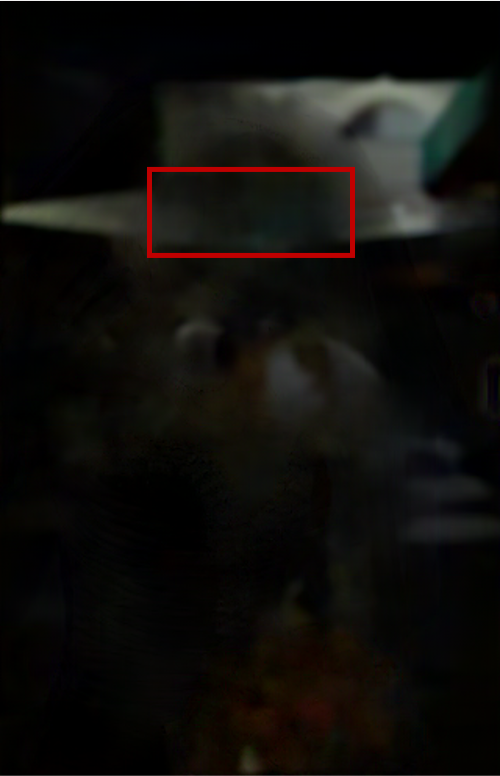}\vspace{1pt}
    \includegraphics[width=1.0\linewidth]{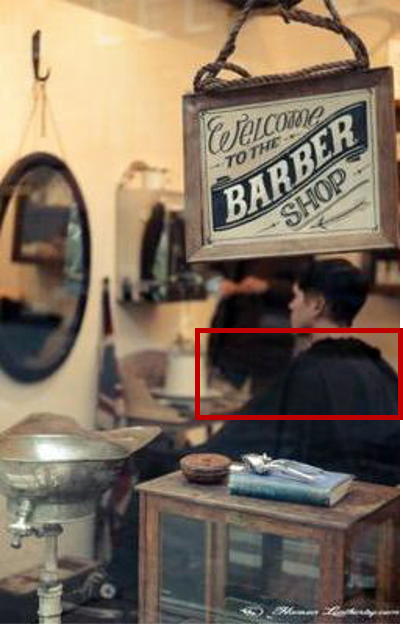}\vspace{0.5pt}
    \includegraphics[width=1.0\linewidth]{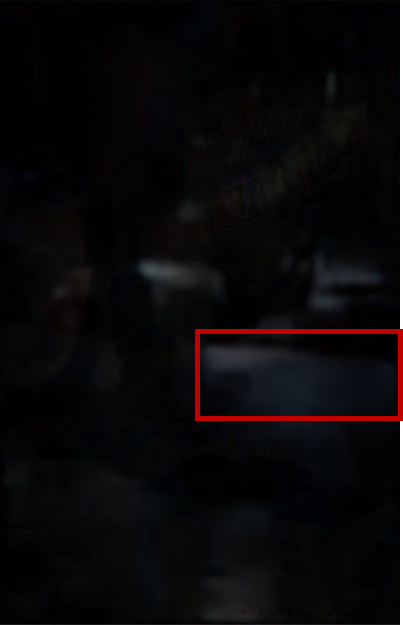}\vspace{1pt}
    \includegraphics[width=1.0\linewidth]{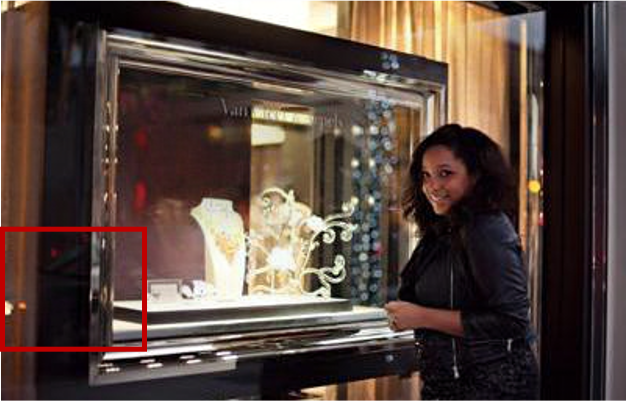}\vspace{0.5pt}
    \includegraphics[width=1.0\linewidth]{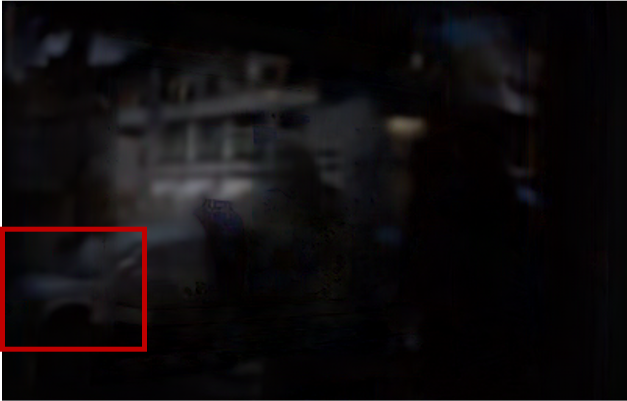}\vspace{1pt}
    \end{minipage}
    }
    \subfigure[YTMT]{
    \begin{minipage}[b]{0.10\textwidth}

    \includegraphics[width=1.0\linewidth]{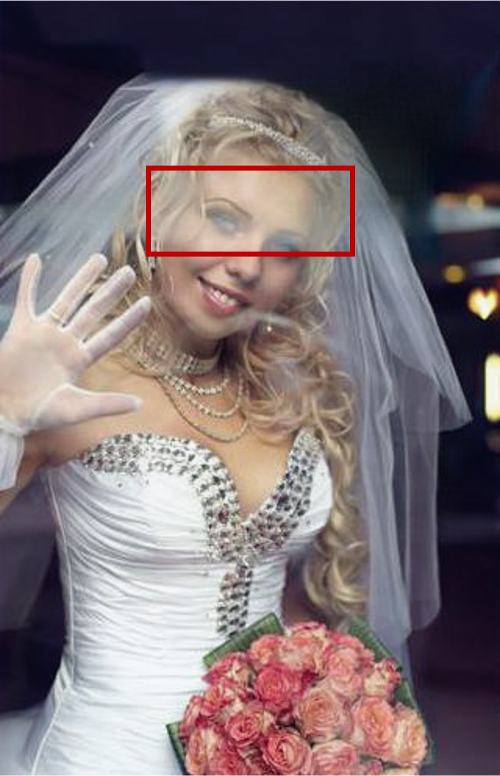}\vspace{0.5pt}
    \includegraphics[width=1.0\linewidth]{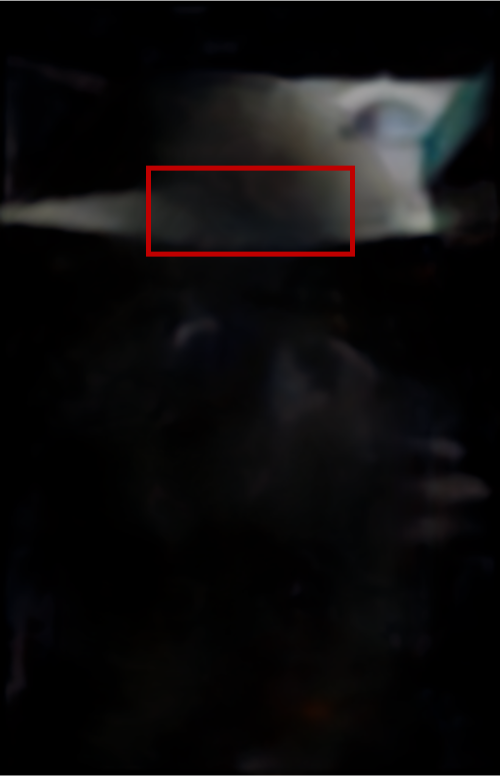}\vspace{1pt}
    \includegraphics[width=1.0\linewidth]{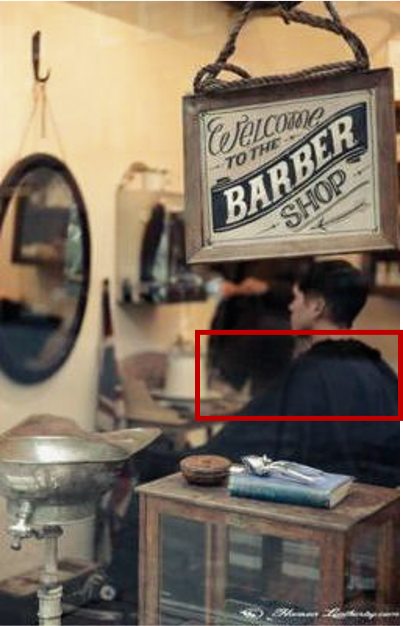}\vspace{0.5pt}
    \includegraphics[width=1.0\linewidth]{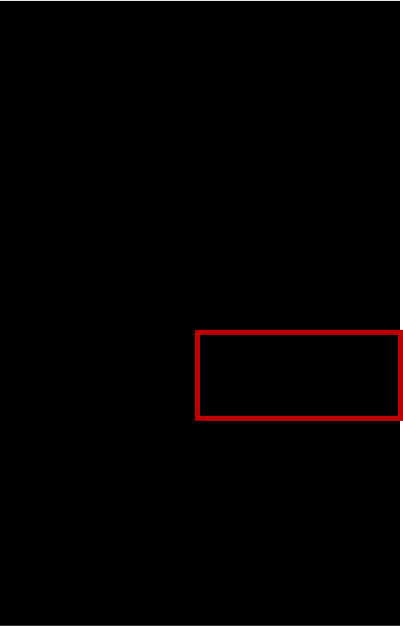}\vspace{1pt}
    \includegraphics[width=1.0\linewidth]{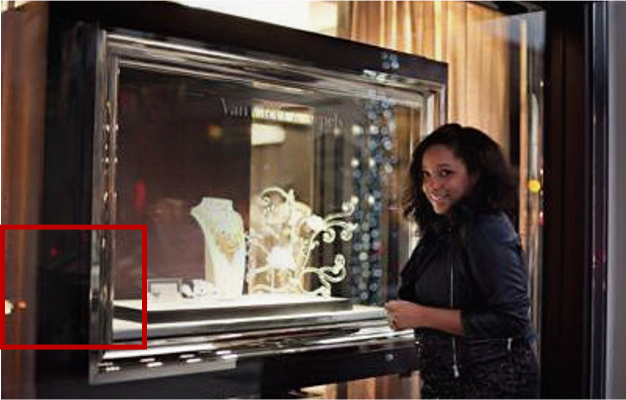}\vspace{0.5pt}
    \includegraphics[width=1.0\linewidth]{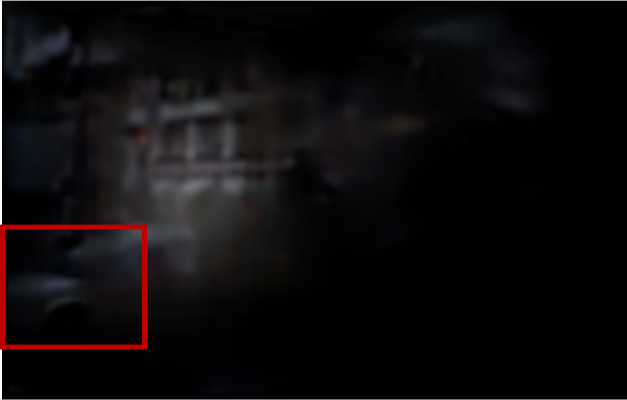}\vspace{1pt}
    \end{minipage}
    }
    \subfigure[DSRNet]{
    \begin{minipage}[b]{0.10\textwidth}
    \includegraphics[width=1.0\linewidth]{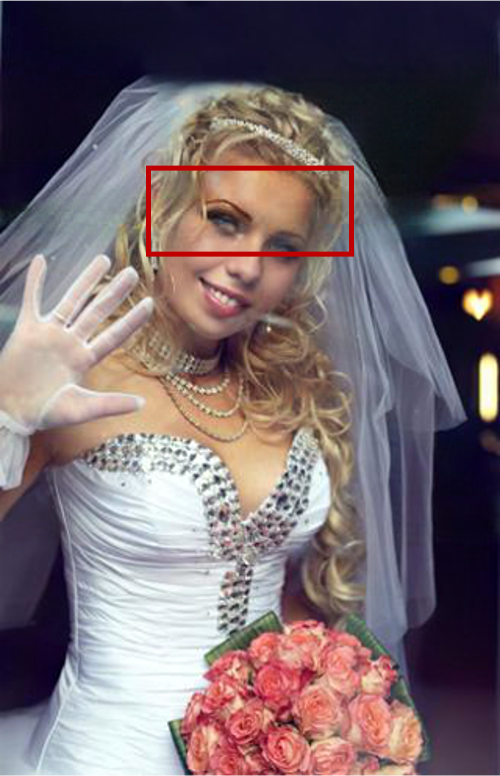}\vspace{0.5pt}
    \includegraphics[width=1.0\linewidth]{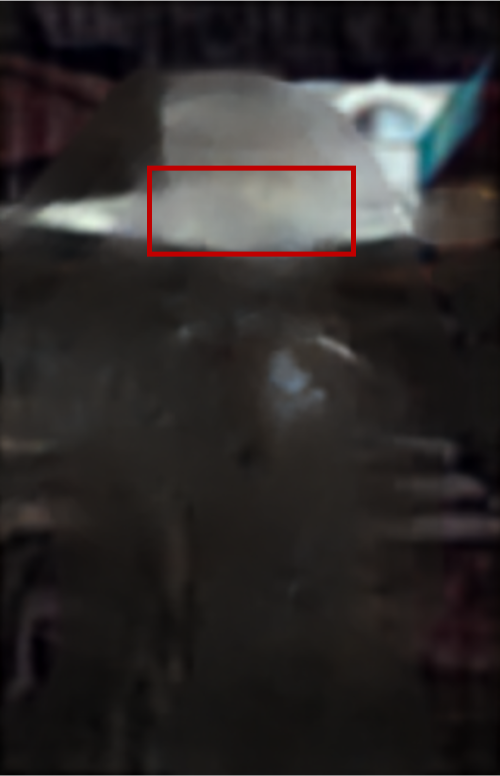}\vspace{1pt}
    \includegraphics[width=1.0\linewidth]{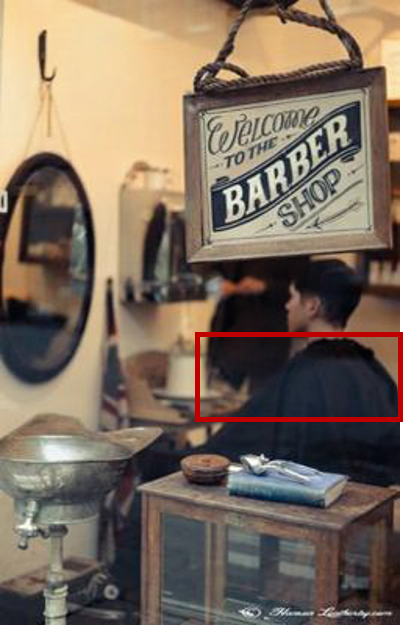}\vspace{0.5pt}
    \includegraphics[width=1.0\linewidth]{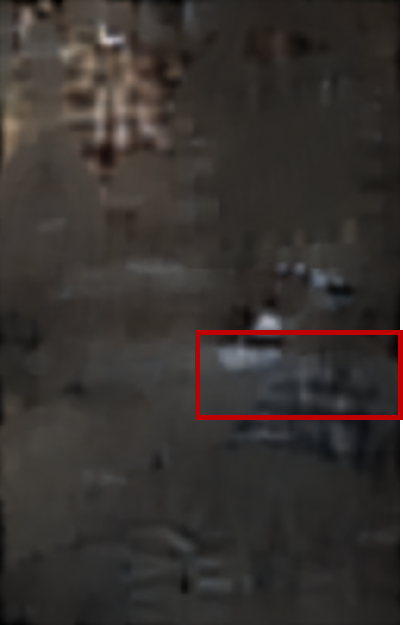}\vspace{1pt}
    \includegraphics[width=1.0\linewidth]{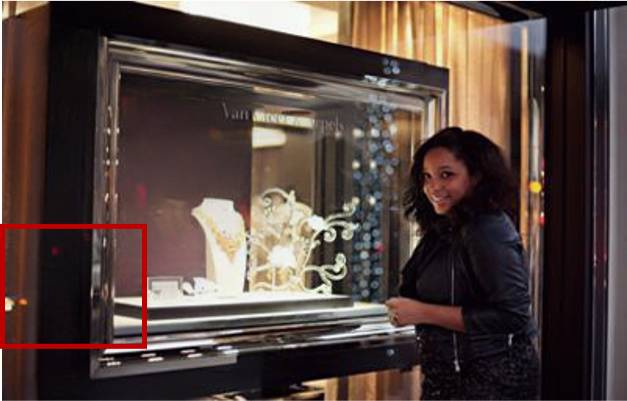}\vspace{0.5pt}
    \includegraphics[width=1.0\linewidth]{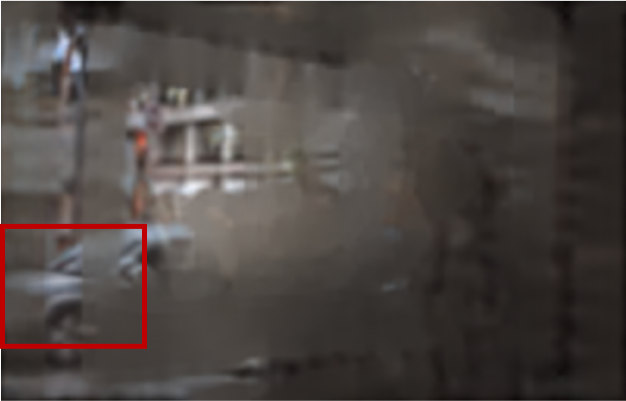}\vspace{1pt}
    \end{minipage}
    }
    \subfigure[DURRNet]{
    \begin{minipage}[b]{0.10\textwidth}
    \includegraphics[width=1.0\linewidth]{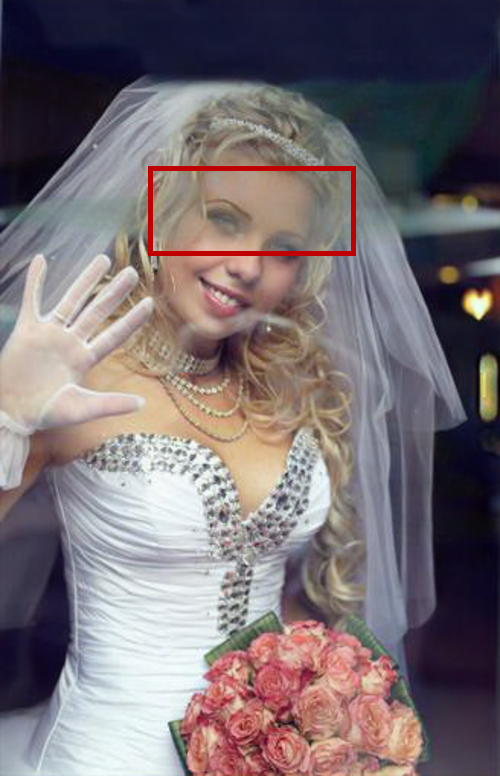}\vspace{0.5pt}
    \includegraphics[width=1.0\linewidth]{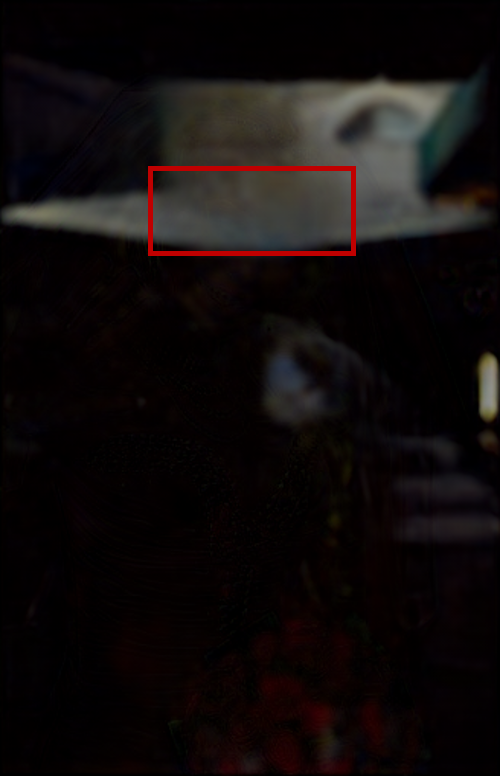}\vspace{1pt}
    \includegraphics[width=1.0\linewidth]{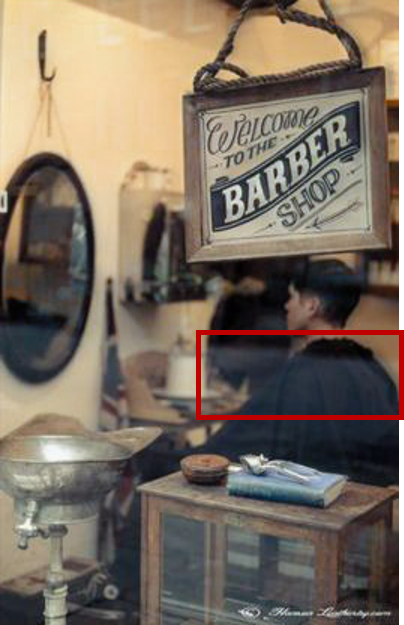}\vspace{0.5pt}
    \includegraphics[width=1.0\linewidth]{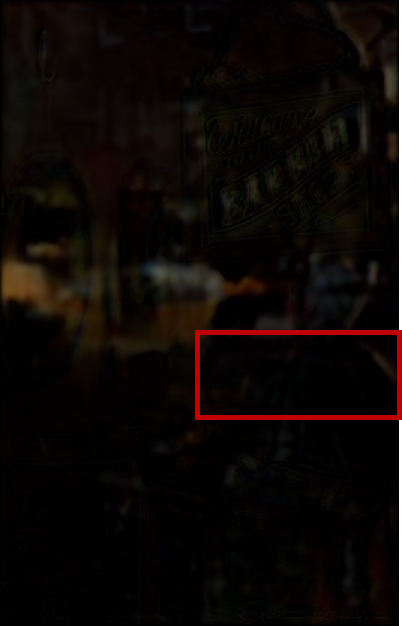}\vspace{1pt}
    \includegraphics[width=1.0\linewidth]{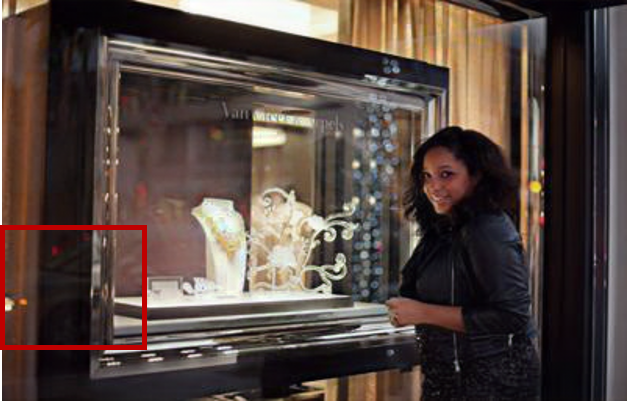}\vspace{0.5pt}
    \includegraphics[width=1.0\linewidth]{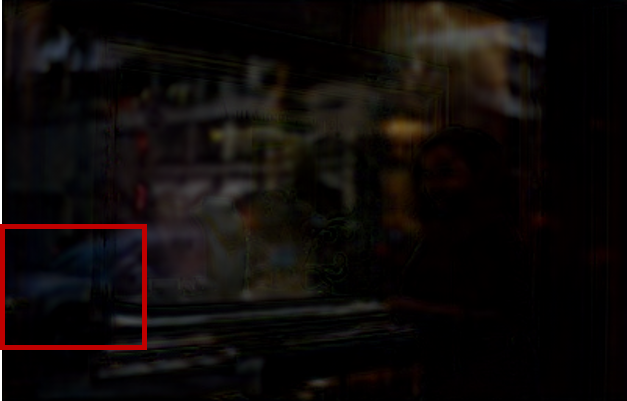}\vspace{1pt}
    \end{minipage}
    }
    \subfigure[DExNet]{
    \begin{minipage}[b]{0.10\textwidth}
    \includegraphics[width=1.0\linewidth]{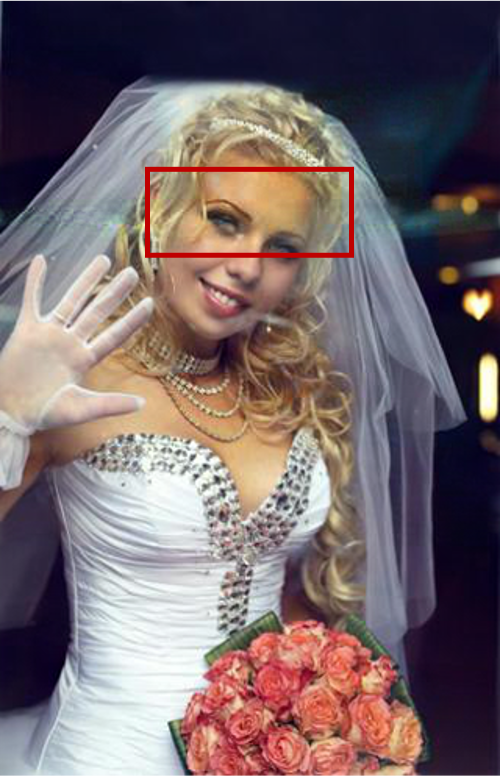}\vspace{0.5pt}
    \includegraphics[width=1.0\linewidth]{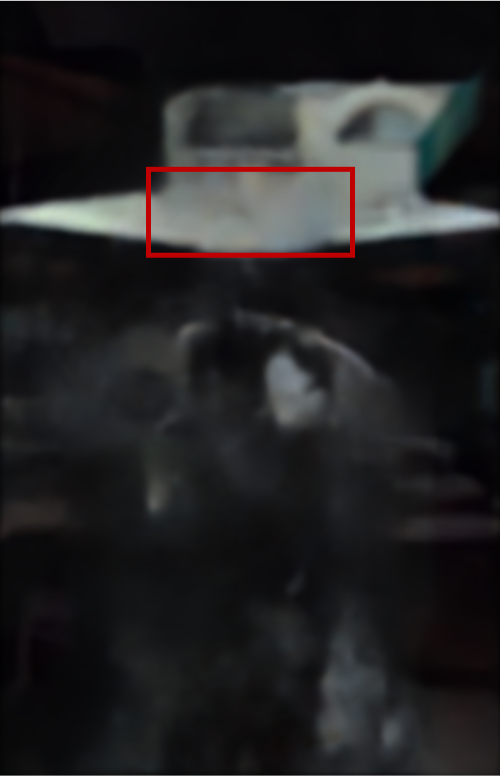}\vspace{1pt}
    \includegraphics[width=1.0\linewidth]{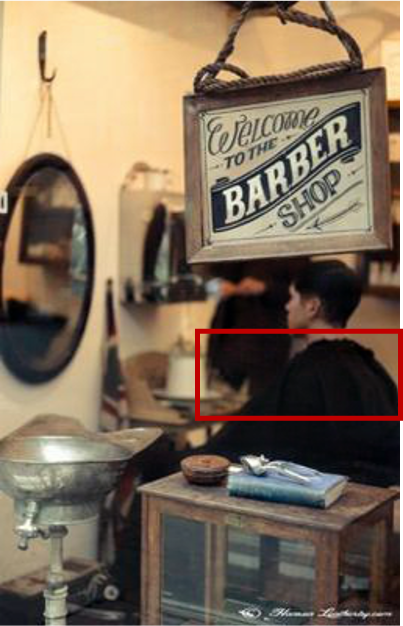}\vspace{0.5pt}
    \includegraphics[width=1.0\linewidth]{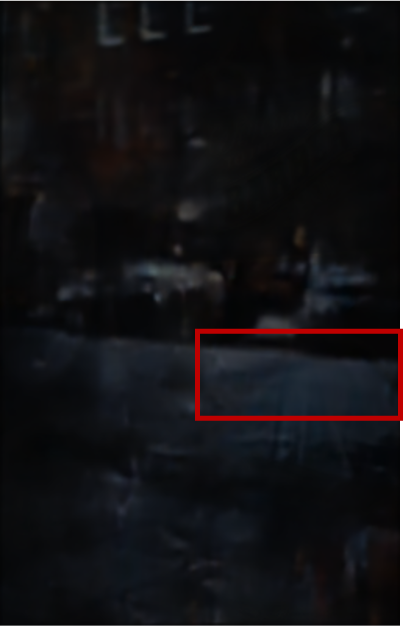}\vspace{1pt}
    \includegraphics[width=1.0\linewidth]{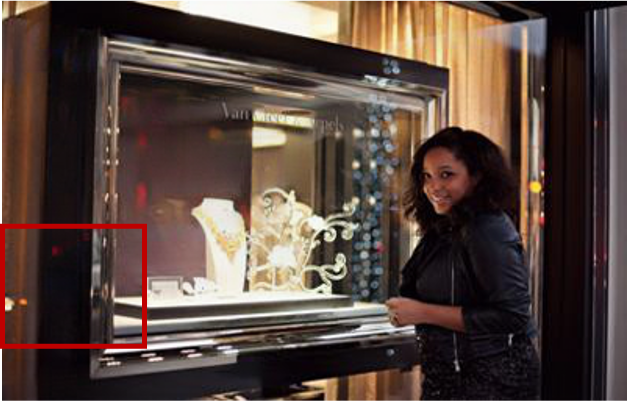}\vspace{0.5pt}
    \includegraphics[width=1.0\linewidth]{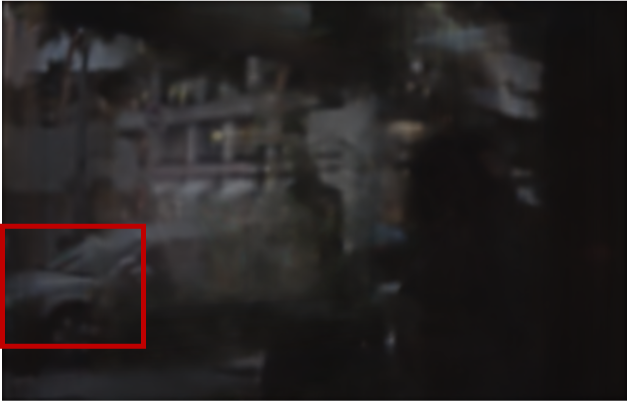}\vspace{1pt}
    \end{minipage}
    }
    \caption{Visual comparisons on \textit{Real45} dataset~\cite{generic_smooth_2017} of different single image reflection methods. The first column shows the input images. The estimated transmission images are shown in the odd rows and the estimated reflection images are shown in the even rows. }
    \label{fig:real45}
\end{figure*}
%-----------------------------------------------------------

%-----------------------------------------------------------

\subsection{Training Loss}

\label{sec:training}
The overall training loss contains a reconstruction loss $\mathcal{L}_{ {r}}$, an auxiliary loss $\mathcal{L}_{{a}}$ evaluated on the auxiliary variable $\mathbf{z}_\mathbf{A}$,  and the perceptual loss $\mathcal{L}_{{p}}$:
\begin{equation}
    \mathcal{L}= \mathcal{L}_{ {r}} + \lambda_{a} \mathcal{L}_{{a}} + \lambda_{p} \mathcal{L}_{ {p}},
\end{equation}
where $\lambda_{e}=0.01$, and $\lambda_{p}=0.1$ are regularization parameters. 
 
The reconstruction loss $\mathcal{L}_{\text {r}}$ is applied to the estimated transmission image $\widehat{\mathbf{T}}$, the estimated reflection image $\widehat{\mathbf{R}}$ and to the reconstructed image based on the final features:
\begin{equation}
\begin{aligned}
    \mathcal{L}_{{r}}&=  {\|\mathbf{T}-\widehat{\mathbf{T}}\|_{2}^{2} +  \|\mathbf{R}-\hat{\mathbf{R}}\|_{2}^{2} +\Vert \mathbf{I} - \widehat{\mathbf{T}} - \widehat{\mathbf{R}} - \widehat{\mathbf{N}} \Vert_1} \\
\end{aligned}
\end{equation}

The auxiliary loss is applied to minimize the auxiliary variable $\mathbf{z}_\mathbf{A}$ so as to minimize the element-wise feature product between two image layers $\left( \mathbf{M}_\mathbf{T} \otimes \mathbf{z}_\mathbf{T} \right) \odot \left( \mathbf{M}_\mathbf{R} \otimes \mathbf{z}_\mathbf{R} \right)$:
\begin{equation}
    \mathcal{L}_{{a}} = \Vert \mathbf{z}_\mathbf{A} \Vert_1.
\end{equation}

The perceptual loss~\cite{perceptual_loss_2018} is used to regularize the estimated images with high perceptual quality:
\begin{equation}
    \mathcal{L}_{ {p}}= \sum \omega_j \|\tau_j(\mathbf{T})-\tau_j(\widehat{\mathbf{T}})\|_{1},
\end{equation}
where $\tau_j(\cdot)$ denotes the features at layer $j$ of the VGG-19 model pre-trained on ImageNet dataset, and $\omega_j$ is the weight of different layers with $\mathbf{\omega} = [0.38; 0.21; 0.27; 0.18; 6.67]$.

\section{Experimental Results}
\label{sec:results}

In this section, we show the implementation details, then perform experiments to compare the proposed DExNet with state-of-the-art single image reflection removal methods, finally conduct ablation studies to investigate properties of DExNet.

\subsection{Implementation Details}

The proposed method is developed with the PyTorch framework\footnote{http://pytorch.org/} and the models are optimized with Adam optimizer. The total number of training epochs is 50, and an early stop strategy is employed. The initial learning rate is set to $10^{-4}$ and is decayed by a factor of 0.5 at epoch 25. The experiments were conducted on a computer equipped with a NVIDIA GeForce RTX 4090 GPU.

By default, the number of scales $S$ in DExNet is set to 4, the number of feature channels $n$ in $\{ \mathbf{D}_{i} \}$ is set to 64 and the number of feature channels $m$ in $\{ \mathbf{M}_{i} \}$ is set to 128.
By varying the number of scales $S$ in DExNet and the number of stages $K$ within the i-SAFU algorithm proposed for unfolding the DExNet, we can achieve different trade-offs between performance and model size. We term DExNet with $S=2$ scales and $K=5$ SAFU modules in each scale as DExNet$_S$ which is with 4.52M parameters and term DExNet with $S=4$ scales and $K=5$ SAFU modules in each scale as DExNet$_L$ which is with 9.66M parameters.

The qualities of image reflection removal are evaluated based on two metrics: Peak Signal-to-Noise Ratio (PSNR) and Structural Similarity Index (SSIM)~\cite{wang2004image}.

\subsection{Datasets}

\noindent \textbf{Training Data.} The training data include the synthetic dataset and real dataset. The synthetic dataset includes 7,643 synthesized image
pairs from the PASCAL VOC dataset~\cite{everingham2010pascal}. 
During each epoch, image pairs are sampled from the synthetic dataset and the \textit{Real89} training dataset~\cite{perceptual_loss_2018} and the \textit{Nature200}~\cite{cascaded_Refine_2020} training dataset with a ratio of 0.6:0.2:0.2 for training. No additional data augmentation is employed.
We follow the synthetic data generation model of DSRNet~\cite{Hu_2023_ICCV}, the synthetic superimposed image $\mathbf{I}_{syn}$ can be expressed as a function of the transmission image $\gamma_1 \mathbf{T}_{syn}$ and reflection image $\gamma_2 \mathbf{R}_{syn}$:
\begin{equation}
    \mathbf{I}_{syn} = \gamma_1\mathbf{T}_{syn} + \gamma_2\mathbf{R}_{syn} - \gamma_1\gamma_2\mathbf{T}_{syn} \odot \mathbf{R}_{syn},
\end{equation}
where $\gamma_1 \in [0.8, 1.0]$ and $\gamma_2 \in [0.4, 1.0]$ are blending parameters.

\noindent \textbf{Testing Data.} Both \textit{Real20}~\cite{perceptual_loss_2018} and \textit{SIR${^2}$} dataset~\cite{wan2017benchmarking} are used to both quantitatively and visually compare different single image reflection methods.
The \textit{Real20}~\cite{perceptual_loss_2018} is with 20 paired transmission and blended images.
The \textit{SIR${^2}$} dataset~\cite{wan2017benchmarking} is with paired transmission, reflection and blended images.
It has 3 subsets including \textit{Postcard} dataset with 199 testing images, \textit{Objects} dataset with 200 testing images and \textit{Wild} dataset with 55 testing images. 
There is also \textit{Real45} dataset~\cite{generic_smooth_2017} which is collected from Internet and is without the ground-truth transmission images. It is used to evaluate visual quality and to validate the generalization ability of different methods.

\subsection{Comparison with State-of-the-arts Methods}

In this section, we quantitatively and visually compare the proposed DExNet with other single image reflection removal methods on both transmission and reflection image estimation. The comparison methods include 
Zhang \textit{et al.}'s method~\cite{perceptual_loss_2018}, BDN~\cite{yang2018seeing}, Zheng \textit{et al.}'s method~\cite{zheng2021single}, CoRRN~\cite{wan2019corrn}, ERRNet~\cite{wei2019single}, IBCLN~\cite{cascaded_Refine_2020}, YTMT~\cite{hu2021trash}, LASIRR~\cite{dong2021location}, DMCN~\cite{FengTIP2021}, RAGNet~\cite{LiAI2023}, and DSRNet~\cite{Hu_2023_ICCV}. 
We also consider three recent deep unfolding based reflection removal methods: MoG-SIRR~\cite{Shao2021MoGSIRR}, CGDNet~\cite{zhang2022content} and DURRNet~\cite{Huang24DURRNet}. However, since MoG-SIRR~\cite{Shao2021MoGSIRR} and CGDNet~\cite{zhang2022content} are not open source, we refer to the transmission image estimation results from the papers and did not evaluate their performances on reflection image estimation.

\subsubsection{Quantitative Comparisons} 
 This section presents the quantitative comparisons on model effectiveness, model size and model complexity based on two commonly used datasets, \textit{i.e.}, \textit{Real20} dataset~\cite{perceptual_loss_2018} and 
 \textit{SIR${^2}$} dataset~\cite{wan2017benchmarking}.

%-----------------------------------------------------------
\begin{figure*}[t]
    \centering
    \subfigure[Ground-truth]{
    \begin{minipage}[b]{0.14\textwidth}
    \includegraphics[width=1\linewidth]{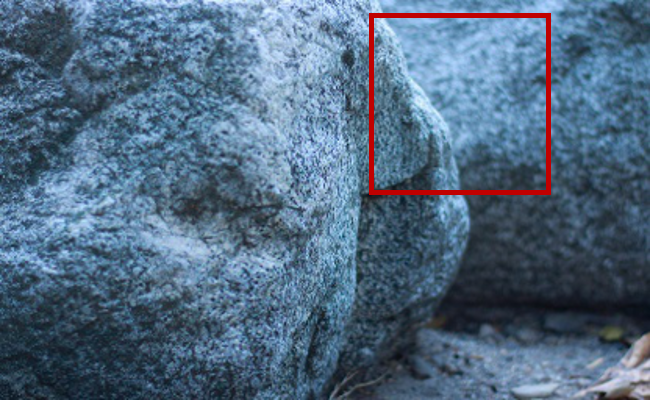}\vspace{1pt}
    \includegraphics[width=1\linewidth]{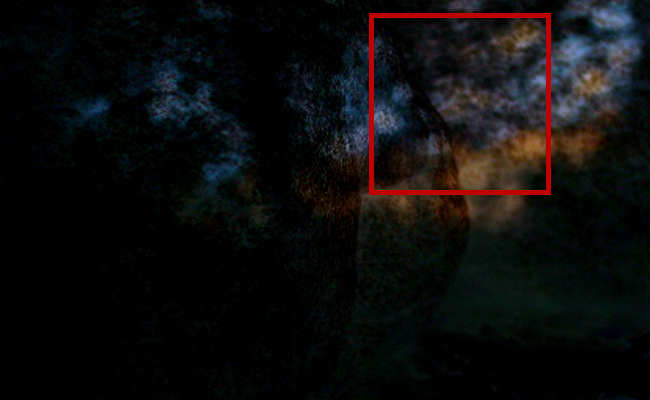}\vspace{2pt}
    \includegraphics[width=1\linewidth]{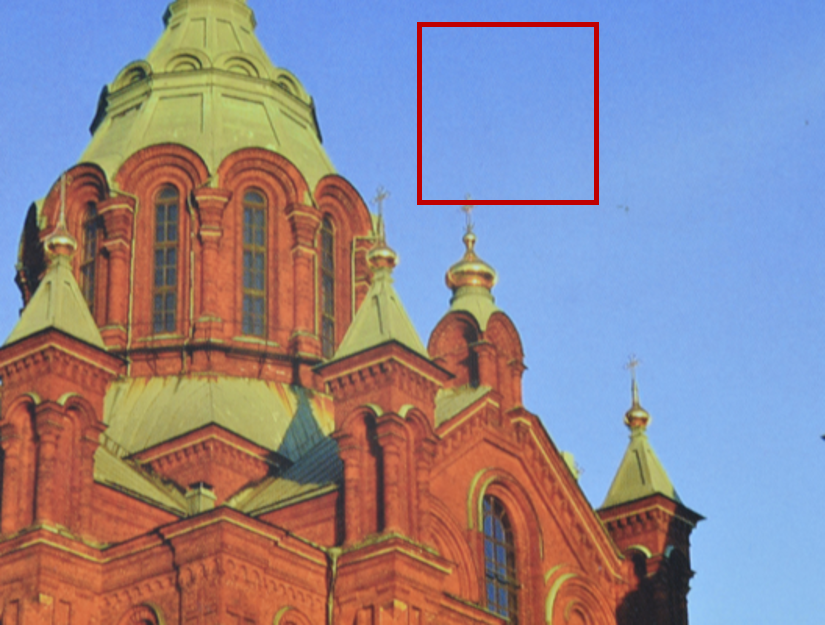}\vspace{1pt}
    \includegraphics[width=1\linewidth]{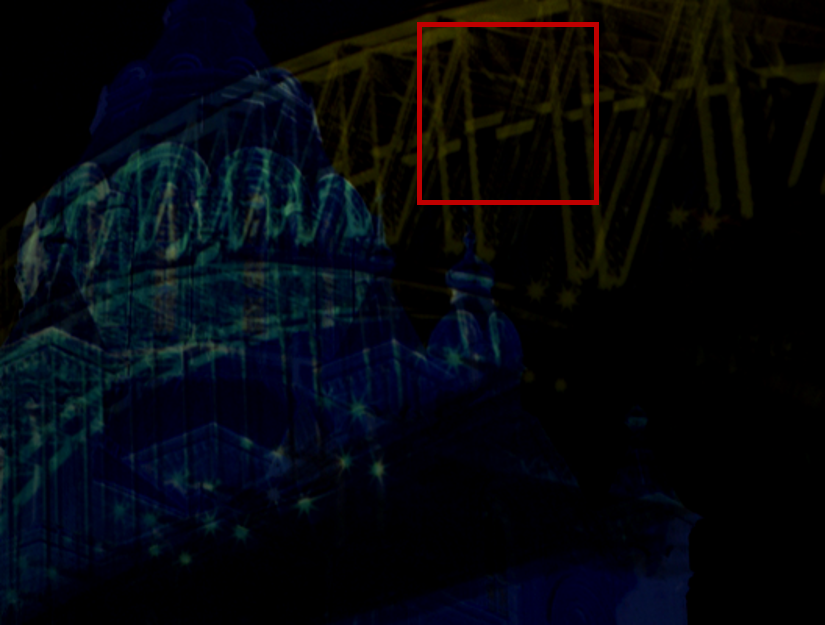}
    \end{minipage}
    }
    \subfigure[DExNet]{
    \begin{minipage}[b]{0.14\textwidth}
    \includegraphics[width=1\linewidth]{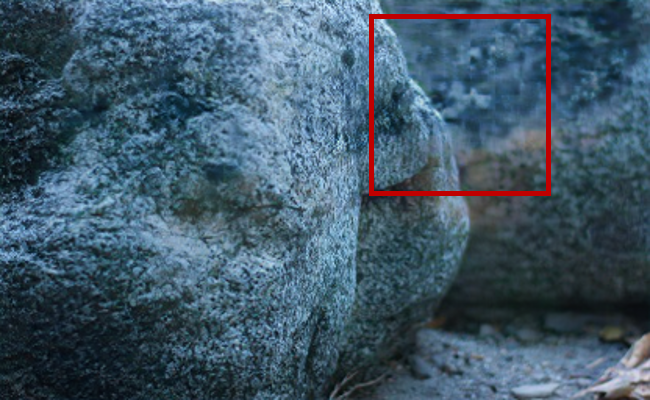}\vspace{1pt}
    \includegraphics[width=1\linewidth]{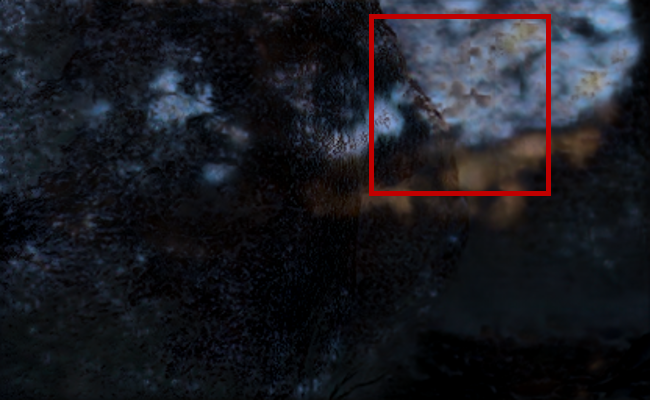}\vspace{2pt}
    \includegraphics[width=1\linewidth]{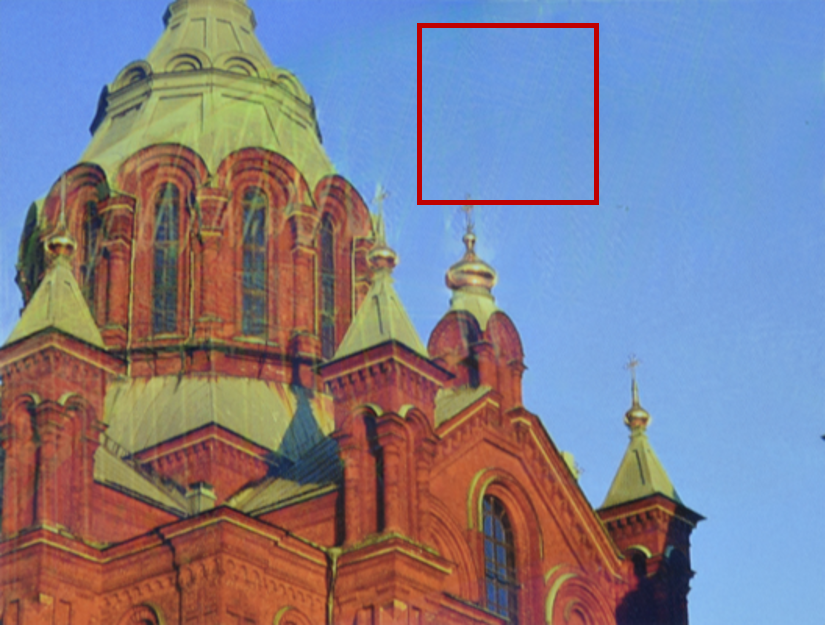}\vspace{1pt}
    \includegraphics[width=1\linewidth]{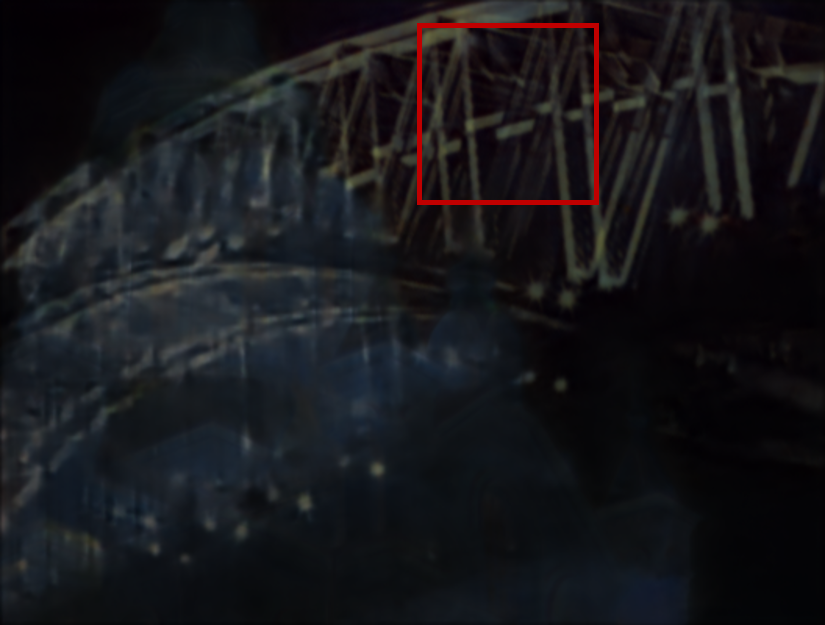}
    \end{minipage}
    }
    \subfigure[w/o AES block]{
    \begin{minipage}[b]{0.14\textwidth}
    \includegraphics[width=1\linewidth]{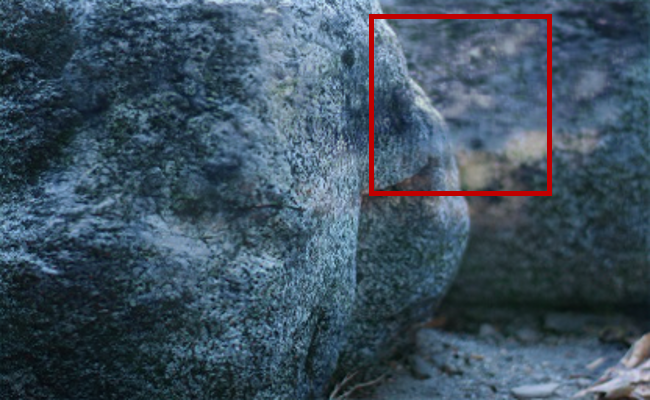}\vspace{1pt}
    \includegraphics[width=1\linewidth]{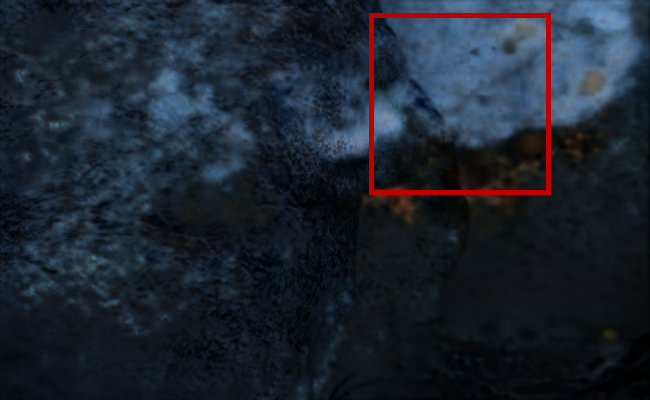}\vspace{2pt}
    \includegraphics[width=1\linewidth]{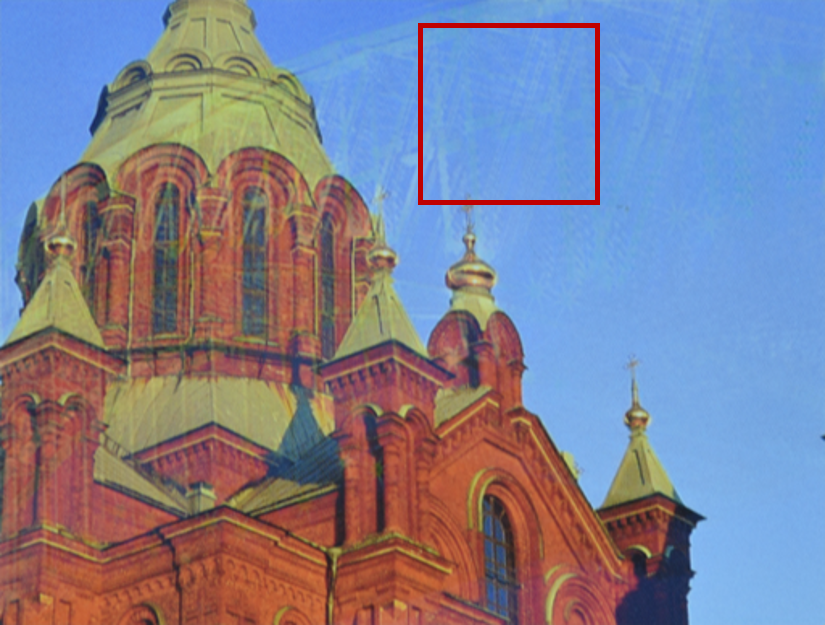}\vspace{1pt}
    \includegraphics[width=1\linewidth]{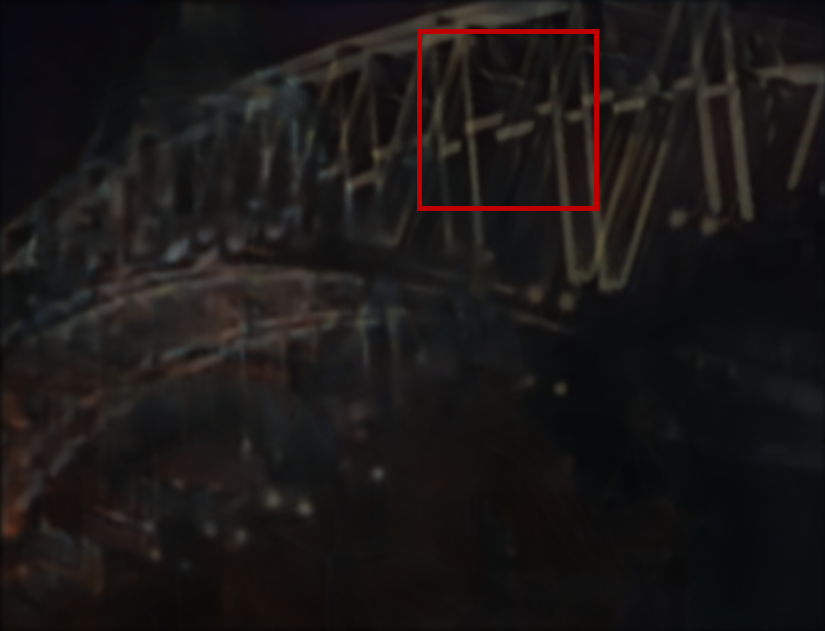}
    \end{minipage}
    }
    \subfigure[w/o AEA block]{
    \begin{minipage}[b]{0.14\textwidth}
    \includegraphics[width=1\linewidth]{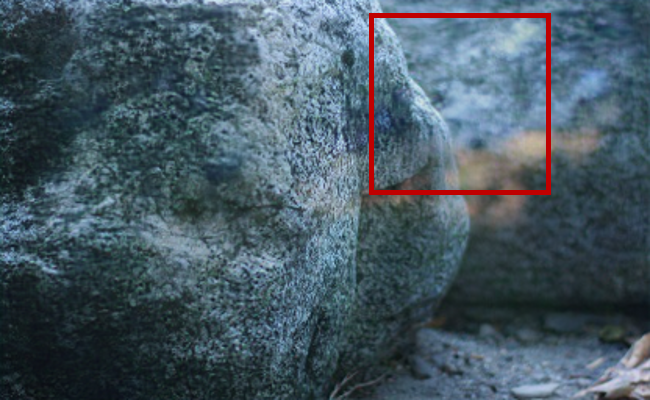}\vspace{1pt}
    \includegraphics[width=1\linewidth]{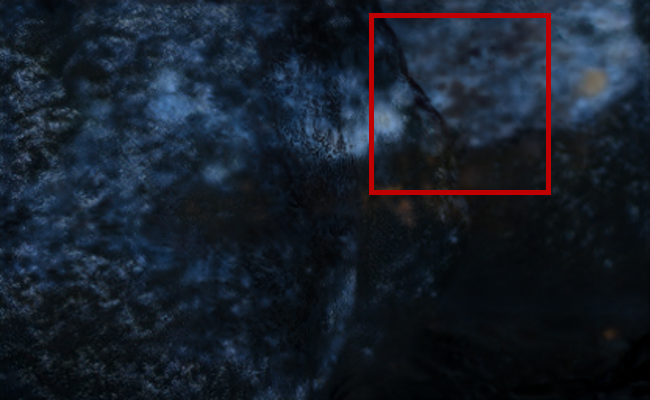}\vspace{2pt}
    \includegraphics[width=1\linewidth]{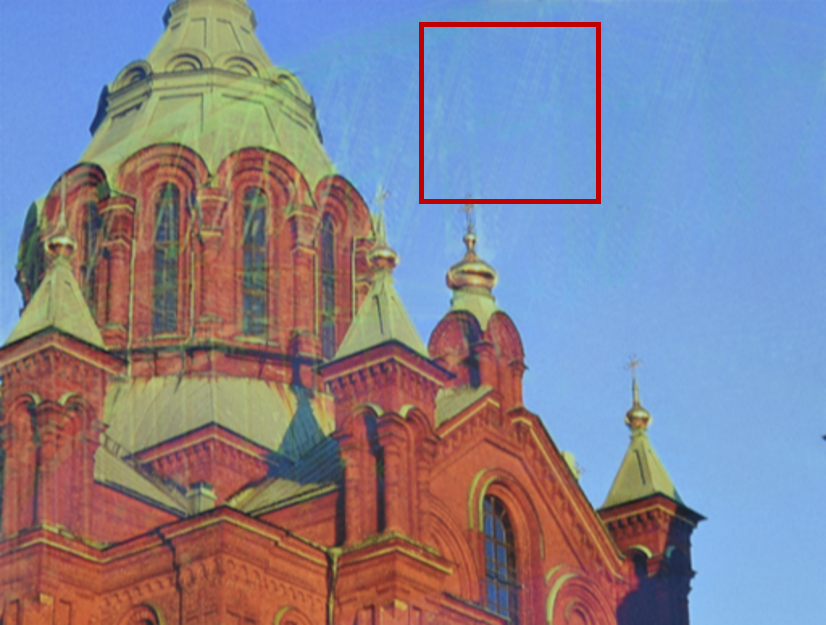}\vspace{1pt}
    \includegraphics[width=1\linewidth]{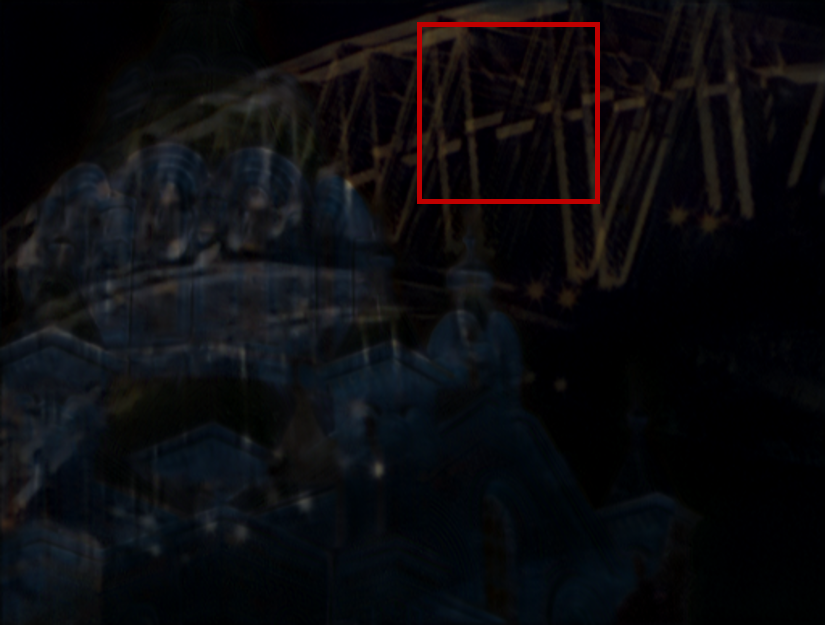}
    \end{minipage}
    }
    \subfigure[w/o {PRE block}]{
    \begin{minipage}[b]{0.14\textwidth}
    \includegraphics[width=1\linewidth]{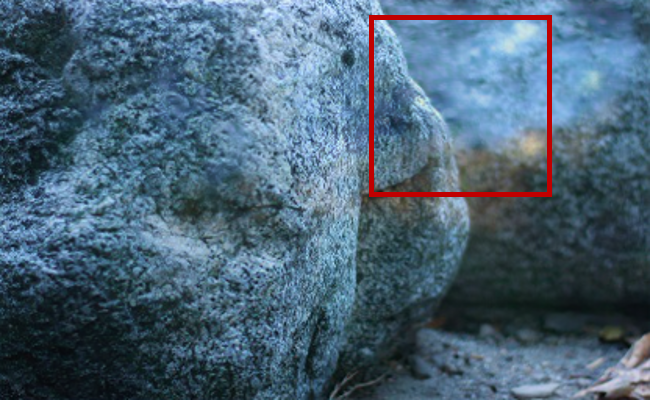}\vspace{1pt}
    \includegraphics[width=1\linewidth]{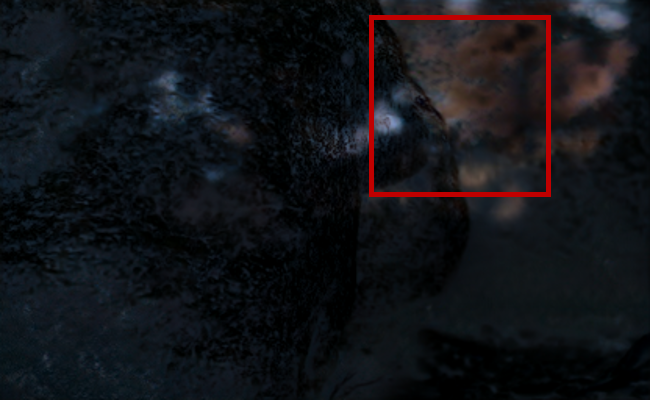}\vspace{2pt}
    \includegraphics[width=1\linewidth]{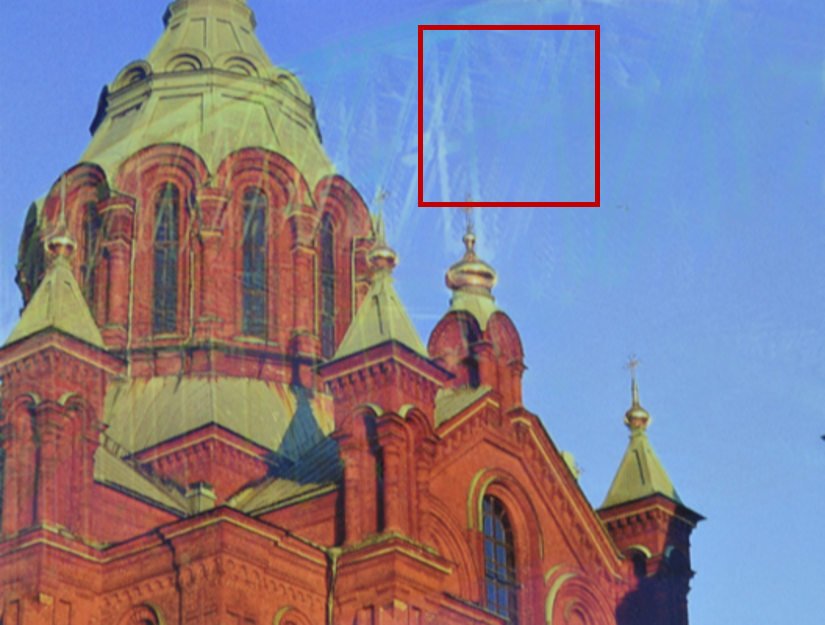}\vspace{1pt}
    \includegraphics[width=1\linewidth]{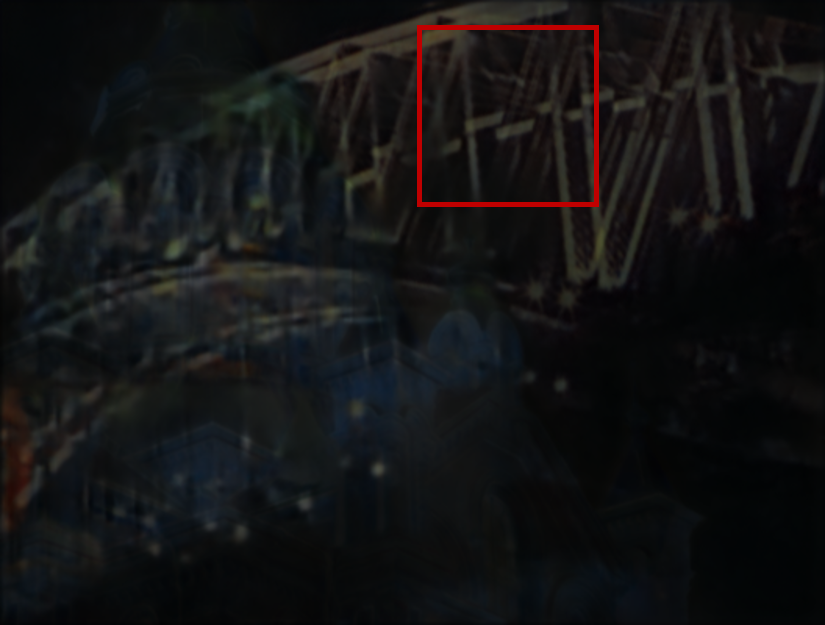}
    \end{minipage}
    }
    \subfigure[w/o {LPN block}]{
    \begin{minipage}[b]{0.14\textwidth}
    \includegraphics[width=1\linewidth]{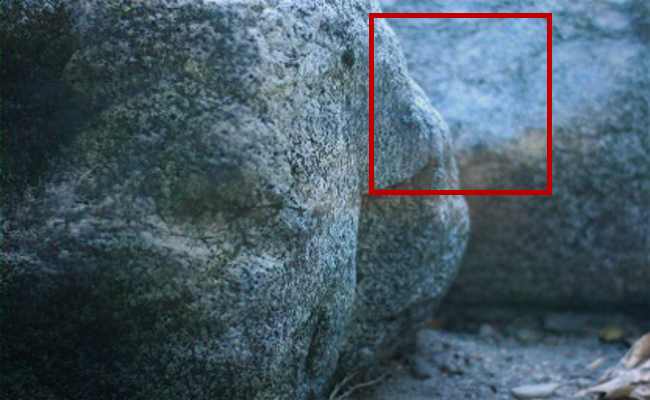}\vspace{1pt}
    \includegraphics[width=1\linewidth]{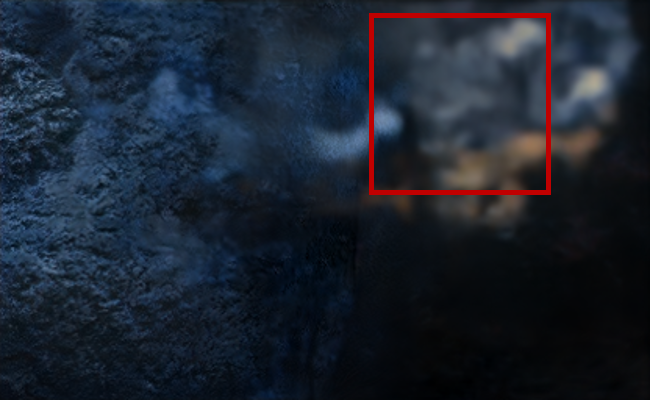}\vspace{2pt}
    \includegraphics[width=1\linewidth]{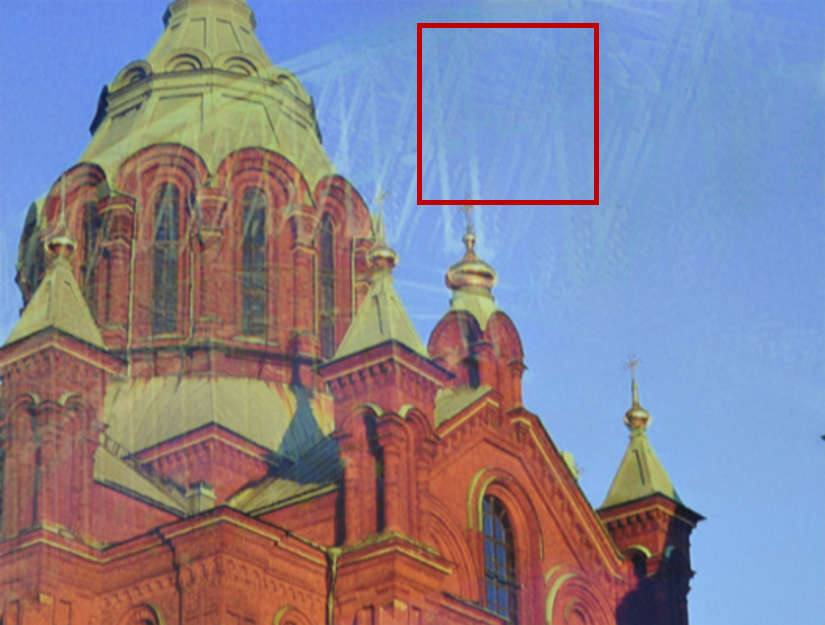}\vspace{1pt}
    \includegraphics[width=1\linewidth]{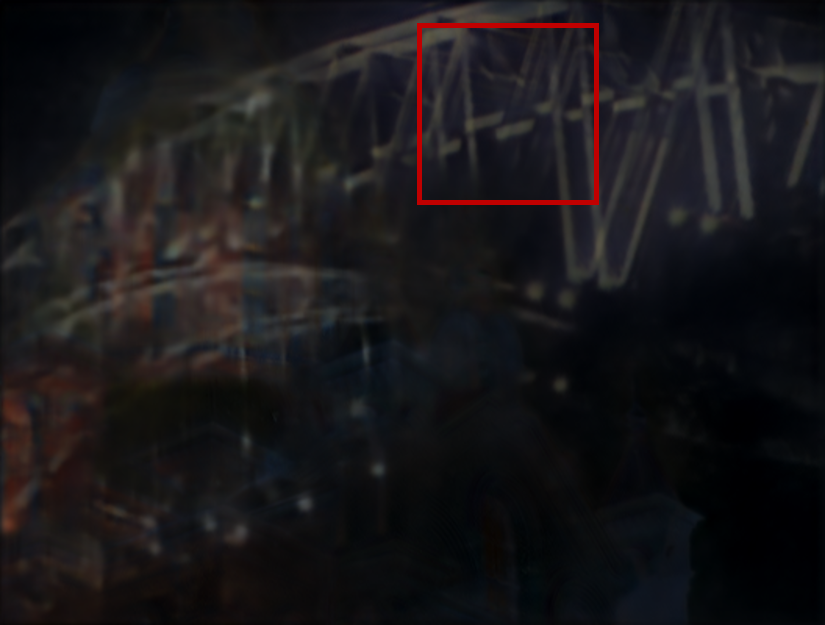}
    \end{minipage}
    }
    
    \caption{Ablation study: visualization of ablation studies on  AES block, AEA block, PRE block and LPN block in the proposed DExNet. The estimated transmission images are shown in the odd rows and the estimated reflection images are shown in the even rows.}
    \label{fig:ablation_Prox}
\end{figure*}
%-----------------------------------------------------------

Table \ref{tab:real20_SIR2} shows the quantitative results of the estimated \textit{transmission image} of different methods evaluated on \textit{Real20} dataset~\cite{perceptual_loss_2018}, and three subsets of the \textit{SIR${^2}$} dataset~\cite{wan2017benchmarking}.
We can see that overall the proposed DExNet achieves superior PSNR and SSIM results among all the comparison methods with a considerably small number of learnable parameters. 
Specifically, DExNet$_L$ with 9.66M parameters achieves an average of 25.96dB in PSNR which is 0.21dB higher compared to that of the SOTA method DSRNet with 124.60M parameters. Moreover, DExNet$_S$ with 4.52M parameters which is with $S=2$ scales of SAFU modules achieves a similar performance as DSRNet. 
Compared to the deep unfolding approaches MoG-SIRR~\cite{Shao2021MoGSIRR} and CGDNet~\cite{zhang2022content}, DExNet is with significantly better results. That could be due to the usage of the general exclusion prior and the novel auxiliary unfolding approach used in DExNet.
Considering that the testing images in \textit{SIR${^2}$} dataset are out-of-domain of the training data and are with diverse scenes, illumination conditions and glass conditions, it is highly challenging and remarkable for DExNet$_L$ to achieve the best average performance in all metrics. 

Table \ref{tab:real20_Reflection} shows the quantitative results of the estimated \textit{reflection image} of different methods evaluated on \textit{Real20} dataset~\cite{perceptual_loss_2018}, and three subsets of the \textit{SIR${^2}$} dataset~\cite{wan2017benchmarking}. We use the ground-truth reflection images of the  \textit{SIR${^2}$} dataset~\cite{wan2017benchmarking} for the reference images. Since there is no ground-truth reflection images in \textit{Real20} dataset~\cite{perceptual_loss_2018},  we use $\tilde{\mathbf{R}} = \mathbf{I} - \mathbf{T}$ for the reference images.
From the table, we can see that the proposed DExNet$_L$ and DExNet$_S$ achieves the best performance (25.70dB/0.574) and the second best performance (25.61dB/0.551) compared with SOTA methods in the task of estimating the reflection images, respectively. In comparison, DSRNet~\cite{Hu_2023_ICCV}, which achieves the second best overall results on recovering the transmission image, obtains an average of 25.00dB/0.507 in PSNR/SSIM on recovering the reflection image which is 0.70dB/0.078 inferior to that of DExNet$_L$. We can also observe that the proposed DExNet achieves a more balanced performance on recovering both the transmission and the reflection image (for instance 25.96dB v.s. 25.70dB), while the other methods tend to recover the transmission image well but are less effective in recovering the reflection image.

Fig. \ref{fig:Performance} compares the effectiveness, size and complexity of different methods on the task of estimating the transmission image. 
The computational complexity is measured on testing images of size $224\times 224$. 
We can clearly see that the proposed DExNet is highly effective and lightweight.
In terms of model effectiveness, DExNet and DSRNet achieve an average of PSNR over 25dB, while other methods scatter within $[23, 24]$dB.
In terms of model size, the two DExNet architectures are among those with the least number of learnable parameters. 
Specifically, DExNet$_L$, DURRNet and LASIRR are with a similar number of parameters (around 10M parameters), while DExNet$_L$ achieves around 1.9dB and 1.7dB higher PSNR than that of DURRNet and LASIRR, respectively.
In terms of model complexity, DExNet$_S$ and DExNet$_L$ are with relatively low computational complexity (around 110 GFLOPs and 170 GFLOPs) compared to other methods which are mostly within the range of $[400, 1000]$ GFLOPs. Compared to the SOTA method DSRNet, DExNet$_L$ only requires around 30\% complexity of that of DSRNet.
DExNet$_S$ and CoRRN are with the lowest complexity (around 110 GFLOPs), while DExNet$_S$ achieves 2.2dB higher PSNR compared to CoRRN. 
This result indicates that the proposed model-inspired DExNet achieves superior performance over SOTA methods with significantly smaller model size and lower computational cost.

Table \ref{tab:runtime} further demonstrates the computational complexity and the average runtime of different methods. We can see that the average runtime of DExNet$_L$ is around 0.15s which is similar to those of the SOTA methods. We also note that the computational complexity results and the runtime results are not entirely consistent. This could be due to different compatibility of the model to hardware.

%-----------------------------------------------------------
\begin{table}[t]
\caption{The computational complexity and the average runtime results of different methods.}
    \centering
    \begin{tabular}{c|c|c} \hline 
        \rowcolor[gray]{.9}
        \toprule[1pt] 
        \textbf{Method} &\textbf{GFLOPs} & \textbf{Runtime (s)} \\
        \hline 
        Zhang\textit{ et al.}~\cite{perceptual_loss_2018}   & 1755.09   & 1.328 \\
        BDN~\cite{yang2018seeing}                           & 164.20    & 0.006 \\
        Zheng \textit{et al.}~\cite{zheng2021single}        & 864.46    & 0.010\\
        CoRRN\cite{wan2019corrn}                            & 117.50    & 0.008 \\
        ERRNet\cite{wei2019single}                          & 679.60    & 0.014\\
        IBCLN~\cite{cascaded_Refine_2020}                   & 601.26    & 0.050\\
        YTMT ~\cite{hu2021trash}                            & 445.60    & 0.099\\
        LASIRR~\cite{dong2021location}                      & 513.30    & 0.076 \\
        DMGN~\cite{FengTIP2021}                             & 1574.91   & 0.015 \\
        RAGNet~\cite{LiAI2023}                              & 592.39    & 0.010 \\
        DSRNet~\cite{Hu_2023_ICCV}                          & 548.08    & 0.120 \\
        DURRNet~\cite{Huang24DURRNet}                       & 333.95    &{0.086}\\
        {DExNet$_S$ (Proposed) }                            & 111.00    & {0.103} \\
        {DExNet$_L$ (Proposed)}                             & 169.43    &{0.150}\\
        \hline
        \toprule[1pt]
 \end{tabular}
    \label{tab:runtime}
\end{table}
%-----------------------------------------------------------

\subsubsection{Visual Comparisons} This section presents the visual comparisons on \textit{Real20} dataset ~\cite{perceptual_loss_2018}, \textit{SIR${^2}$} dataset~\cite{wan2017benchmarking} and \textit{Real45} dataset~\cite{generic_smooth_2017}.

\noindent\textbf{Real20 Dataset:}
Fig. \ref{fig:real20} shows the estimated transmission and reflection images by different methods on three exemplar images from the \textit{Real20} dataset~\cite{perceptual_loss_2018}. The last column shows the ground-truth transmission images and the reflection images for reference.
We can see that the proposed DExNet can recover natural looking transmission and reflection images simultaneously. This could be due to the fact that the information of the image formation model and priors have been embedded properly into the network architecture. 
For comparison methods, Zhang \textit{et al.}'s method~\cite{perceptual_loss_2018} can well separate most reflections in the input, but sometimes generates images with visible artifacts.
IBCLN~\cite{cascaded_Refine_2020} struggles to separate well large overlapping reflections and  strong transmission image contents remain in the estimated reflection image.
ERRNet~\cite{wei2019single} achieves good separation results on the first two images but does not recover well the strong bright reflections.
YTMT~\cite{hu2021trash} and DSRNet~\cite{Hu_2023_ICCV} did not handle the complex reflections and image content details accurately and sometimes could be less effective in estimating the reflection images.
DURRNet~\cite{Huang24DURRNet} is less effective on separating detailed structures.

\noindent\textbf{SIR$^2$ Dataset:} Fig.~\ref{fig:sir2} shows three exemplar reflection separation results of different methods
from \textit{SIR$^2$} dataset~\cite{wan2017benchmarking}. The first and the last columns show the input blended images and the ground-truth transmission images, respectively. 
There are clearly visible reflection elements in the estimated transmission results of Zhang \textit{et al.}~\cite{perceptual_loss_2018}, IBCLN~\cite{cascaded_Refine_2020}, ERRNet~\cite{wei2019single}, YTMT~\cite{hu2021trash} and DURRNet~\cite{Huang24DURRNet}.
We also note that DSRNet~\cite{Hu_2023_ICCV} demonstrates better separation results, but still leaves noticeable shared contents in the estimated transmission and reflection images. Compared with the other methods, our proposed approach achieves more accurate separation of reflection content and transmission images without noticeable shared contents. 

\noindent\textbf{Real45 Dataset:}
Fig. \ref{fig:real45} further shows the visualization results of three exemplar images from the \textit{Real45} dataset~\cite{generic_smooth_2017} of which the ground-truth images are unavailable. Therefore, it is a good testing venue for the model generalization ability. The first column shows the input images, and from column 2 to column 6, the odd and the even rows show the estimated transmission and reflection images, respectively. We can see that Zhang \textit{et al.}'s method~\cite{perceptual_loss_2018} in general can separate the transmission and reflection images but the separated reflection images usually still contain information from the transmission layer. 
IBCLN~\cite{cascaded_Refine_2020} does not generalize well on the \textit{Real45} dataset and cannot well separate two image layers. ERRNet~\cite{wei2019single} also does not recover effectively the reflection layer. YTMT~\cite{hu2021trash} does not produce reflection images with intact contents.
DURRNet~\cite{Huang24DURRNet} can well handle strong reflections, while is less effective on finer details.
DSRNet~\cite{Hu_2023_ICCV} shows better separation capability but the estimated reflection images are with large blurred areas. 
In the last column, we can see that the proposed DExNet can properly separate the reflection image content from the input reflection-contaminated image. The reflection images contain more intact reflection contents and have little information coming from the transmission image.

\subsection{Ablation Studies}
\label{sec:ablation}

In this section, we perform ablation studies to investigate the performance improvement in terms of network architectures and loss functions.

%-----------------------------------------------------------
\begin{table}[t]
\center
\caption{Ablation study: the effectiveness of AES block, AEA block, PRE block and LPN block. The number of scales and the number of stages is set to $S=4$  and $K=5$, respectively.}

\begin{tabular}{c c c c|c|c|c}
\rowcolor[gray]{.9}
\toprule[1pt] 
\textbf{AES} &\textbf{AEA} & \textbf{PRE} & \textbf{LPN} & \textbf{Model Size} & \textbf{PSNR} $\uparrow$  & \textbf{SSIM} $\uparrow$   
\\ \hline  
\Checkmark &\Checkmark      & \Checkmark    & \Checkmark         & 9.66M   & 25.96 & 0.912\\ 
\textcolor{gray}{\XSolidBrush} &\Checkmark     & \Checkmark    & \Checkmark   & 9.28M        &25.28 & 0.904\\ 
\Checkmark &\textcolor{gray}{\XSolidBrush}     & \Checkmark    & \Checkmark   & 9.66M        &25.26& 0.907\\ 
\textcolor{gray}{\XSolidBrush} &\textcolor{gray}{\XSolidBrush}     & \Checkmark    & \Checkmark  & 8.91M  &  25.09  & 0.903\\ 
\Checkmark &\Checkmark      & \textcolor{gray}{\XSolidBrush}   & \Checkmark   &9.45M        &25.56& 0.907\\ 
\Checkmark &\Checkmark      & \Checkmark    & \textcolor{gray}{\XSolidBrush}  &5.43M        & 24.50 & 0.900 \\ 
\bottomrule[1pt]

\end{tabular}
\label{tab:ablation-net}
\end{table}
%-----------------------------------------------------------

\subsubsection{Effectiveness of SAFU Module} 
Table \ref{tab:ablation-net} shows the ablation study on the key network components of SAFU module including AES, AEA, PRE and LPN blocks evaluated on 474 image from \textit{Real20} dataset~\cite{perceptual_loss_2018} and \textit{SIR${^2}$} dataset~\cite{wan2017benchmarking}. 
From the table, we can see that the model-inspired AES, AEA and PRE blocks only require less than 4\% of the overall parameters, but can significantly improve the model performance. 
Among them, AES block and AEA block impose the exclusion condition by performing element-wise multiplication between features of transmission and reflection image, and enable identification of the shared feature contents. The AEA blocks use the same parameter as AES blocks.
We can see that both AES block and AEA block are essential to the proposed SAFU module. Removing either block will only save a marginal number of parameters, but lead to around 0.7dB performance degradation in PSNR. When both AES blocks and AEA blocks are removed, this is equivalent to a model without the exclusion prior and there is around 0.9dB drop in PSNR.
This confirms that introducing modules that handle the interplay between common features across ${\mathbf{T}}$ and ${\mathbf{R}}$ is essential
to the performance of DExNet. The PRE block only requires around 2\% of the parameter and imposes the reconstruction condition for the estimated transmission, reflection and nonlinear residual images. It helps to stabilize the training process and improve average PSNR by around 0.4dB. The LPN block is used to learn and impose the sparse prior and general exclusion prior. We can see that applying LPN blocks for prior learning can improve the model performance by around 1.5dB.
To further visualize the functionality of AES block, ACN block, PRE block and LPN block, Fig. \ref{fig:ablation_Prox} shows the single image reflection removal results of the complete model of DExNet, and the DExNet w/o AES block, w/o ACN block, w/o PRE block and w/o LPN block.

\subsubsection{Effectiveness of Multi-scale Architecture} 
The proposed DExNet consists of $S$ scales of SAFU modules to mimic the multi-scale exclusion loss and progressively estimate the transmission image and the reflection image from low-resolution scales to high-resolution scales. 
In Table \ref{tab:ablation-scales}, we further analyze the effectiveness of the number of scales $S$ in DExNet by fixing the number of stages in each scale at $K=5$. From the table, we can observe that there is a boost in model performance by increasing the number of scales in DExNet. Moreover, the computational cost is marginally increased since the size of the feature maps are halved for each lower scale.
This indicates that the multi-scale architecture can effectively and efficiently integrate information from different scales and lead to better performance. 
In Table \ref{tab:ablation-scales}, we show the ablation study results with different number of stages in each scale by fixing the scale number $S=4$. From the table, we can see that by increasing the number of stages from 1 to 5, there is a close to linear increase in model performance.
Therefore, we set the number of scales $S=4$ and the number of stages $K=5$ by default.

%-----------------------------------------------------------
\begin{table}[t]
\center
\caption{Ablation study: the number of scales. The number of stages is set to $K=5$.}

\begin{tabular}{c|C{1.8cm}|C{1.5cm}|C{1.5cm}}
\rowcolor[gray]{.9}
\toprule[1pt] 
\textbf{Scales} & \textbf{Model Size} & \textbf{PSNR} $\uparrow$ & \textbf{SSIM} $\uparrow$
\\ \hline 
1 & 2.18 M  & 24.87 & 0.899\\
2 & 4.52 M  & 25.66 & 0.907\\
3 & 6.87 M  & 25.31 & 0.905\\
4 & 9.66 M  & 25.96 & 0.912\\
\bottomrule[1pt]
\end{tabular}
\label{tab:ablation-scales}
\end{table}
%-----------------------------------------------------------

%-----------------------------------------------------------
\begin{table}[t]
\center
\caption{Ablation study: the number of stages. The number of scales is set to $S=4$.}

\begin{tabular}{c|C{1.8cm}|C{1.5cm}|C{1.5cm}}
\rowcolor[gray]{.9}
\toprule[1pt] 
\textbf{Stages} & \textbf{Model Size} & \textbf{PSNR} $\uparrow$ & \textbf{SSIM} $\uparrow$
\\ \hline 
1 & 5.26 M &  25.09  &0.903\\
2 & 6.55 M & 25.34 & 0.907\\
3 & 7.58 M &  25.53& 0.909\\
4 & 8.62 M & 25.65 & 0.908\\
5 & 9.66 M  & 25.96 & 0.912\\
\bottomrule[1pt]
\end{tabular}
\label{tab:ablation-scales}
\end{table}
%-----------------------------------------------------------

%-----------------------------------------------------------
\begin{figure}[t]
    \centering
    \includegraphics[width=1\linewidth]{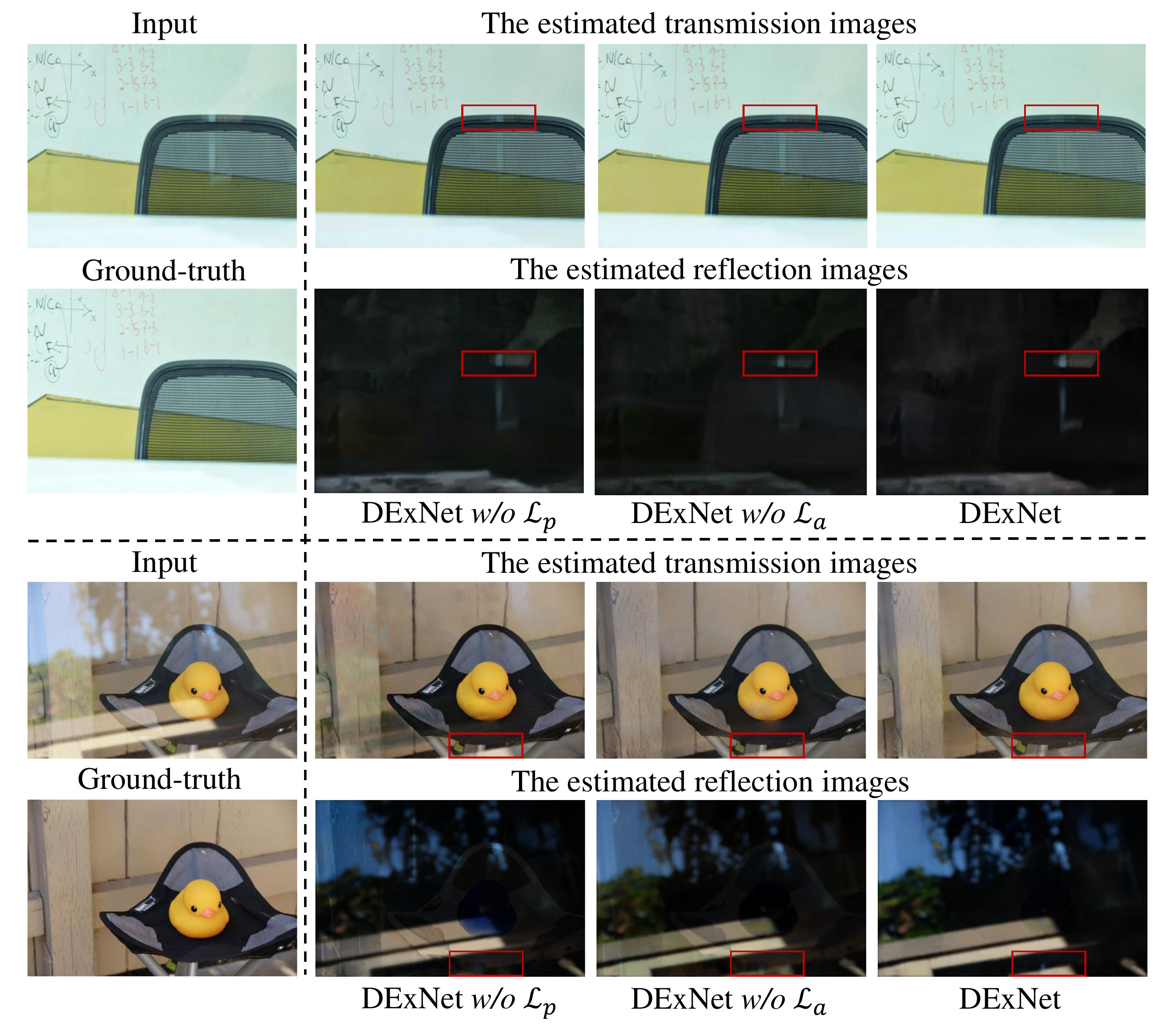}    
    \caption{Ablation study on loss functions.}
    \label{fig:ablation_loss}
\end{figure}

%-----------------------------------------------------------
\begin{table}[t]
\center
\caption{Ablation study: the effectiveness of the loss functions. The number of scales and the number of stages is set to $S=4$  and $K=5$, respectively.
}

\begin{tabular}{c c c |C{1.5cm}|C{1.5cm}}
\rowcolor[gray]{.9}
\toprule[1pt] 
$\mathcal{L}_r$ &$\mathcal{L}_p$  & $\mathcal{L}_a$ & \textbf{PSNR} $\uparrow$  & \textbf{SSIM} $\uparrow$  
\\ \hline  
\Checkmark      & \Checkmark         & \Checkmark    &  25.96    & 0.912\\ 
\Checkmark      & \textcolor{gray}{\XSolidBrush}         & \Checkmark   &   24.73   & 0.894 \\
\Checkmark      & \Checkmark         & \textcolor{gray}{\XSolidBrush}  &  25.35    & 0.906 \\
\bottomrule[1pt]
\end{tabular}
\label{tab:ablation-loss}
\end{table}
%-----------------------------------------------------------

\subsubsection{Effectiveness of Loss Functions} 
We perform ablation study to analyze the effectiveness of different loss functions.
The proposed DExNet utilizes the reconstruction loss $\mathcal{L}_r$, the perceptual loss $\mathcal{L}_p$ and the auxiliary loss $\mathcal{L}_a$ for training. 
Since the reconstruction loss cannot be removed, we focus on analyzing the effectiveness of perceptual loss and auxiliary loss.
The ablation results are shown in Table~\ref{tab:ablation-loss}. We can find that the perceptual loss can effectively improve the PSNR performance by around 1.2dB. This validates that the feature loss from the VGG network can help improve both perceptual and quantitative performance. 
The auxiliary loss can help to improve the PSNR by around 0.6dB. 
Furthermore, Fig.~\ref{fig:ablation_loss} visualizes two examples of the ablation study results on loss functions. We can see that the perceptual loss can help to identify and remove semantically difficult cases, and the auxiliary loss can help to further remove the fine-grained reflection contents.

\section{Conclusions}
\label{sec:conclusion}

In this paper, we introduced Deep Exclusion  unfolding Network (DExNet), a novel and lightweight model-inspired network for single image reflection removal. The proposed DExNet integrates a simple yet accurate image formation modelling and a general exclusion prior into the deep architecture which enables accurate joint reflection and transmission image estimation and rich feature interactions between two image layers to effectively diminish the presence of common features. 
We first formulate the reflection removal problem as a model-based Convolutional Sparse Coding problem which further includes a general exclusion prior to identify and penalize the commonalities between features of two image layers so as to facilitates the separation of two image layers. Our contribution extends to the introduction of the iterative Sparse and Auxiliary Feature Update (i-SAFU) algorithm. By unfolding and parameterizing the i-SAFU algorithm, the proposed DExNet is constructed using multiple scales of Sparse and Auxiliary Feature Update (SAFU) Modules which has exact one-to-one relationship with the iterative algorithm, ensuring a seamless integration of model-based and learning-based approaches. The SAFU module comprises three key components: Auxiliary Equality for Sparse features (AES) blocks, Auxiliary Equality for Auxiliary features (AEA) blocks, and Projected Reconstruction Error (PRE) blocks. These components are crucial to the network's functionality and operate with a marginal parameter footprint. By combing the merits of model-based and learning-based paradigms, DExNet achieves state-of-the-art performance on single image reflection removal. Notably, DExNet stands out for its lightweight nature and enhanced interpretability.

In the future, it would be of great interests to apply and adapt the proposed DExNet to a broader spectrum of blind image and signal separation tasks, such as image deraining, image dehazing and hyperspectral image unmixing.

\ifCLASSOPTIONcaptionsoff
  \newpage
\fi

\bibliographystyle{IEEEtran}
\bibliography{bibs}

\end{document}